%% file: main.tex
\documentclass[aps,prd,twocolumn,superscriptaddress,preprintnumbers]{revtex4-2}
\usepackage{blindtext}

\usepackage[utf8]{inputenc} % allow utf-8 input
\usepackage[T1]{fontenc}    % use 8-bit T1 fonts
\usepackage{url}            % simple URL typesetting
\usepackage{booktabs}       % professional-quality tables
\usepackage{amsfonts}       % blackboard math symbols
\usepackage{nicefrac}       % compact symbols for 1/2, etc.
\usepackage{microtype}      % microtypography
\usepackage{lipsum}
\usepackage{graphicx}
\usepackage{hyperref}       % hyperlinks
\usepackage{subcaption}
\usepackage{amssymb}
\usepackage{amsthm}
\usepackage{mathtools}
\usepackage{float}
\begin{document}
%\pacs{}

%\title{The measurement of proton-carbon forward scattering at $20$, $30$, and $120$ GeV/c in EMPHATIC experiment}
\title{A measurement of proton-carbon forward scattering in a proof-of-principle test of the EMPHATIC spectrometer}
\input{authors}

%\author{\include{authors}}
%\author{test}
\preprint{FERMILAB-PUB-21-273-E-ND-SCD}

\begin{abstract}
The next generation of long-baseline neutrino experiments will be capable of precision measurements of neutrino oscillation parameters, precision neutrino-nucleus scattering, and unprecedented sensitivity to physics beyond the Standard Model.  %enter a precision era. %One of the limiting factors in the neutrino measurements comes from the poor knowledge of the neutrino flux. 
Reduced uncertainties in %Improvements in %knowledge of the 
neutrino fluxes are necessary to achieve high precision and sensitivity in these future precise neutrino measurements.
%and Neutrinos are decay products of hadrons created in hadron collisions with nuclei in production targets in accelerators or the atmosphere. 
New measurements of hadron-nucleus interaction cross sections are needed to reduce uncertainties of neutrino fluxes.  We report measurements of the differential cross section as a function of scattering angle for proton-carbon interactions with a single charged particle in the final state %produced at angles less than $20\:$mrad with respect to the incident proton 
at beam momenta of $20$, $30$, and $120\:$GeV/c. These measurements are the result of a beam test for EMPHATIC, a hadron-scattering and hadron-production experiment.  The total, elastic and inelastic cross-sections are also extracted from the data and compared to previous measurements. These results can be used in current and future long-baseline neutrino experiments, and demonstrate the feasibility of future measurements by an upgraded EMPHATIC spectrometer.  
\end{abstract}

\maketitle

\section{Introduction}

Measurements of hadron interactions spanning two orders of magnitude ($1-100\:$GeV/c) of incident particle momenta are of crucial importance for reducing the neutrino production modeling uncertainty in accelerator-based and atmospheric neutrino experiments.  The neutrino flux uncertainty is the dominant uncertainty in many neutrino measurements, including neutrino nucleus cross-section measurements, sterile neutrino searches, and CP violation measurement in atmospheric neutrinos.
Long-baseline neutrino experiments are entering a precision era with the future Hyper-Kamiokande ~\cite{H2KDesignReport} and DUNE~\cite{DUNETDR} projects. %In these experiments, the neutrino oscillation probability is extracted by comparing near and far detector event rates. The energy dependence of the neutrino event rates is necessarily different between near and far detectors due to neutrino oscillations. 
Uncertainties in the energy dependence of the neutrino flux and cross section are amongst the most challenging systematic uncertainties in these neutrino experiments. 

Neutrinos are produced by the decays of hadrons produced in proton interactions in nuclei. Since it is at best extremely difficult and time consuming to measure the neutrino beam flux as a function of energy, Monte Carlo simulations based on hadron interactions and decays are used to make a-priori predictions of the neutrino flux.
%Neutrino flux prediction is also essential in measuring the neutrino cross sections in the near detector.
%The most direct way to predict the neutrino flux is to use the 
%Monte Carlo simulation based on the hadron interaction and decays. 
This approach is limited by the sparse hadron interaction data often with significant errors, and hadron interactions are the dominant systematic uncertainty in the neutrino flux prediction. 
Interpolation and extrapolation of the hadron interaction cross section using phenomenological models introduces additional uncertainties.
Measurements of hadron interactions are used to constrain or scale the models to provide a more precise prediction of the neutrino flux. A good example of how data are used in neutrino flux simulations can be found in ~\cite{Abe:2012av}. 

%The other methods are generally used to constrain our MC predictions.  For example, in-situ measurements of the processes with well predicted cross sections~\cite{Valencia:2019mkf} like neutrino-electron elastic scattering cross scattering. Unfortunately, the neutrino-electron cross section is very small and it is challenging to accumulate sufficient statistics for the long baseline neutrino experiments. Neutrino interactions with low energy transfer to charged leptons (Low$\nu$) provide another approach. The challenge of the Low$\nu$ method is to control the remaining systematic uncertainty in the neutrino cross section. Even these methods require accurate neutrino beam simulations.

%\subsection{Existing hadron production data}
Many of the hadron interaction data relevent to GeV-energy neutrino flux predictions were taken in the second half of twentieth century. Although these data are very valuable, they are insufficient for the precise neutrino flux predictions due to low precision and lack of error covariances. In more recent years, experiments like NA61/SHINE~\cite{Abgrall:2015hmv, Abgrall:2016jif, Aduszkiewicz:2019xna, Berns:2018tap}, HARP~\cite{Apollonio:2009bu, Apollonio:2009en} or MIPP~\cite{MIPPNuMI, Lebedev:2007zz, Seun:2007zz} took valuable data with the direct requests and input from the neutrino experiments. These data include cross-section and hadron production measurement for various targets and beam momenta. 
Even with these data, the typical neutrino flux uncertainty in the current generation of the accelerator-based neutrino experiments is between $5\%$ and $15\%$ due to limited phase space coverage and other sources of systematic uncertainties.

%\subsection{A case for the additional data}

The neutrino flux uncertainty directly affects all measurements done in a single (near) detector where far-to-near detector cancellation is not possible.
However the reduction of the neutrino flux uncertainty is also important for neutrino oscillation measurements, in particular measurements of CP violation in the lepton sector. 
Constraining the $\nu_e/\nu_{\mu}$ ratio and measuring $\nu_e$ cross-section is of utmost importance for reducing the systematic uncertainty in CP violation measurements in the Hyper-Kamiokande and DUNE experiments \cite{H2KDesignReport,DUNESensStudies}. These cross sections  will be measured in the intermediate water Cherenkov detector (IWCD) in Hyper-Kamiokande and the DUNE near detector. However, these measurement is limited by the neutrino flux uncertainty and reduction of the flux uncertainty to levels of $3\%$ are necessary. 
Another example is the measurement of CP violation in atmospheric neutrinos, which is limited by the neutrino flux uncertainty coming from the sub-$20\:$GeV/c cosmic ray interactions with the atmosphere.

Several missing pieces of data are necessary to reduce the neutrino flux uncertainty in current and future accelerator-based and atmospheric neutrino experiments: 
\begin{enumerate}
	\item hadron production in sub-$10\:$GeV pion and kaon interactions on carbon, aluminium, titanium and iron,
	\item hadron production in sub-$20\:$GeV  proton-air interactions (or equivalent targets),
	\item measurements of coherent elastic and quasi-elastic interactions of hadrons for carbon, aluminium, titanium and iron targets between $1\:$GeV/c and $120\:$GeV/c,
	\item measurements of strange hadron production in proton-carbon interactions to validate older measurements.  
\end{enumerate}
The detailed explanation of each point is out of scope of this paper and can be found in ~\cite{Akaishi:2019dej}. This paper describes a measurement of (3) above which is possible using data from an early beam test of EMPHATIC (Experiment to Measure the Production of Hadrons At a Testbeam in Chicagoland). Future EMPHATIC measurements will address the remaining items in the list.

\section{EMPHATIC experiment}
EMPHATIC is designed to study hadron interactions in the $2-120\:$GeV/c range at the Fermilab Test Beam Facility (FTBF). The physics program of EMPHATIC covers previously listed requirements for the improvement of the neutrino flux in the upcoming long-baseline neutrino oscillation experiments~\cite{Akaishi:2019dej}.

The EMPHATIC design exploits \textasciitilde$ 10\:\mu$m spatial resolution of silicon strip detectors that results in a compact hadron spectrometer. The dipole magnet is a custom-built Halbach array of NdFeB magnets with a $\int{Bdl}$ of approximately $0.25\:$Tm.
Beam particle identification is done using gas and aerogel threshold Cherenkov detectors. Identification of secondary particles is done by time-of-flight measurements in resistive plate chambers (RPCs) and by measuring the Cherenkov angle in aerogel ring imaging detector (ARICH) based on the Belle-II design~\cite{Burmistrov:2020dvn}. A lead calorimeter at the downstream end of the experiment enables separation of electrons, muons and hadrons. The total length of the spectrometer is approximately $2\:$m. 

%Before building the EMPHATIC experiment we have tested the silicon strip detectors and prove that they can replace large hadron spectrometers used in previous experiments.  

In this paper, we present the results of an EMPHATIC test-beam measurement done in January 2018 using only silicon strip detectors to record particle trajectories.  No magnet and no detectors for secondary particle identification were used in the test-beam setup.
The results include differential cross-section for p+C$\rightarrow$X$^{\pm}$ at $20$, $30$, and $120\:$GeV/c, where X$^{\pm}$ is a single charged particle  within 20 mrad, the acceptances of the tracking detectors used in this measurement. This enables the measurements of forward scattering (coherent elastic and quasi-elastic interactions). 
The event topology with a single forward charged particle includes coherent-elastic interactions, quasi-elastic interactions and some inelastic interactions. Coherent elastic interactions are defined as scattering off the whole nucleus. Both the incident particle and the target nucleus survive in this case. In quasi-elastic interactions, the target nucleus is fragmented while the incident hadron survives. And finally, inelastic interactions are those in which at least one new meson is produced. 
We have also fitted a simple model to our data to extract useful quantities such as total and elastic cross-section. Results are compared to both Monte Carlo predictions and other measurements.

%The structure of this paper is as follows: in the section~\ref{sec:setup}, experimental setup is summarized. In section~\ref{sec:ana}, data analysis is discussed. Results are presented in section~\ref{sec:res} followed by the summary.

\section{Experimental setup}
\label{sec:setup}
%\subsection{Beam}
The EMPHATIC test-beam measurements were done at the FTBF. The facility provides a primary $120\:$GeV/c proton beam from the Main Injector, or a secondary beam with momentum above $2\:$GeV/c.% and $80\:$GeV/c. 
The beam is delivered in $4\:$s spills every minute. The intensity is tunable from $1\:$kHz to $100\:$kHz and the typical beam spot area is $2\:$cm$^2$. The momentum resolution of the secondary beam ($\Delta p/p$) is approximately $2\%$. 
The FTBF provides a set of gas threshold Cherenkov detectors for secondary beam particle identification. Pion identification is possible above $5\:$GeV/c, while kaon identification is only possible above $18\:$GeV/c. 
The pressure in the first Cherenkov detector was tuned to detect positrons, muons and pions in the secondary beam. The trigger includes signals from the first Cherenkov detector in the anti-coincidence with two scintillators to remove all particles except kaons and protons. The pressure in the second gas Cherenkov detector was tuned to detect kaons, which can be separated from the protons during the data analysis.
%\subsection{Tracking}

\begin{figure*}[!t]
\begin{center}
 %\begin{subfigure}[t]{1\textwidth}
% 		\includegraphics[trim= 20 40 60 20, clip, %width=1\textwidth]{figures/setup.png}
%    %\caption{ }
%  \end{subfigure}
%  \begin{subfigure}[t]{1\textwidth}
 		\includegraphics[trim= 20 20 60 20, clip, width=1\textwidth]{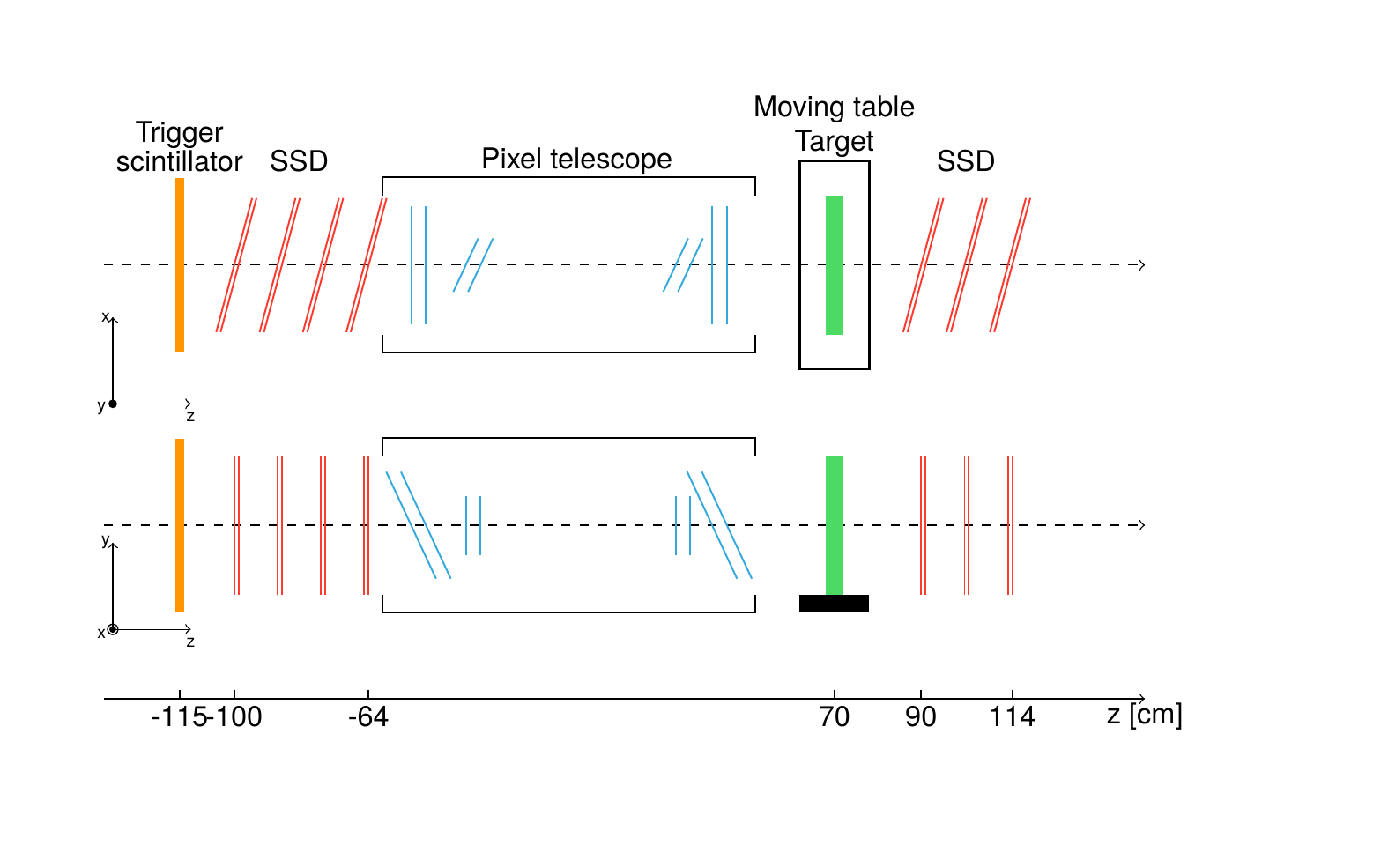}
    %\caption{ }
 % \end{subfigure}
    \end{center}
%	\caption{A photo of the experimental setup used for this measurement (a) and a schematic overview of the setup (b). The first trigger scintillator and gas Cherenkov detectors are located several meters upstream, and are not shown.}\label{fig:setup}
	\caption{Schematic overview of the experimental setup. The first trigger scintillator and gas Cherenkov detectors are located several meters upstream, and are not shown.}\label{fig:setup}
\end{figure*}

The FTBF also provides a set of silicon strip detectors (SSDs) with effective area of $3.8\times 3.8\:$cm$^2$ and the strip pitch of $60\:\mu$m. Each detector has two silicon strip planes for measuring two independent dimensions. For the EMPHATIC beam test, four detectors were placed upstream of the target and three detectors were placed downstream of the target. In addition to silicon strip detectors, the FTBF provides a silicon pixel telescope consisting of eight pixel planes located between four upstream SSDs and the target. Four of the planes have sensitive area of $3.24\times 1.62\:$cm$^2$ and the other four have an area of $1.62\times 2.43\:$cm$^2$. Due to inefficiencies with the pixel data acquisition and the smaller effective area, data from the pixels are not used for this measurement. Instead, the pixel telescope is treated as a passive material in the beamline and data are corrected for energy losses in this material. The schematic overview of the setup with defined coordinate system is shown in Fig.~\ref{fig:setup}.

%\subsection{Target}

The target is made of Toyo Tanso IG-$43$ graphite which is the target material of choice in the T2K beamline. The same graphite was used for NA61/SHINE measurements. The target thickness is $2\:$cm which is approximately $5\%$ of the interaction length. The measured target density is $1.83\pm 0.04\:$g/cm$^3$. 

\section{Simulation and Data analysis}\label{sec:ana}
\begin{table}
  \centering
  \begin{tabular}{ccc}
    \toprule
    p [GeV/c] & Target & Number of triggers [$10^6$] \\
    \midrule
    $20$ & carbon & $0.463$ \\
    $20$ & empty & $0.410$ \\
    $30$ & carbon & $1.031$ \\
    $30$ & empty & $0.197$ \\
    $120$ & carbon & $1.013$ \\
    $120$ & empty & $1.068$ \\
    \bottomrule
  \end{tabular}
  \caption{Collected number of triggers}\label{tab:events}
\end{table}
The data collected with graphite and empty targets are summarized in Tab.~\ref{tab:events}. The empty target data are used for silicon strip alignment and for estimating background interactions in the data analysis. The alignment is done by selecting $10000$ empty target events with a single hit per silicon strip plane. A simple line fit is used to fit a track in each event. The tracking plane positions and angles are determined by minimizing the sum of $\chi^2$ values for all $10000$ track fits. Any position misalignment is at sub-micron level and any angular misalignment is below $0.01\:$mrad.

After the alignment, angular resolution is determined for each dataset by fitting upstream and downstream tracks separately and calculating the angle between them in x-z and y-z planes. A Gaussian is fit to the angular distributions and the width divided by $\sqrt{2}$ is taken as the effective angular resolution. The effective angular resolution includes intrinsic angular resolution of the detector and smearing effect caused by multiple scattering in the detector material.

\begin{figure*}[htb]
\begin{center}
 \begin{subfigure}[t]{0.48\textwidth}
 		\includegraphics[trim= 20 20 60 20, clip, width=1\textwidth]{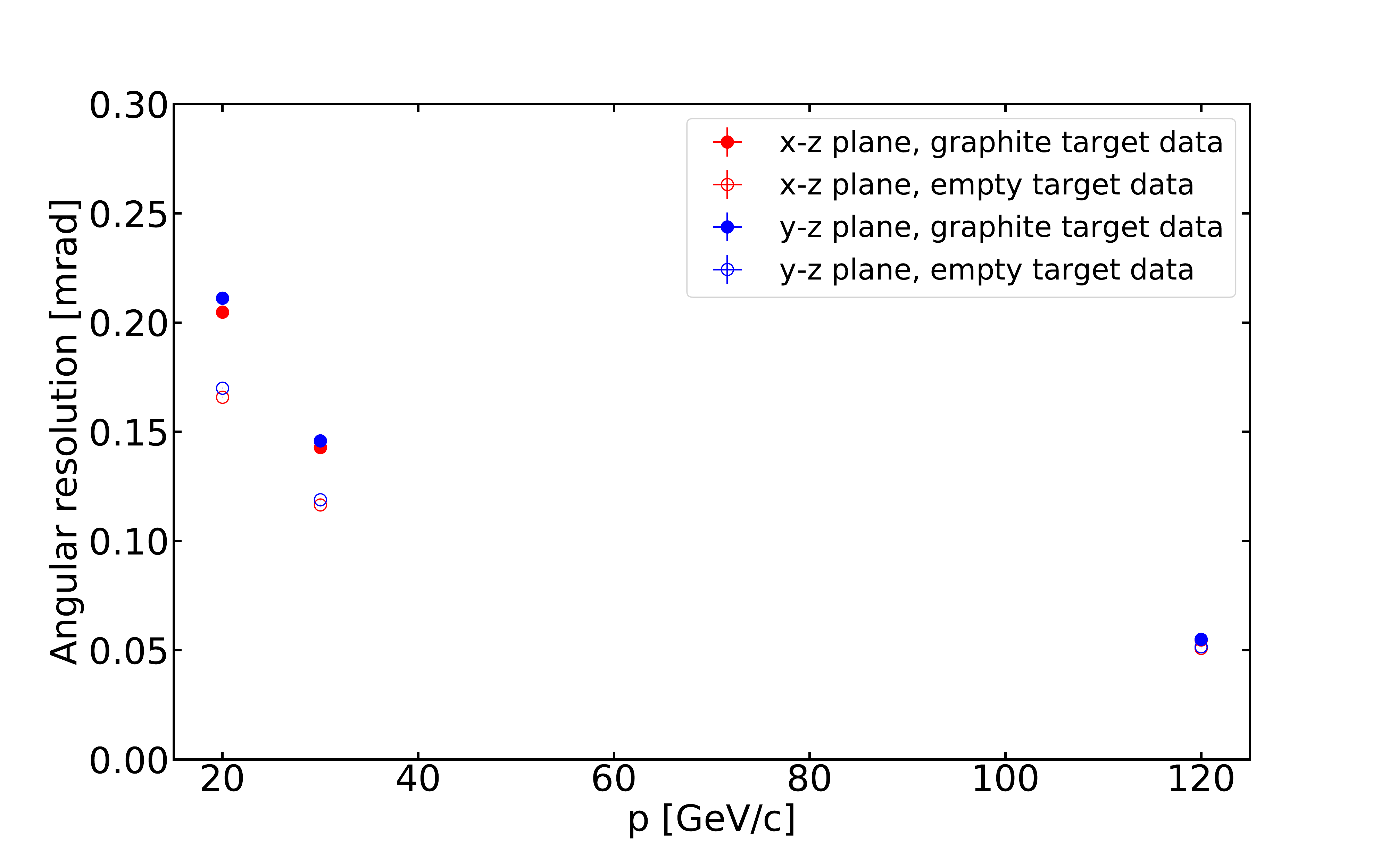}
    \caption{}
  \end{subfigure}
  \begin{subfigure}[t]{0.48\textwidth}
 		\includegraphics[trim= 20 20 60 20, clip, width=1\textwidth]{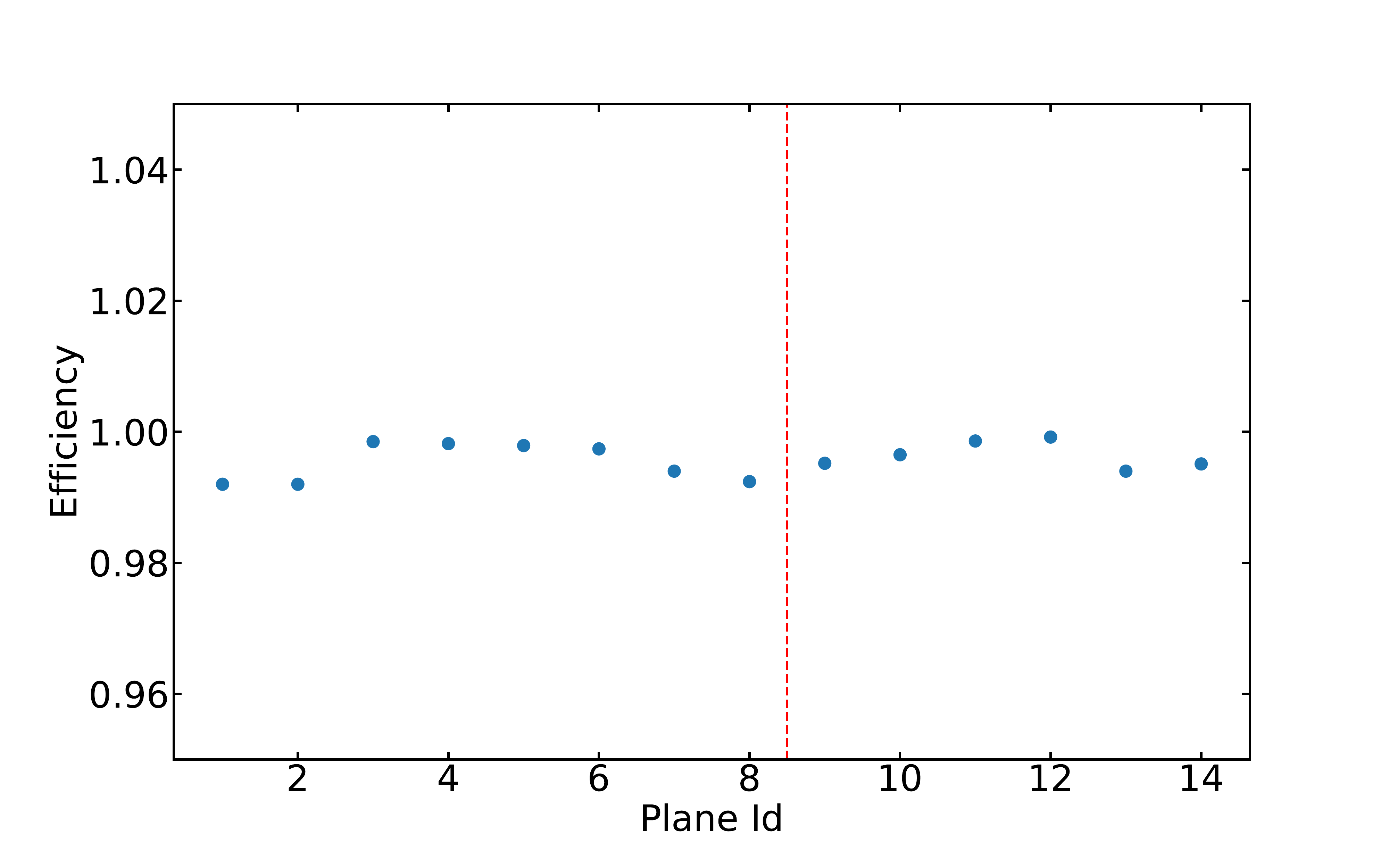}
    \caption{}
  \end{subfigure}
    \end{center}
	\caption{The effective angular resolution vs. beam momentum (a) and mean silicon plane efficiencies (b). The angular resolution is shown for both, the empty and the carbon target data. Multiple scattering in the target makes the effective angular resolution in the carbon target data worse at low momentum. The vertical red line in the efficiency plot separates upstream and downstream planes.}\label{fig:SSDperf}
\end{figure*}

Silicon strip efficiencies are also calculated based on the empty target data. For a given silicon strip plane, events are selected by requiring a single cluster per plane in all other planes. A track is fitted for each event and extrapolated to the selected plane and checked if there is a cluster present within $\pm 60\:\mu$m. The efficiencies are better than $99\%$. Both, the angular resolution for different datasets and silicon plane mean efficiencies are presented in Fig.~\ref{fig:SSDperf}.

The positions and efficiencies of the silicon strip planes obtained from these studies as well as beam-profile measurements from data are used as input parameters in Geant4-based Monte Carlo simulation\cite{Geant4}.  Angular distributions from the data are used to validate the simulation. The simulation includes silicon strip and pixel planes, target and trigger scintillators. In total, $10$ million beam protons are simulated for each beam settings with FTFP\_BERT and QGSP\_BERT physics lists from Geant4.10.05.p01. The simulation is used to estimate detector, acceptance and reconstruction inefficiencies and corresponding corrections are applied to data. Additionally, simulated events are used for some of the systematic studies described in Sec.~\ref{sec:res}.

The measurement of the forward differential cross-section without final-state momentum measurement is based on the assumption that four-momentum transfer $t$ is approximately equal to
%\begin{equation}\label{eq:tapprox}
$t \approx -p_\mathrm{b}^{2}\theta^{2}$
%\end{equation}
for small t, where $p_\mathrm{b}$ is the incident beam momentum, and $\theta$ is the scattering angle. This approximation is valid for coherent elastic and quasi-elastic interactions. However, it cannot be used for the inelastic scattering. 
The measurements presented here include proton-carbon differential cross-section for events with a single charged particle emitted from the target within $\pm 20\:$mrad with respect to the beam particle. Such interactions also include inelastic scattering where for example a neutron and pion are emitted in the forward detection. Therefore, the approximation for $t$ does not hold for these events. It is possible to remove inelastic events by applying an undesirable model-dependent correction. Instead we report $d\sigma/d(p_\mathrm{b}^{2}\theta^{2})$: 
\begin{equation}
	\left(\frac{d\sigma}{d(p_\mathrm{b}^2\theta^2)}\right)_i = \frac{1}{N_{pot}}\frac{N_{i}}{nd\cdot\Delta(p_\mathrm{b}^2\theta^2)_{i}},
\end{equation}
where $N_{pot}$ is the number of protons on target, $N_{i}$ is the corrected number of events with a single downstream track in a $p_\mathrm{b}^2\theta^{2}$ bin $i$, $nd=0.000179\:$mb$^{-1}$ is the target number density multiplied by the target length and $\Delta(p_\mathrm{b}^2\theta^2)_{i}$ is $i$-th bin width. A set of cuts is applied to the data to remove any background interactions.
%%JMP: I think the paragraph above could use some re-work.  Eg, I would start with simply stating what we are reporting, and then point out that this reduces to $t$ for coherent elastic and quasi-elastic interactions.

\subsection{Event selection}
The event selection is divided into upstream and downstream track selection. The purpose of the former is to select a pure proton sample and to remove any interactions upstream from the target.  The purpose of the latter is to select a pure sample of events with a single scattered charged particle downstream of the target. 

\begin{figure*}[ht]
    \begin{center}
        \includegraphics[trim= 0 0 0 0, clip, width=0.7\textwidth]{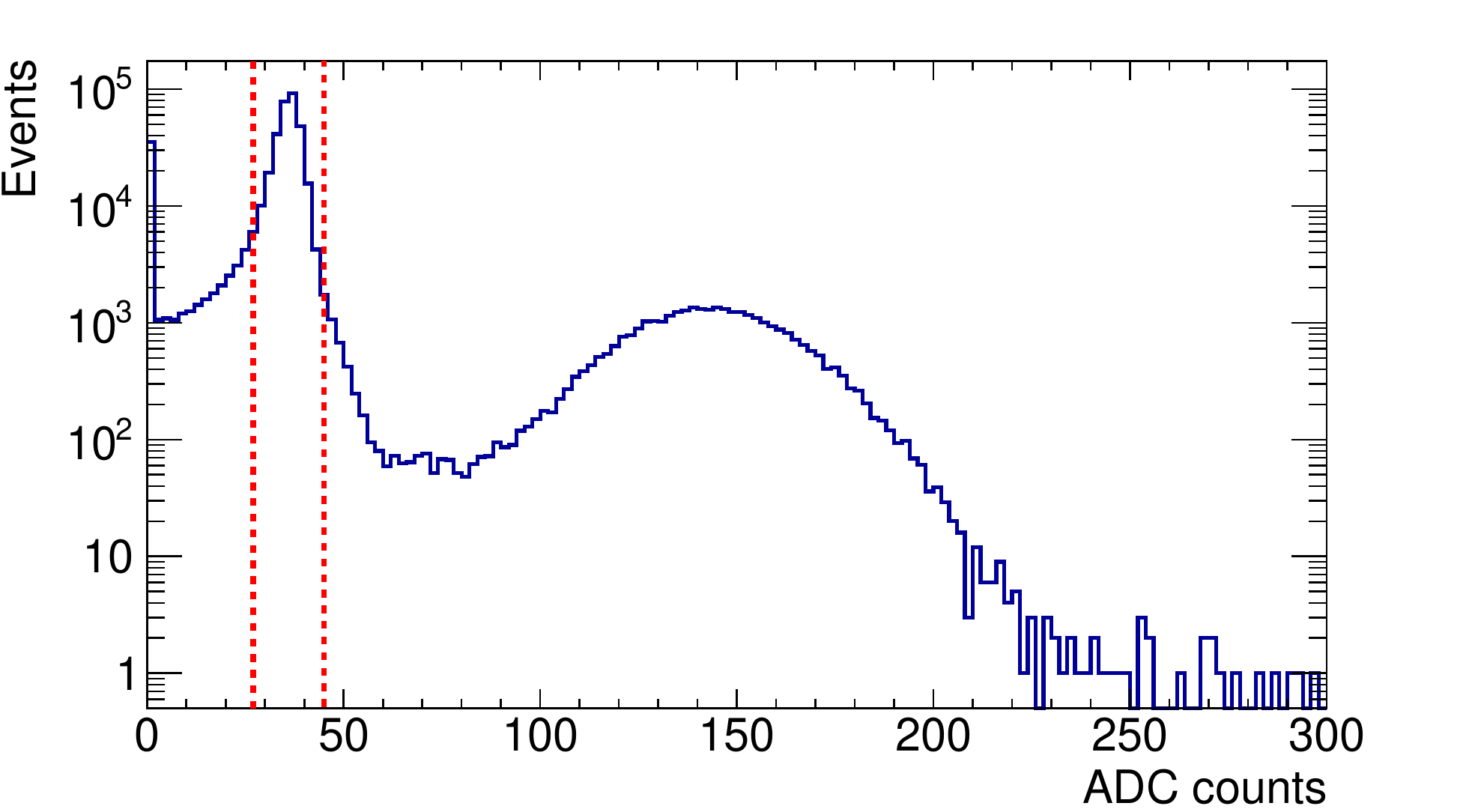}
    \end{center}
	\caption{The ADC count distribution for the PMT in the second gas Cherenkov detector for $30\:$GeV/c data. Selected events are located between dashed red lines.}\label{fig:CherenkovADC}
\end{figure*}

%\subsubsection{Upstream selection}
Beam particle identification is done by using gas Cherenkov detector available in FTBF. One detector used in anti-coincidence with the trigger scintillator is used to remove pions and electrons in the beam. The gas pressure in the second detector was set above the kaon threshold. Kaons and any electrons and pions remaining are removed by cutting out the signal in the recorded ADC distribution as shown in Fig.~\ref{fig:CherenkovADC}. Protons are part of the pedestal since they did not produce any Cherenkov light. The average number of photoelectrons detected from kaons passing through the detector is around $30$. Any non-proton contamination after the cut is estimated by fitting the kaon signal and pedestal to be much smaller than $1\%$. The $120\:$GeV/c beam is a pure proton beam since it comes directly from the Main Injector and Cherenkov detectors are not necessary in this case. 

Additional cuts are applied to remove most of the interactions in the upstream silicon strips and trigger scintillator. Only events with a single upstream track with a hit cluster in each plane and a sufficiently low $\chi^{2}$ are selected. Additionally, events with tracks in the tails of the beam divergence ($dx/dz$ and $dy/dz$) distributions are removed since these are mostly coming from the upstream interactions. The estimated number of the upstream interactions remaining after the selection is below $0.1\%$. However, these do not include interactions in the silicon pixel layers that are between the upstream strip layers and the target. 
Finally, a cut is applied to the incoming beam particle so that scattered particles within $\pm 20\:$mrad always falls within the acceptance of the downstream silicon strip layers.

\begin{figure*}[ht]
    \begin{center}
        \includegraphics[trim= 20 90 50 60, clip, width=1\textwidth]{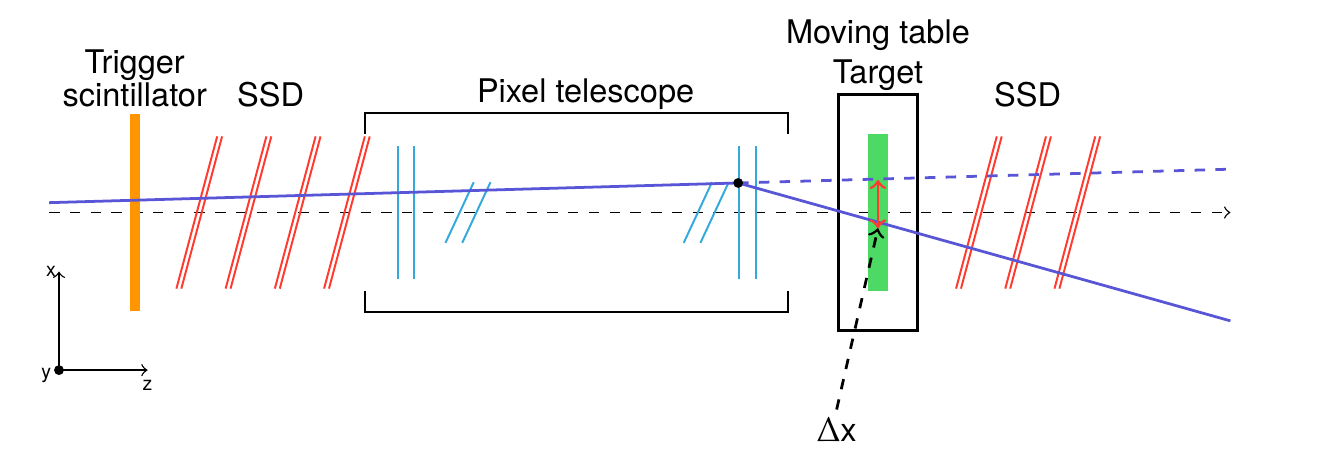}
    \end{center}
	\caption{A schematic of the $x$ and $y$ cut. If interaction happens in a pixel plane, $x$ and $y$ distances between upstream and downstream tracks at target z position will be significantly different than zero.}\label{fig:dxyCut}
\end{figure*}

%\subsubsection{Downstream selection} 
The purpose of the downstream selection is to identify forward scattered beam protons and remove hard inelastic interaction in the pixel telescope, target, and downstream SSDs. Selected events have only one reconstructed track with a hit cluster in each downstream plane  and the $\chi^2$ value below $6$. An additional cut is applied to remove interactions in silicon pixel layers. The upstream and downstream tracks in an event are extrapolated toward the center of the target. A cut is applied on the $x$ and $y$ distances between the tracks. If an interaction happened somewhere outside of the target, the difference in $x$ and $y$ positions would be larger. The cut on the $x$ and $y$ distances is defined as:

\begin{equation}
	|x_{up} - x_{down}| > 3\cdot \sigma_x + 3\cdot\sigma_{\theta_{x}}\cdot|z_{vert}-z_{targ}|
\end{equation}
where $\sigma_x$ is a width of the $x$ distance distribution and $\sigma_{\theta_{x}}$ is a width of $\theta_x$ distribution. A schematic of the $x$ and $y$ cut is shown in Fig.~\ref{fig:dxyCut}. 

\begin{figure*}[ht]
  \begin{subfigure}[t]{0.48\textwidth}
        \includegraphics[trim= 0 0 0 0, clip, width=1\textwidth]{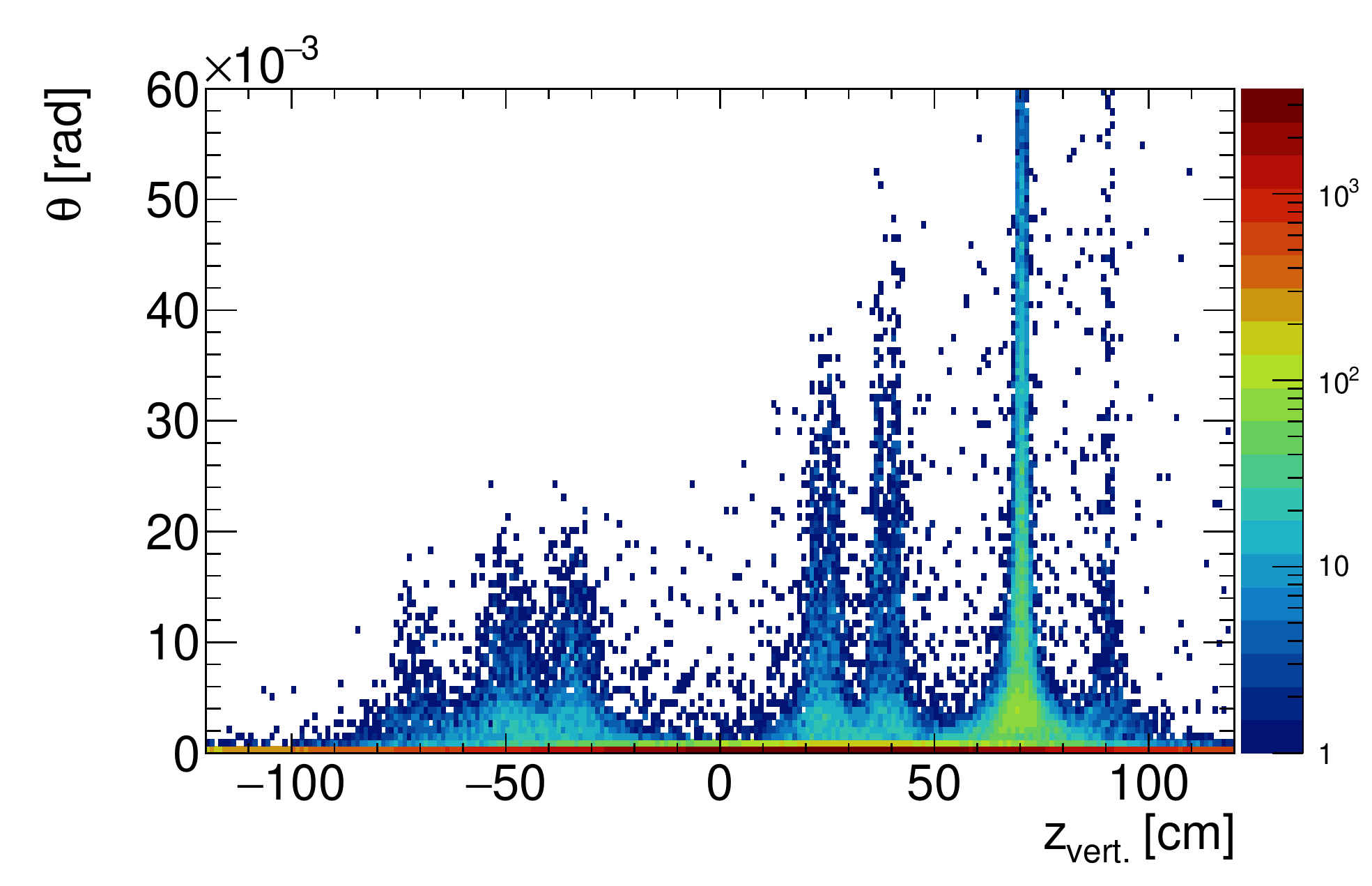}
        \includegraphics[trim= 0 0 0 0, clip, width=1\textwidth]{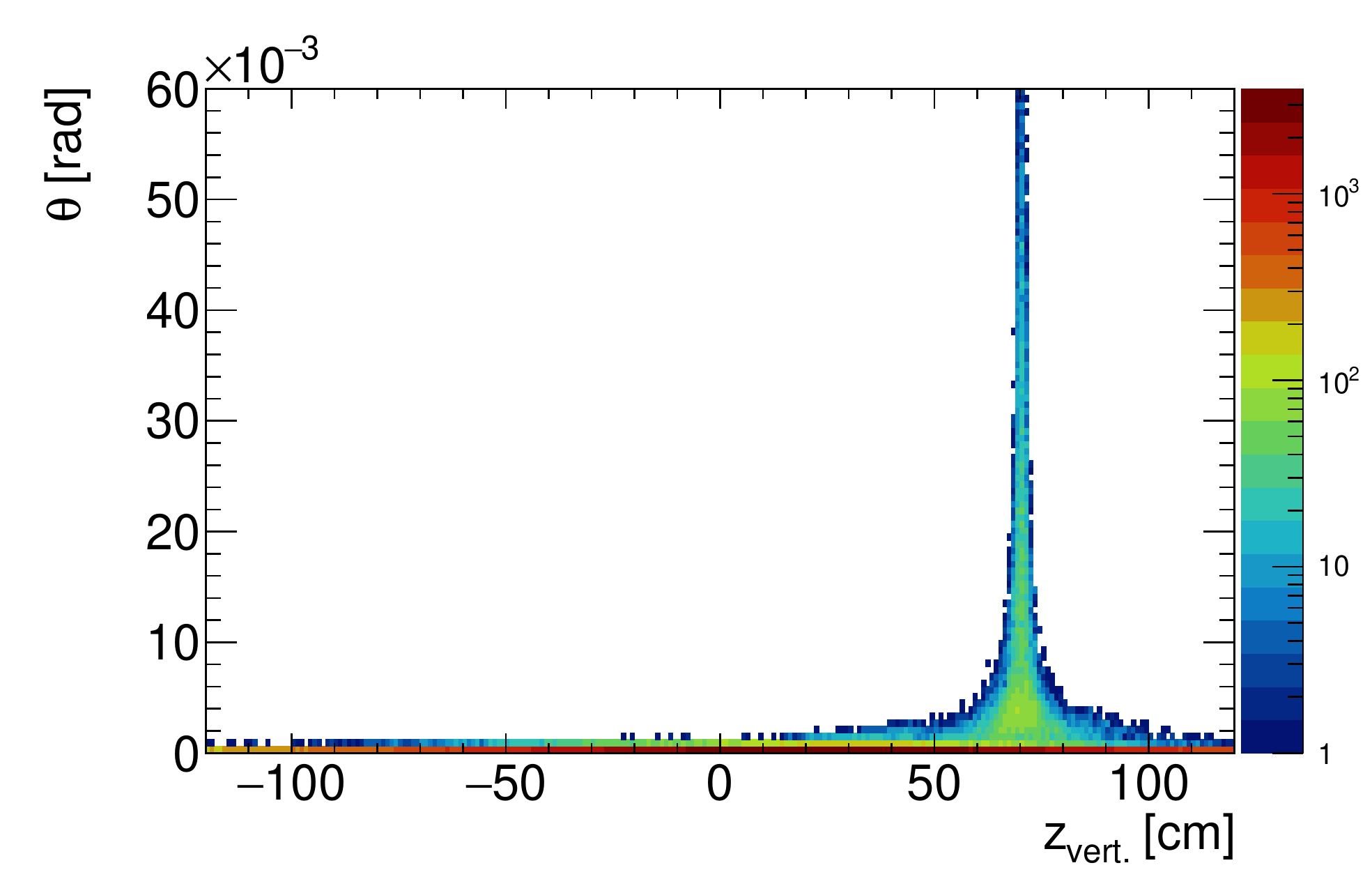}
    \caption{Data}
  \end{subfigure}
  \begin{subfigure}[t]{0.48\textwidth}
        \includegraphics[trim= 0 0 0 0, clip, width=1\textwidth]{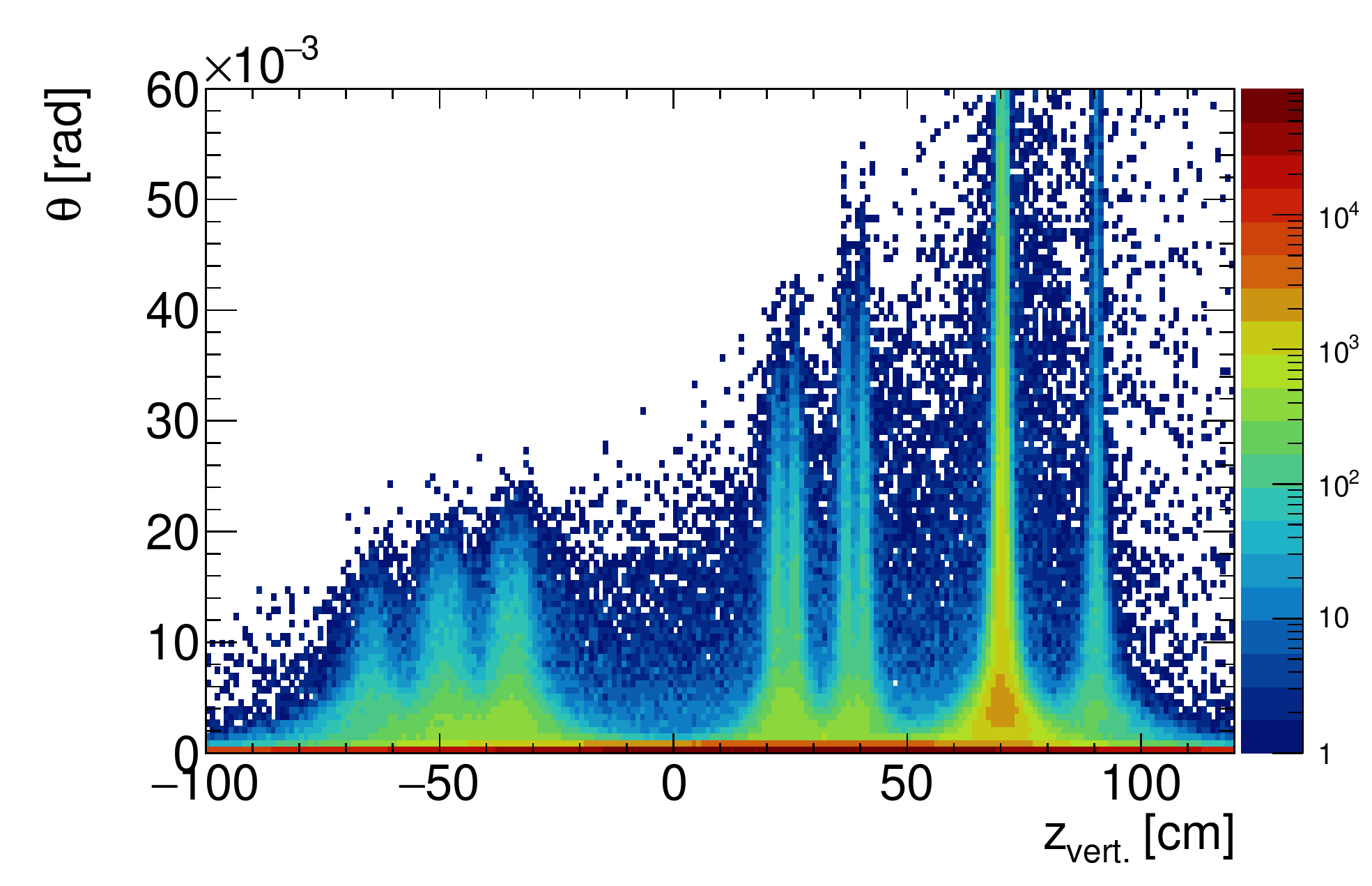}
        \includegraphics[trim= 0 0 0 0, clip, width=1\textwidth]{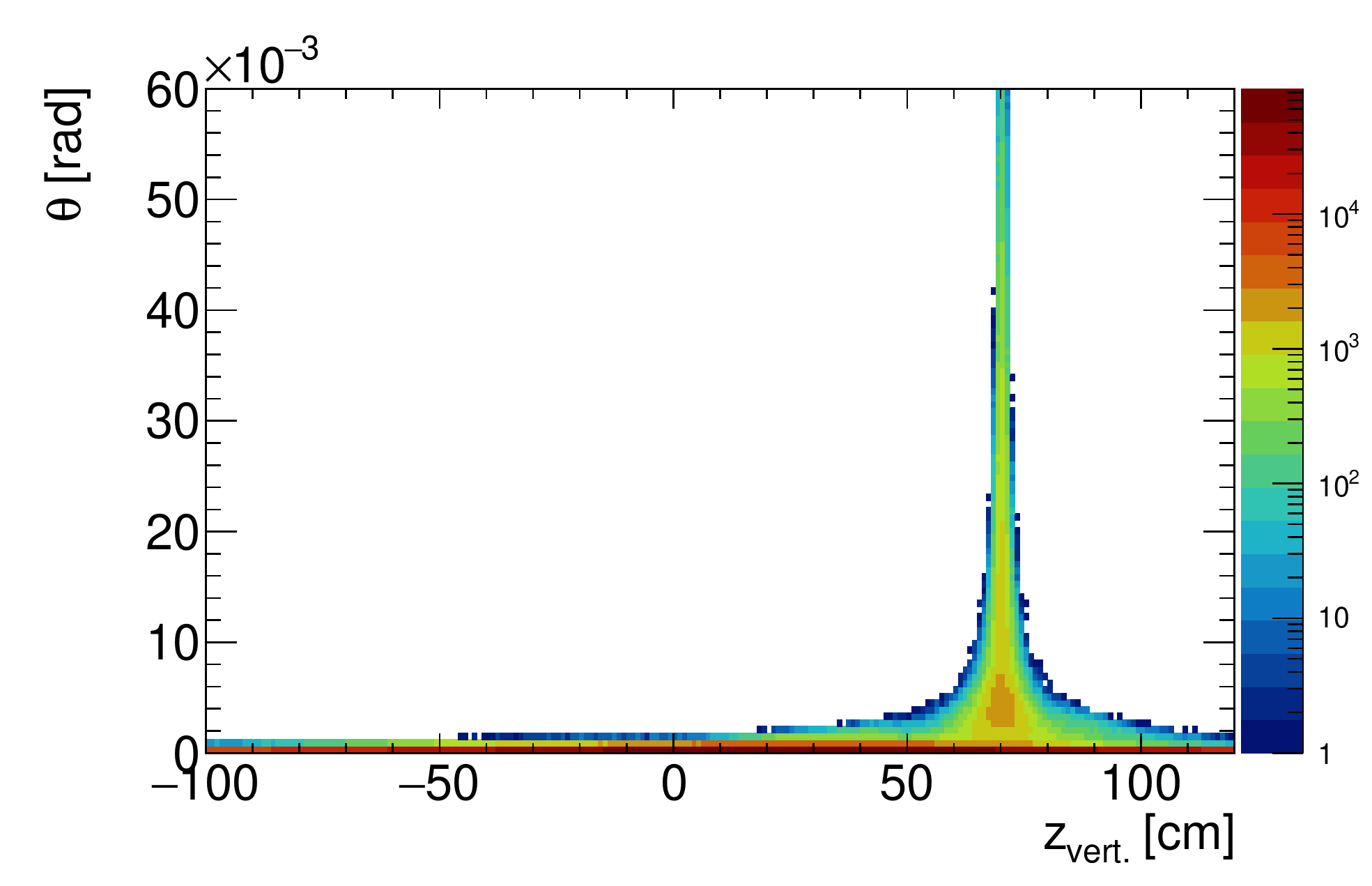}
    \caption{Simulation}
  \end{subfigure}
	\caption{Reconstructed angle vs. reconstructed vertex $z$ position for proton-carbon data (a) and FTFP\_BERT G4.10.04.p2 Monte Carlo (b) at $30\:GeV/c$ before (top) and after (bottom) $x$ and $y$ cuts. First peak at $z=-70\:$cm shows interactions in the last upstream silicon plane. The next four peaks are in fact four double peaks (clearly visible in simulation) and they show interactions in eight pixel planes. The large peak at $z=70\:$cm includes interactions in the target, and the last peak includes interactions in the first downstream silicon strip detector.}\label{fig:thvsz30GeV}
\end{figure*}

To illustrate the effect of the $x$ and $y$ cuts, a reconstructed angle vs. reconstructed vertex position distribution is shown in Fig.~\ref{fig:thvsz30GeV} before and after applying the cuts. Several peaks are visible in the distribution. The first peak from the left corresponds to the last upstream silicon strip plane. The next four peaks correspond to the eight pixel planes (pixel planes come in pairs, which are only $5\:$cm apart. The largest peaks shows interactions in the target. The last peaks correspond to the first downstream silicon strip plane. After $x$ and $y$ cuts are applied, inelastic interactions outside of the target are removed in both data and simulation.

%\subsection{Corrections}
The raw differential cross-section needs to be corrected for various inefficiencies such as: SSD inefficiencies, reconstruction inefficiencies, selection inefficiencies and interactions outside of the target. The efficiency corrections can be grouped into two groups: corrections based on Monte Carlo simulation and corrections based on the data.

\subsection{Monte Carlo correction factors}
\begin{figure*}[ht]
  \begin{subfigure}[t]{0.48\textwidth}
        \includegraphics[trim= 0 0 0 0, clip, width=1\textwidth]{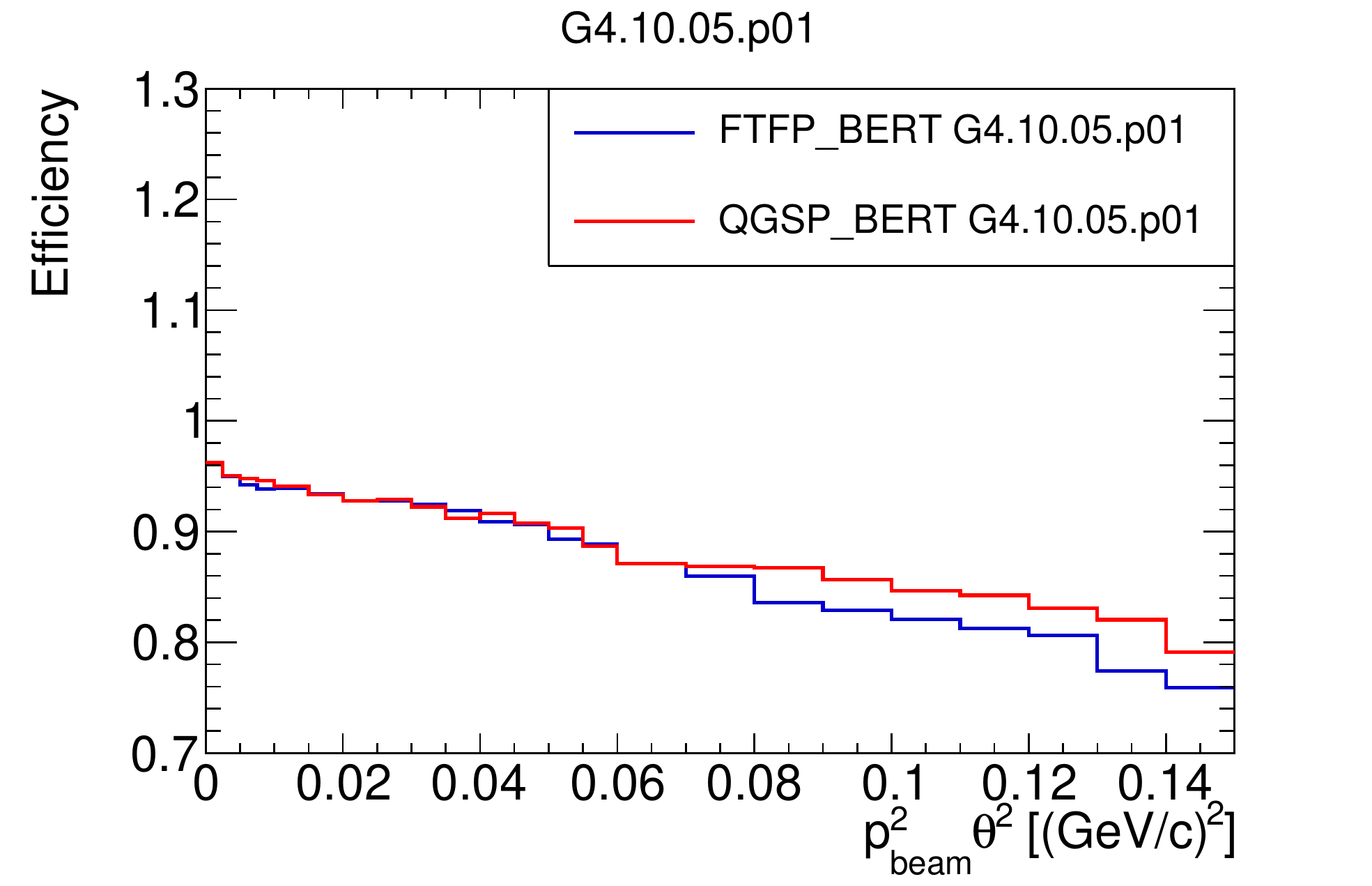}
    \caption{}
  \end{subfigure}
  \begin{subfigure}[t]{0.48\textwidth}
        \includegraphics[trim= 0 0 0 0, clip, width=1\textwidth]{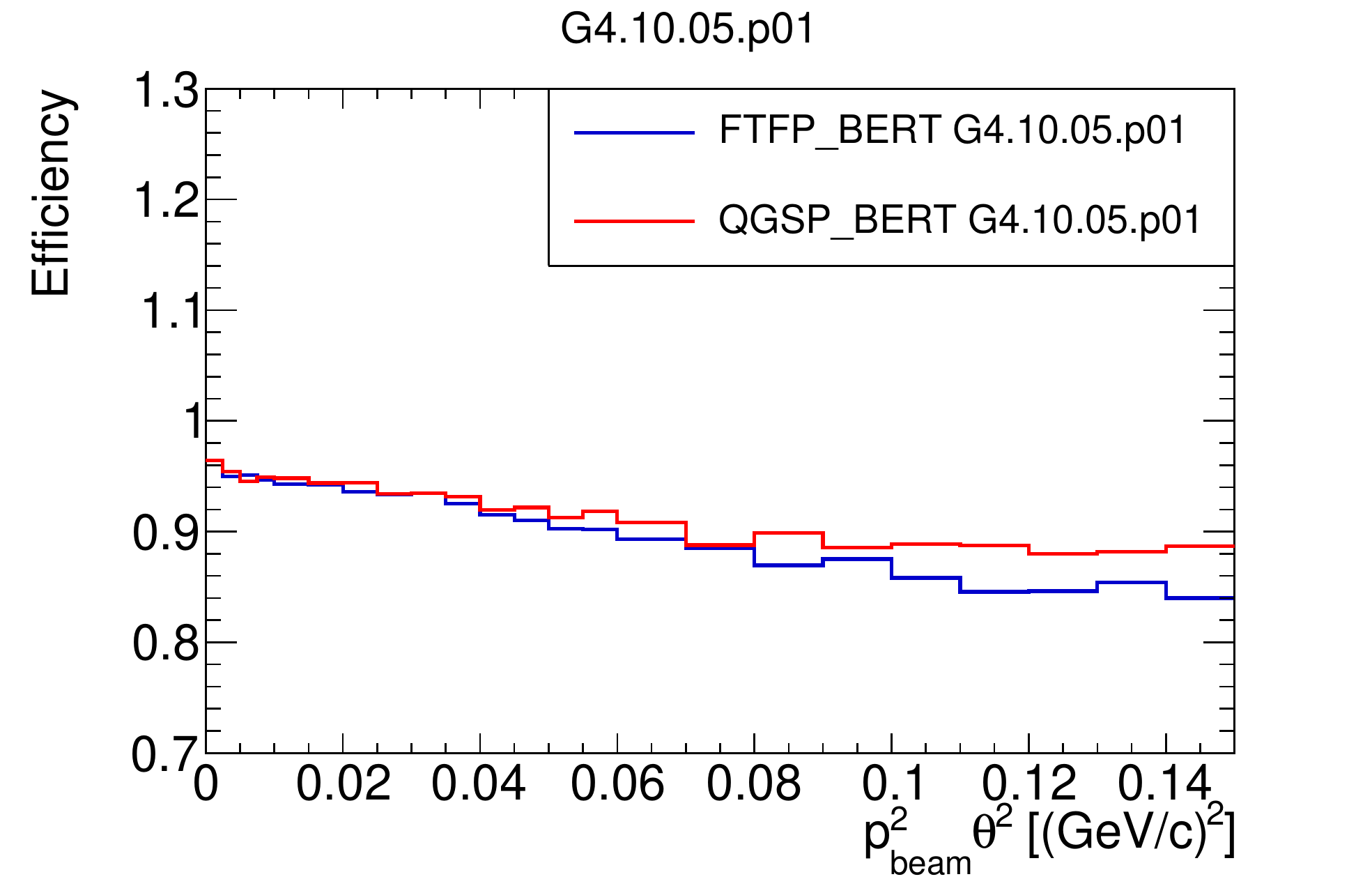}
    \caption{}
  \end{subfigure}
  \begin{subfigure}[t]{0.48\textwidth}
        \includegraphics[trim= 0 0 0 0, clip, width=1\textwidth]{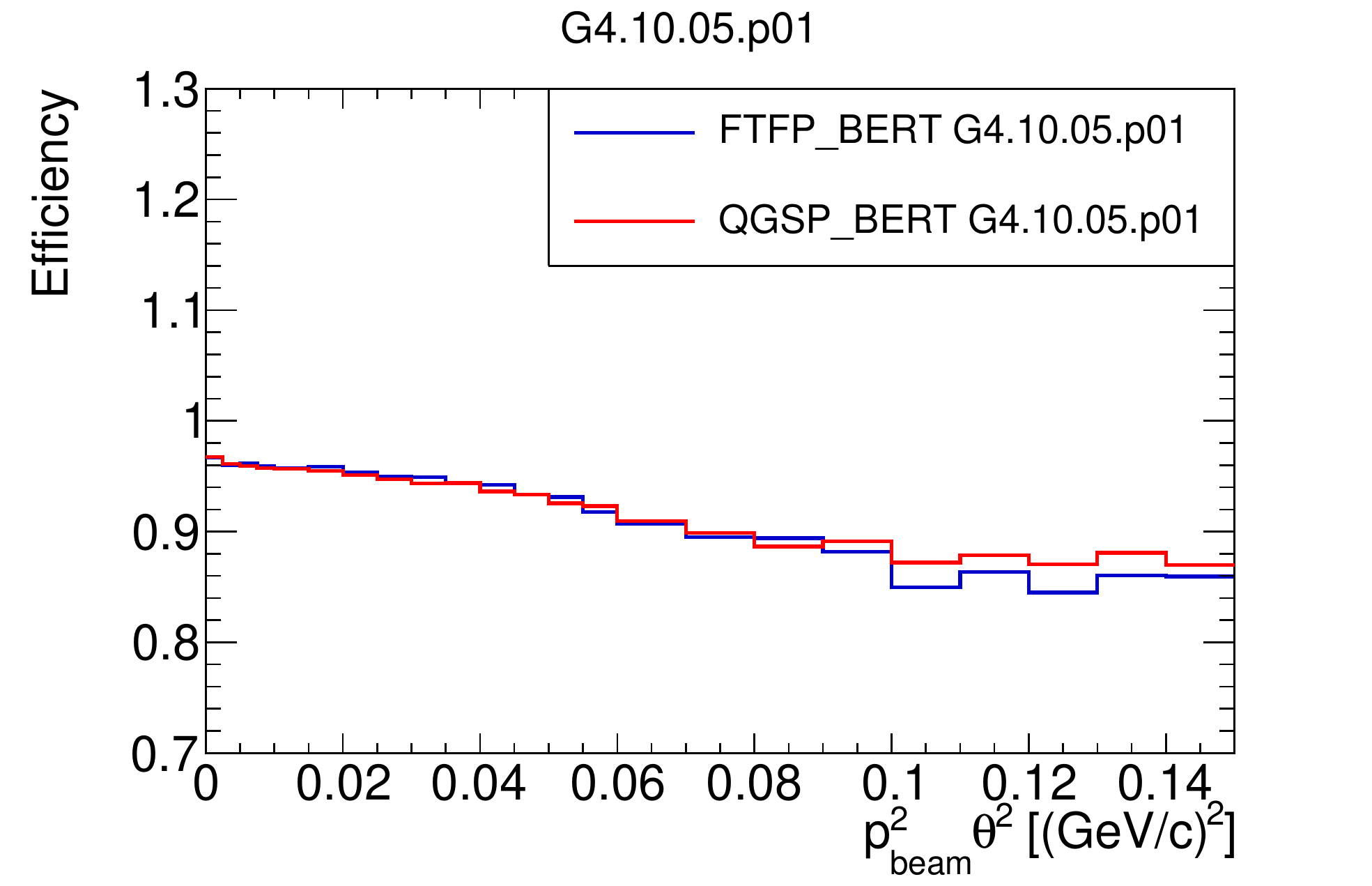}
    \caption{}
  \end{subfigure}  
	\caption{Downstream selection efficiency for proton-carbon data at $20\:$GeV/c (a), $30\:$GeV/c (b), and $120\:$GeV/c (c) for FTFP\_BERT and QGSP\_BERT G4.10.05.p1 physics lists.}\label{fig:pCEff}
\end{figure*}

Monte Carlo efficiency includes silicon strip efficiency, reconstruction efficiency and selection efficiency. However, it does not include smearing effects caused by detector resolution. The efficiency factor is defined as:
\begin{equation}
\epsilon_{i} = \frac{N_{i, sel, true}}{N_{i, true}},
\end{equation}
where $i$ is the $p^{2}\theta^{2}$ bin number, $N_{i, sel, true}$ is the number of selected downstream tracks in the true bin $i$, and $N_{i, true}$ is the number of true tracks in the bin $i$ before the selection. The efficiency is calculated only for beam particles hitting the target without any prior interaction. The correction factor is defined as an inverse efficiency. A Geant4 based Monte Carlo simulation has been used to calculate the corrections. The simulation includes the target, silicon strip layers, silicon pixel layers and the trigger scintillator. Two physics lists are used to calculate the correction factor: FTFP\_BERT and QGSP\_BERT from Geant4.10.05.p1. The Monte Carlo efficiency is presented in Fig.~\ref{fig:pCEff}. The differences in efficiency between datasets come from the differences in the fraction of inelastic events. Selected events with a single forward low momentum charged proton or pion tend to have lower efficiency. The probability of scattering in downstream tracking layers and multiple scattering increases for low momentum particles. Therefore the $\chi^{2}$ of these tracks tends to be higher on average, resulting in lower efficiency. 

\subsection{Data correction factors}
Pixel interactions removed by $\Delta x$ and $\Delta y$ cuts are elastic interactions with higher four-momentum transfer and inelastic interactions. Since they happen before the beam hits the target, any interacting beam particle is lost and needs to be removed from the number of protons on target. 
A normalization correction is applied to the number of incoming beam particles. The correction is estimated from the empty target data:
\begin{equation}
N_{POT, cor} = N_{tout,rem}\cdot C\cdot \frac{N_{tin, POT}}{N_{tout, POT}},
\end{equation}
where $N_{tout,rem}$ is the number of removed pixel interactions in the empty target data, $C$ is the purity correction based on simulation, and $\frac{N_{tin, POT}}{N_{tout, POT}}$ is the ratio of number of selected events after upstream selection in the carbon target and empty target data. The purity correction is a ratio of the true removed pixel interactions and the total number of removed events. Empty target data is used for this correction to avoid any bias from including interactions in the target. The normalization correction is $2.0\%$, $2.3\%$, and $3.0\%$ for $20$, $30$, $120\:$GeV/c data respectively.

\section{Systematic uncertainties}
Several systematic contributions are considered in the analysis: beam purity, the number of interactions in upstream detectors (beam loss), target density and thickness uncertainties, and efficiency variations.
As previously mentioned, kaon contamination in the proton beam after gas Cherenkov cut is estimated to be negligible. Beam loss is caused by interactions in the pixel telescope as explained in the previous section. Since the beam loss is between $2\%$ and $3\%$ and the purity correction is around $95\%$, any systematic uncertainty is going to be small. The beam loss systematic contribution includes the statistical uncertainty from the empty target data and purity variation estimated by using FTFP\_BERT and QGSP\_BERT physics lists. Values are below $1\%$ for all three datasets.
A normalization uncertainty from the measured target density and thickness is estimated to be $2\%$ and it is the dominant contribution at low $p^2\theta^2$.

The efficiency uncertainty includes contributions from MC statistics, SSD plane efficiencies, differences in $\chi^2$ distributions between data and MC, differences in angular resolution, and model differences. Silicon efficiencies in simulation are reduced in all planes by their uncertainties, and the efficiency is reevaluated. The difference between the nominal and reevaluated efficiency is taken as a systematic uncertainty. 

A similar approach is used for the variation of cut parameters. The $\chi^2$ distribution from the data has a longer tail compared to the distribution from simulation. This is caused by a small number of track clusters with multiple active strips in data. These clusters are created by a passing charged particle and emitted delta electron. Delta electrons induce signals in the neighbouring strips and create a systematic shift in the measured position of the original particle. Reconstructed tracks with such clusters will have increased $\chi^{2}$ value. The $\chi^2$ value used in MC cut is adjusted so that the fraction of removed events is the same as the one in the data. The efficiency is reevaluated after using the new cut and the difference is taken as a systematic uncertainty.

Similarly, the angular resolution in Monte Carlo simulation can be from $3\%$ to $7\%$ different from the data, depending on the dataset. Angular resolution parameters in  $\Delta x$ and $\Delta y$ cuts are varied within $\pm 7\%$ and the difference is taken as a systematic uncertainty. 

Finally, differences in efficiency estimated with FTFP\_BERT and QGSP\_BERT are used as a systematic uncertainty. These differences are caused by the variation in the number of events with a single low momentum pion in the forward direction. Low momentum pions ($<5\:$GeV/c) have lower selection efficiency compared to elastically scattered beam particles. Effective angular resolution for these pions is worse due to increase in multiple scattering. Therefore, these pions will be overrepresented in the tails of the angular distribution or they will have increased $\chi^{2}$ values. The difference between models is largest for higher $p^2\theta^2$ values because of higher fraction of low momentum pions. 

Two co-dominant contributions in the efficiency uncertainty are coming from SSD efficiency variations and angular resolution differences and are between $1\%$ and $2\%$.  
The total efficiency uncertainty is between $2\%$ and $3\%$ and it is the dominant systematic uncertainty contribution at higher $p^2\theta^2$.

\section{Differential cross section results}\label{sec:res}

\begin{figure*}[ht]
  \begin{subfigure}[t]{0.48\textwidth}
        \includegraphics[trim= 0 0 0 0, clip, width=1\textwidth]{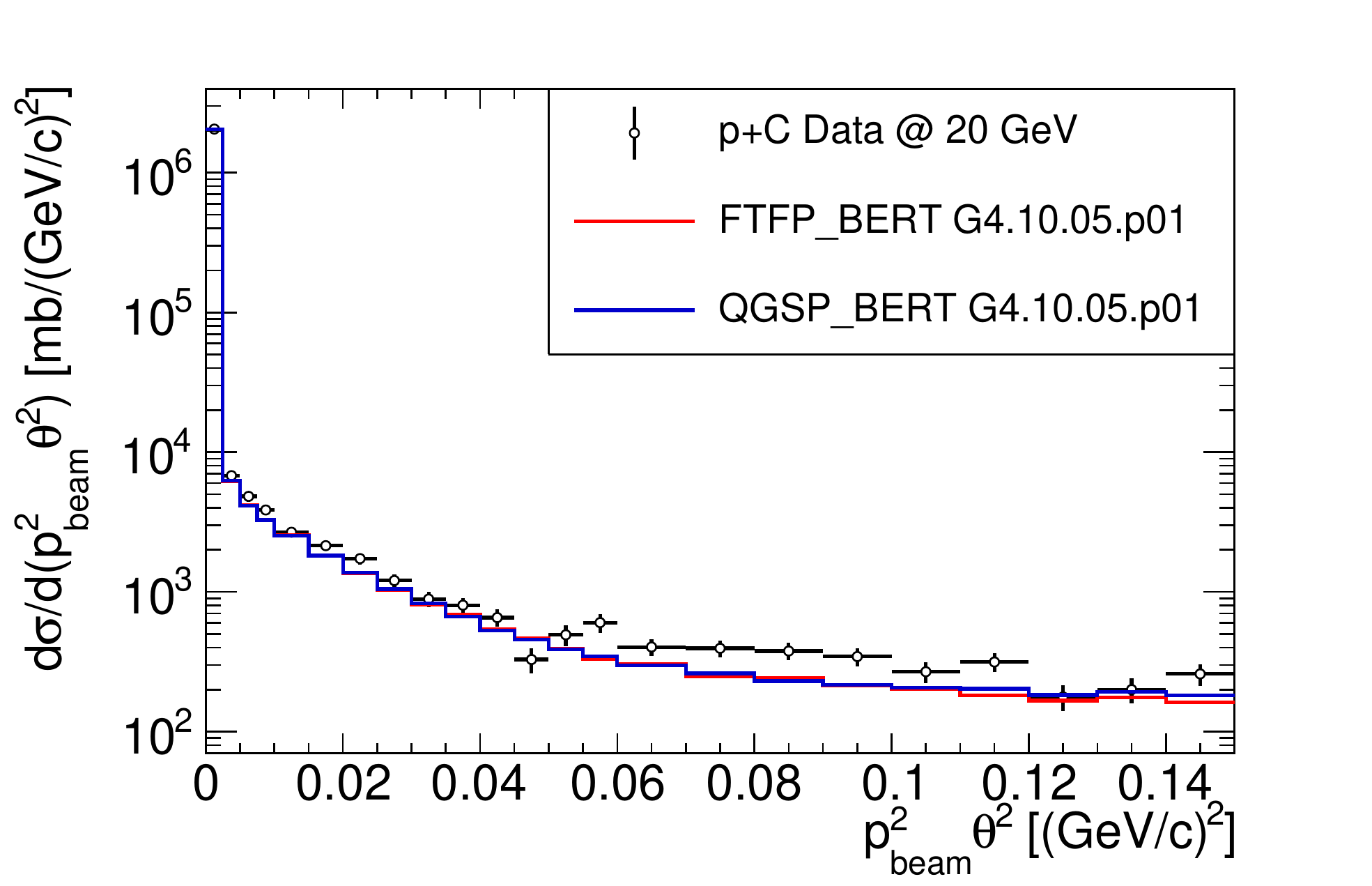}
    \caption{}
  \end{subfigure}
  \begin{subfigure}[t]{0.48\textwidth}
        \includegraphics[trim= 0 0 0 0, clip, width=1\textwidth]{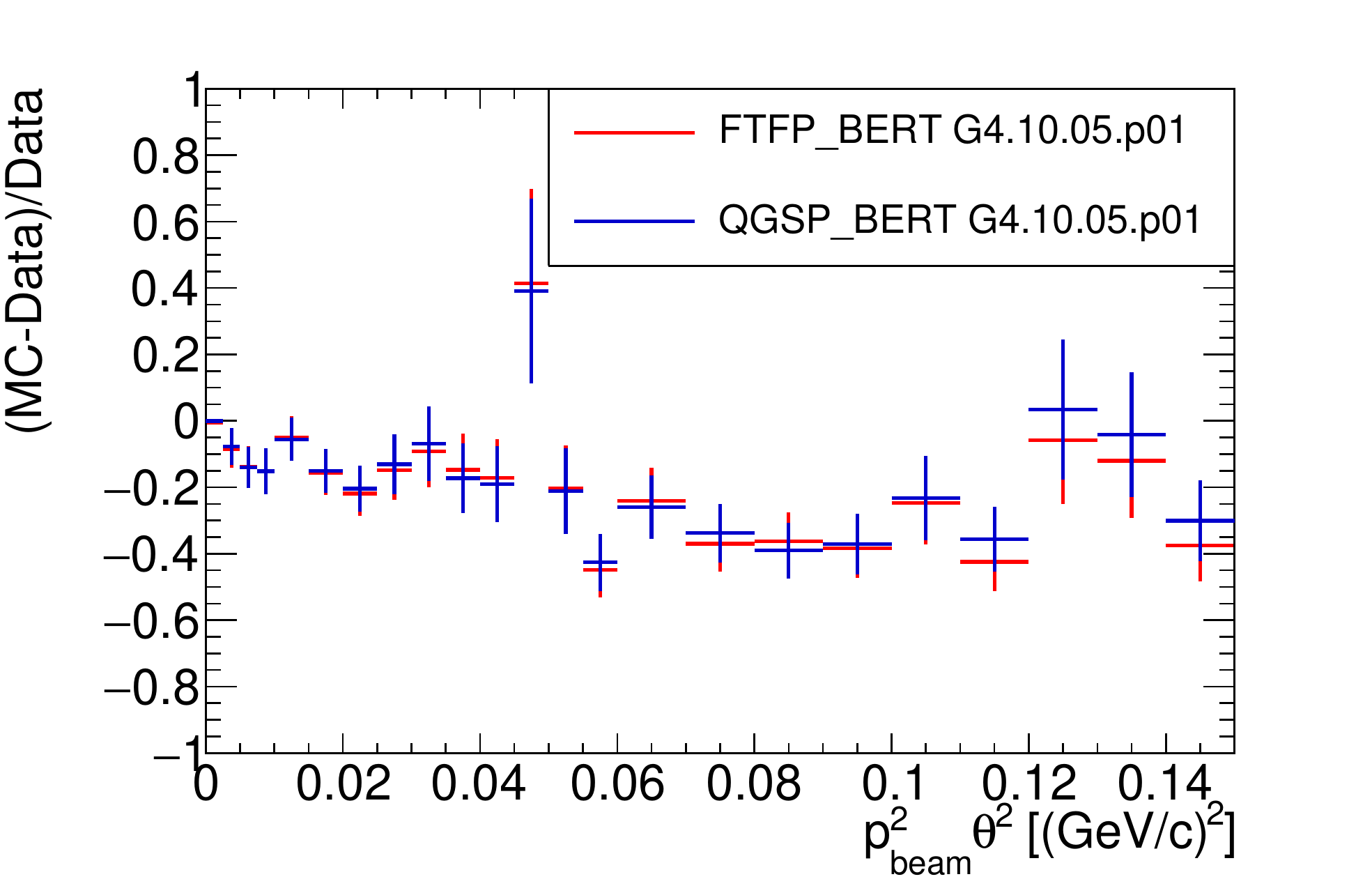}
    \caption{}
  \end{subfigure}
    
  \begin{subfigure}[t]{0.48\textwidth}
        \includegraphics[trim= 0 0 0 0, clip, width=1\textwidth]{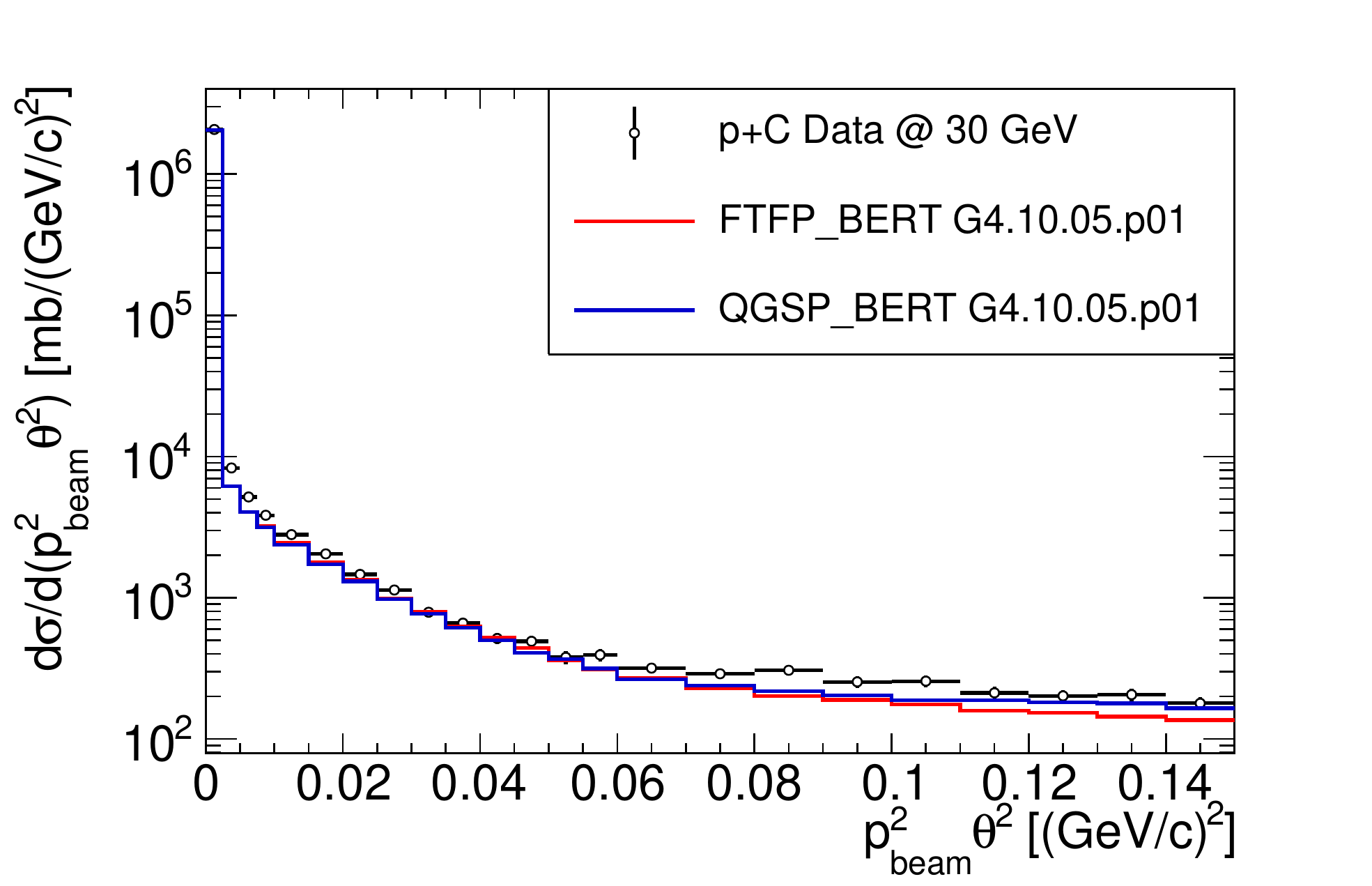}
    \caption{}
  \end{subfigure}
  \begin{subfigure}[t]{0.48\textwidth}
        \includegraphics[trim= 0 0 0 0, clip, width=1\textwidth]{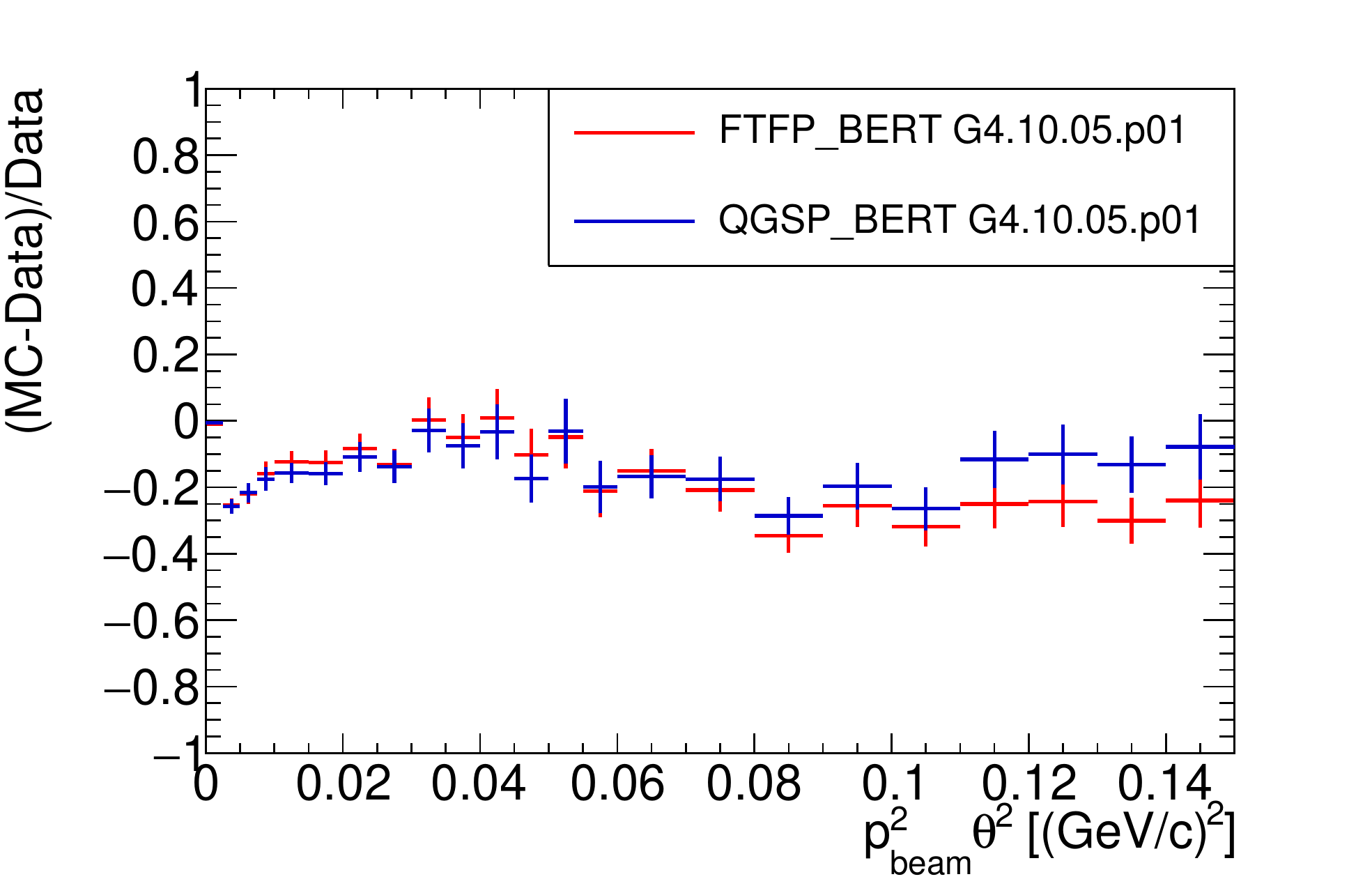}
    \caption{}
  \end{subfigure}
    
  \begin{subfigure}[t]{0.48\textwidth}
        \includegraphics[trim= 0 0 0 0, clip, width=1\textwidth]{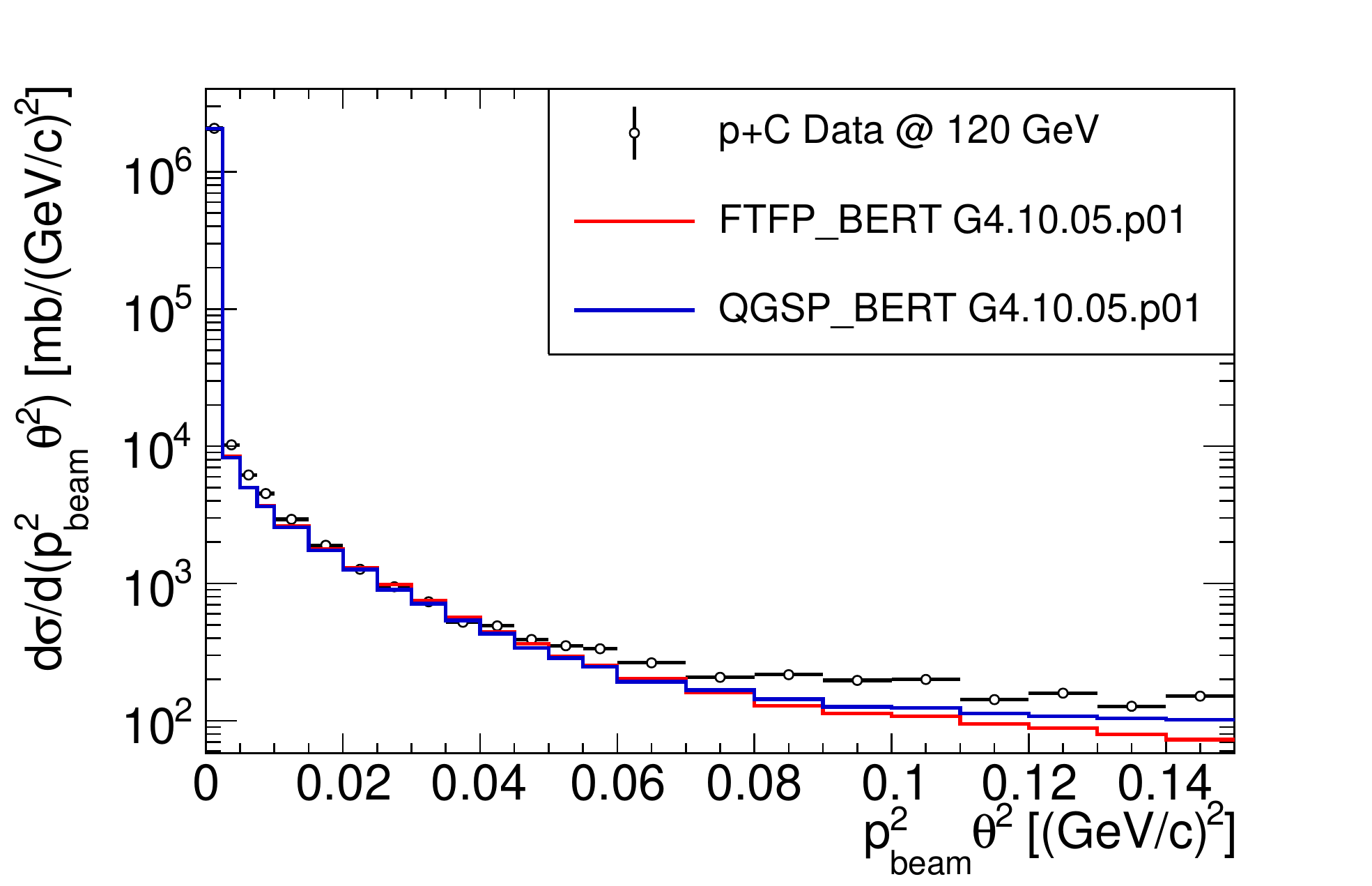}
    \caption{}
  \end{subfigure}
  \begin{subfigure}[t]{0.48\textwidth}
        \includegraphics[trim= 0 0 0 0, clip, width=1\textwidth]{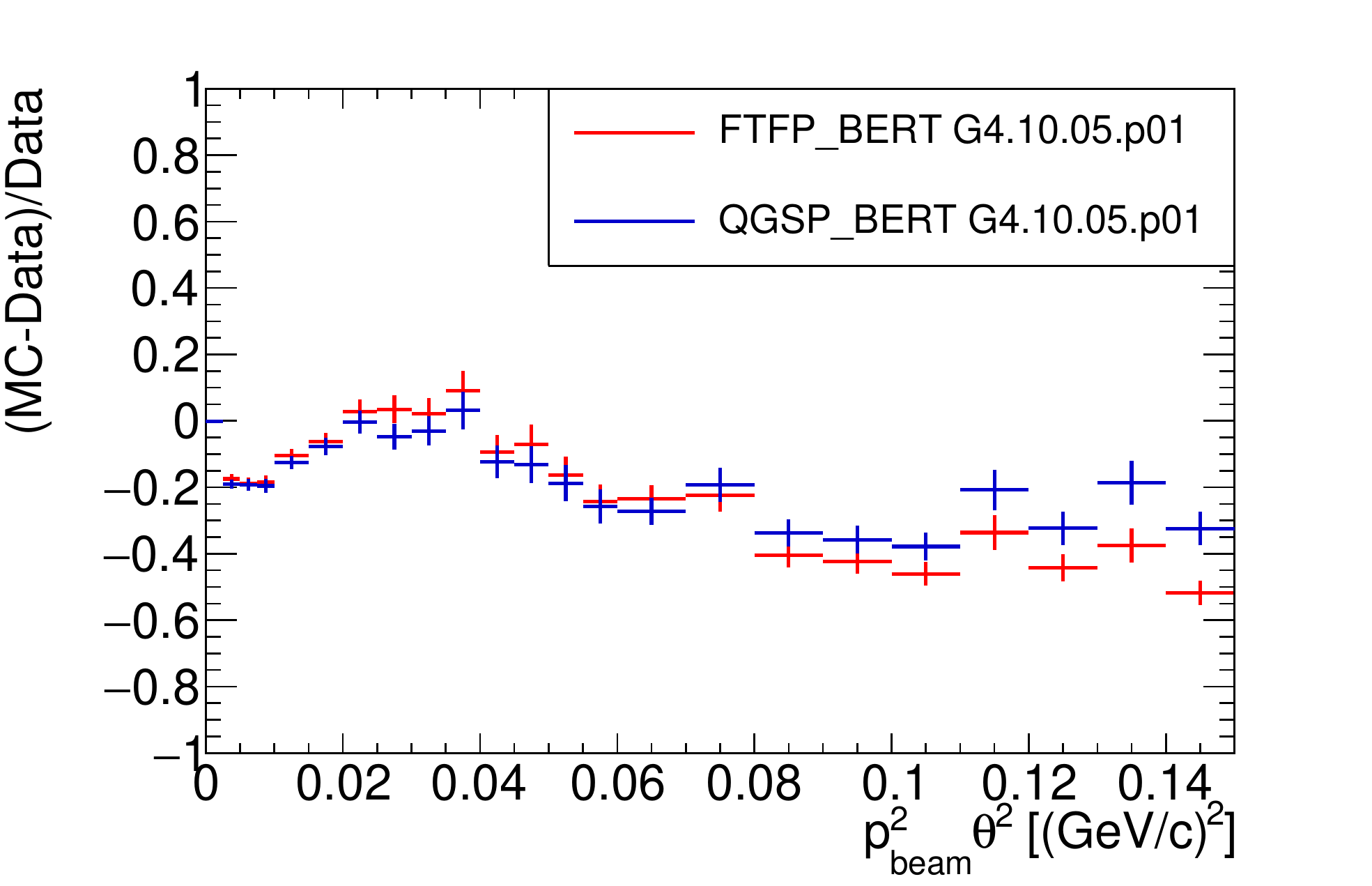}
    \caption{}
  \end{subfigure}
    
	\caption{The p+C differential cross-section at $20\:$GeV/c (a), $30\:$GeV/c (c), and $120\:$GeV/c (e),  and their corresponding comparisons to the FTFP\_BERT and QGSP\_BERT models from Geant 4.10.05.p01 (b), (d), and (f).}\label{fig:pCRes}
\end{figure*}

\begin{figure*}[ht]
  \begin{subfigure}[t]{0.48\textwidth}
        \includegraphics[trim= 0 0 0 0, clip, width=1\textwidth]{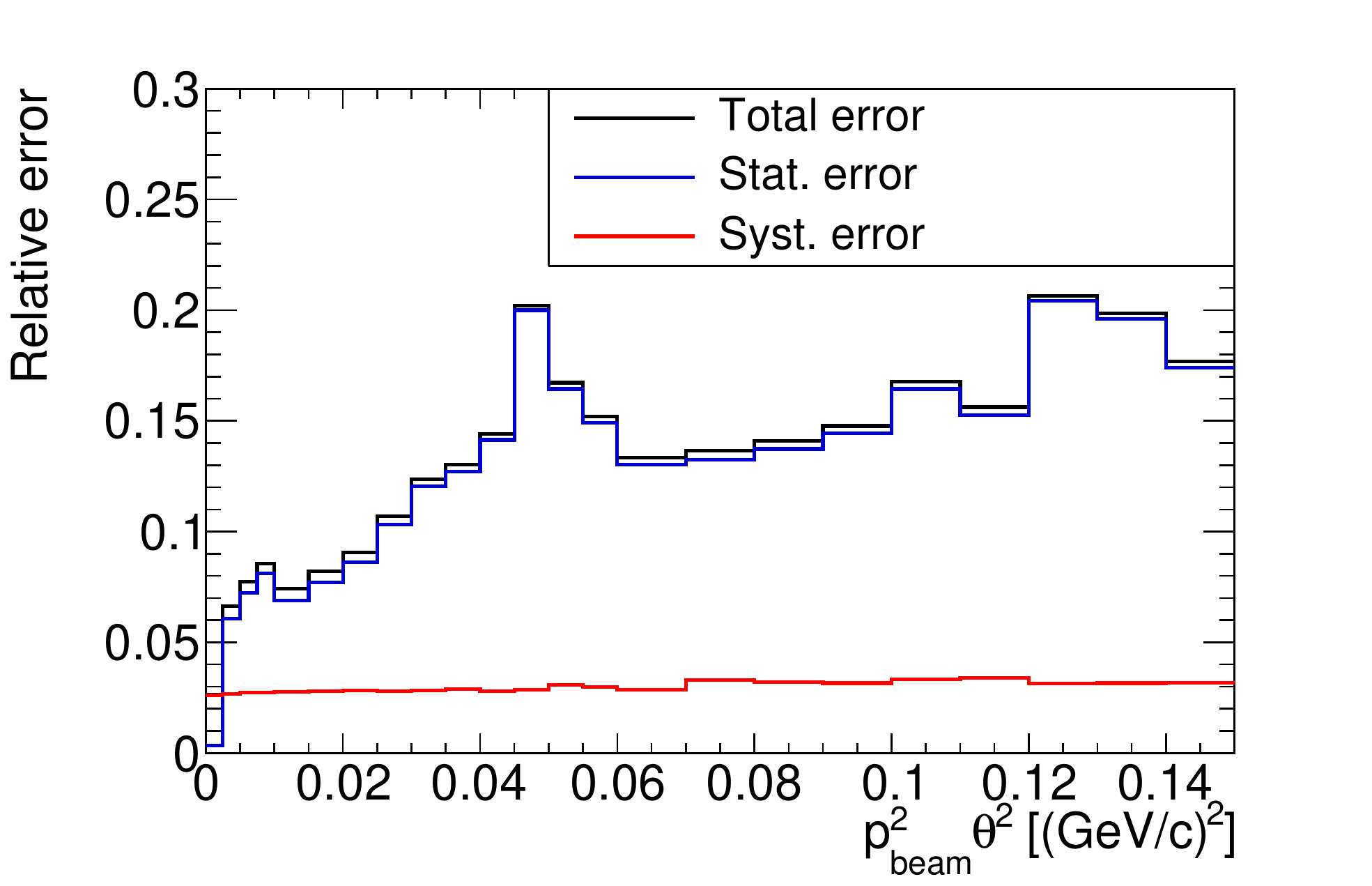}
    \caption{}
  \end{subfigure}
  \begin{subfigure}[t]{0.48\textwidth}
        \includegraphics[trim= 0 0 0 0, clip, width=1\textwidth]{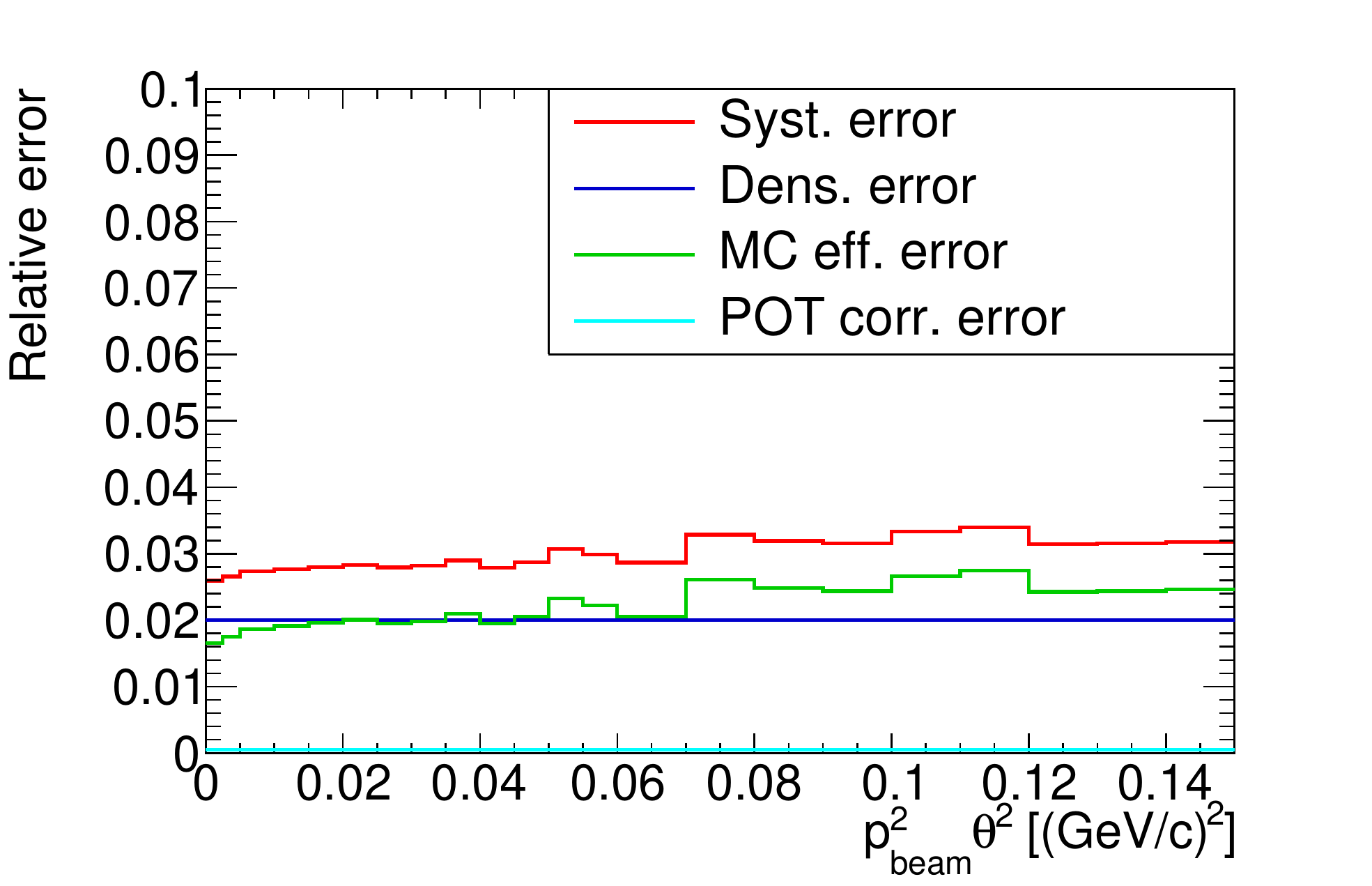}
    \caption{}
  \end{subfigure}

  \begin{subfigure}[t]{0.48\textwidth}
        \includegraphics[trim= 0 0 0 0, clip, width=1\textwidth]{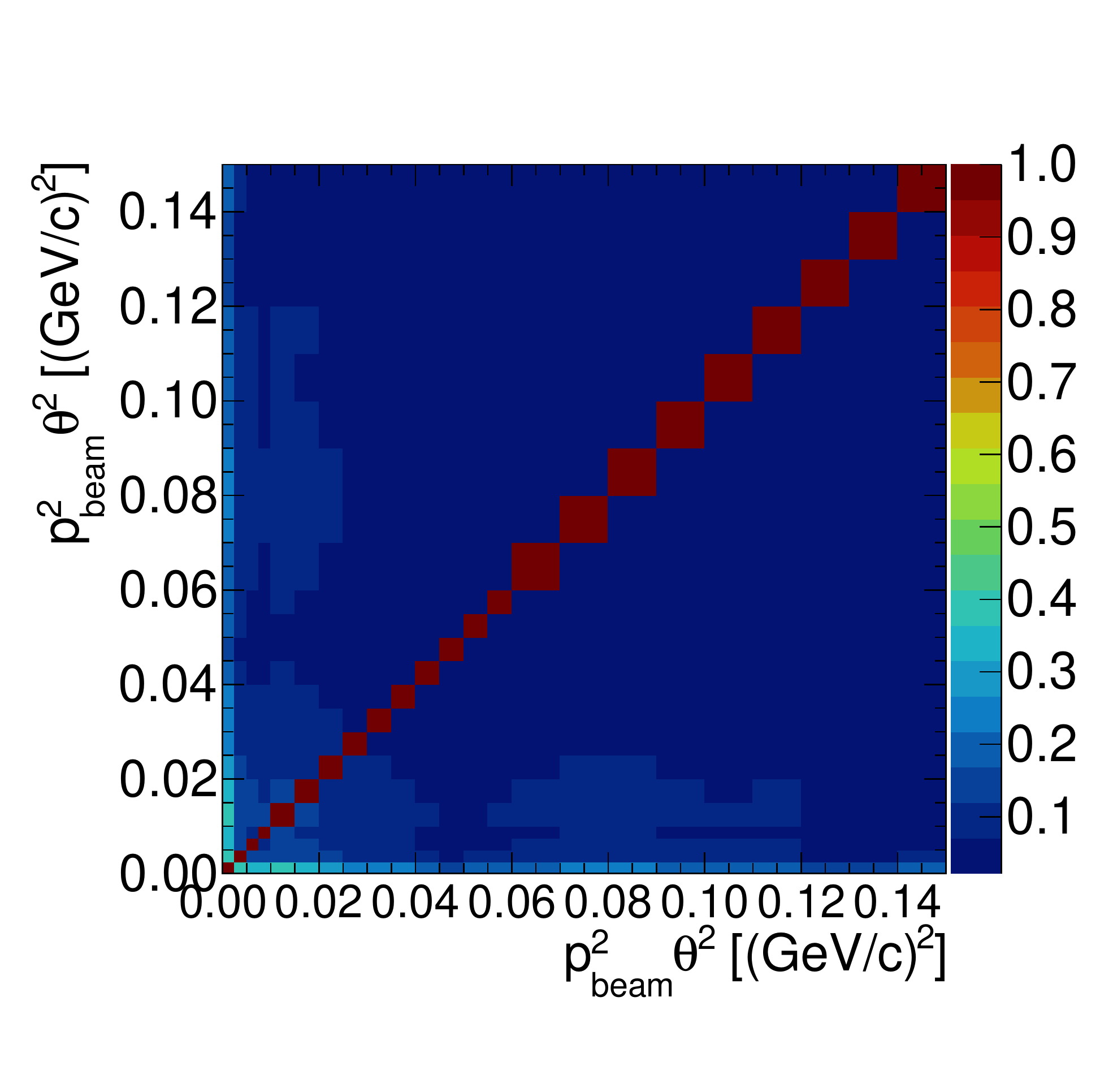}
    \caption{}
  \end{subfigure}
      
	\caption{The total, statistical and systematic uncertainty for p+C differential cross-section at $20\:$GeV/c (a) and different contributions to the systematic uncertainty (b), correlation matrix (c).}\label{fig:pC20GeVErr}
\end{figure*}

\begin{figure*}[ht]
  \begin{subfigure}[t]{0.48\textwidth}
        \includegraphics[trim= 0 0 0 0, clip, width=1\textwidth]{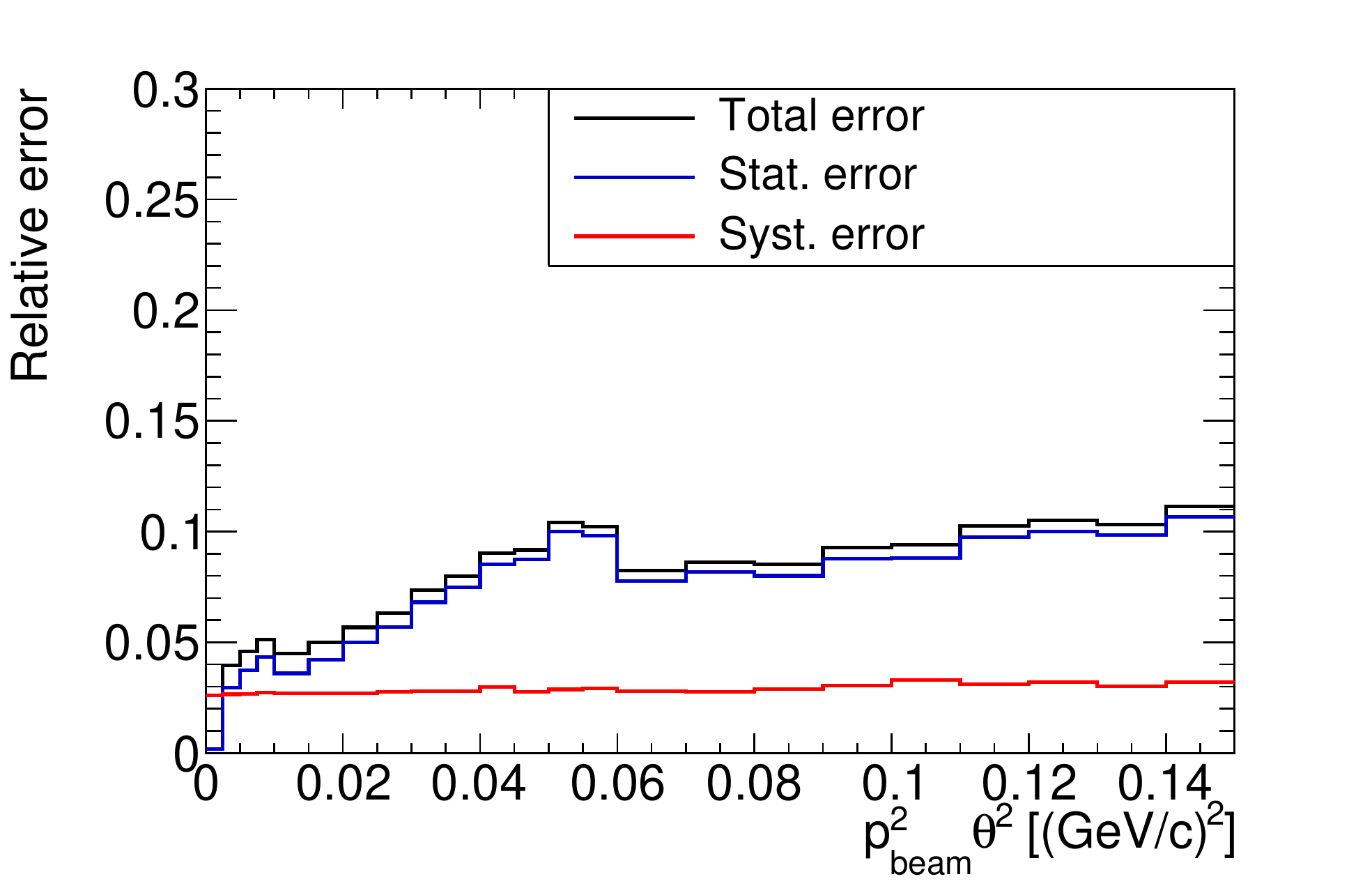}
    \caption{}
  \end{subfigure}
  \begin{subfigure}[t]{0.48\textwidth}
        \includegraphics[trim= 0 0 0 0, clip, width=1\textwidth]{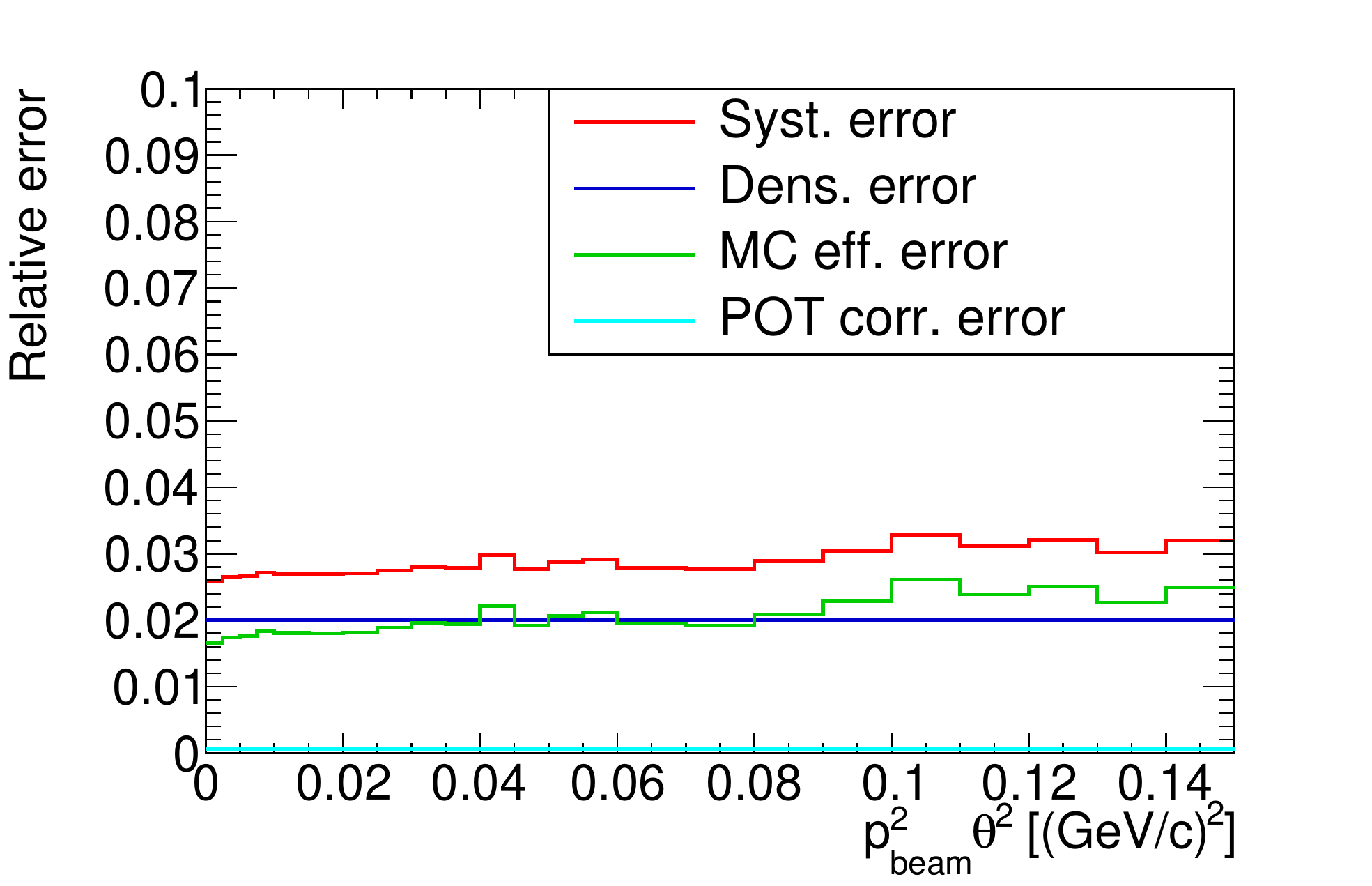}
    \caption{}
  \end{subfigure}

  \begin{subfigure}[t]{0.48\textwidth}
        \includegraphics[trim= 0 0 0 0, clip, width=1\textwidth]{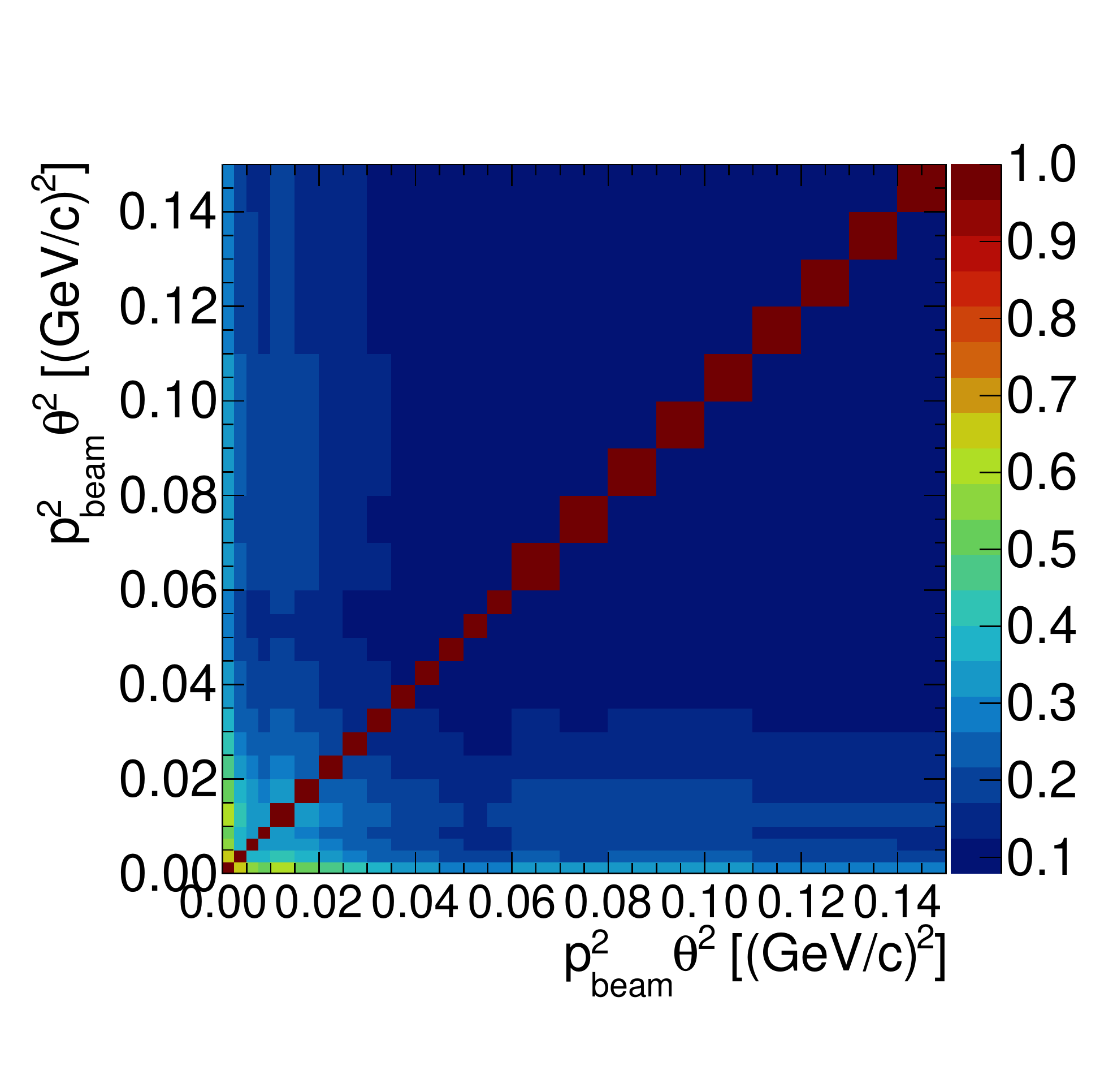}
    \caption{}
  \end{subfigure}
      
	\caption{The total, statistical and systematic uncertainty for p+C differential cross-section at $30\:$GeV/c (a) and different contributions to the systematic uncertainty (b), and correlation matrix (c).}\label{fig:pC30GeVErr}
\end{figure*}

\begin{figure*}[ht]
  \begin{subfigure}[t]{0.48\textwidth}
        \includegraphics[trim= 0 0 0 0, clip, width=1\textwidth]{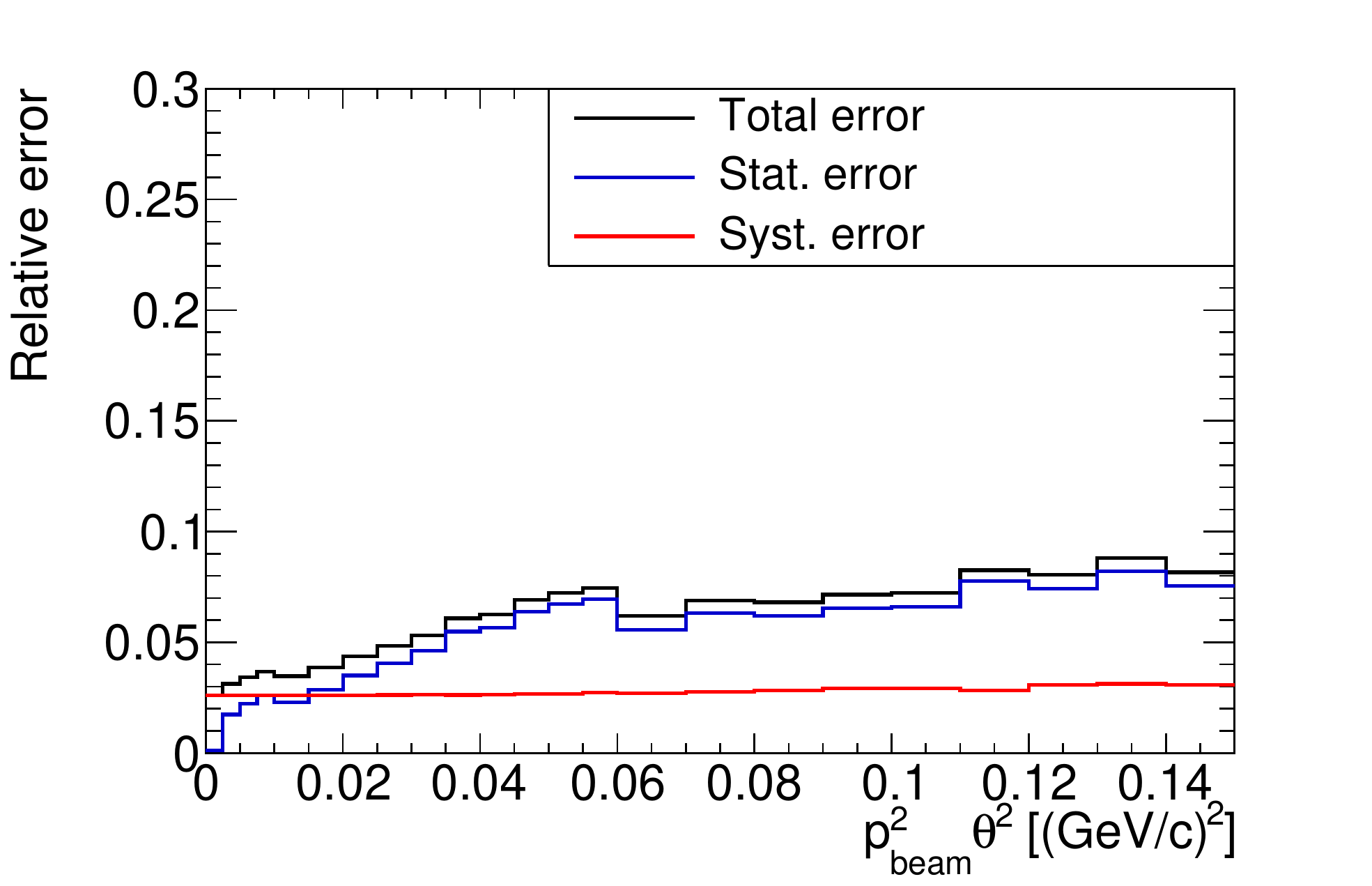}
    \caption{}
  \end{subfigure}
  \begin{subfigure}[t]{0.48\textwidth}
        \includegraphics[trim= 0 0 0 0, clip, width=1\textwidth]{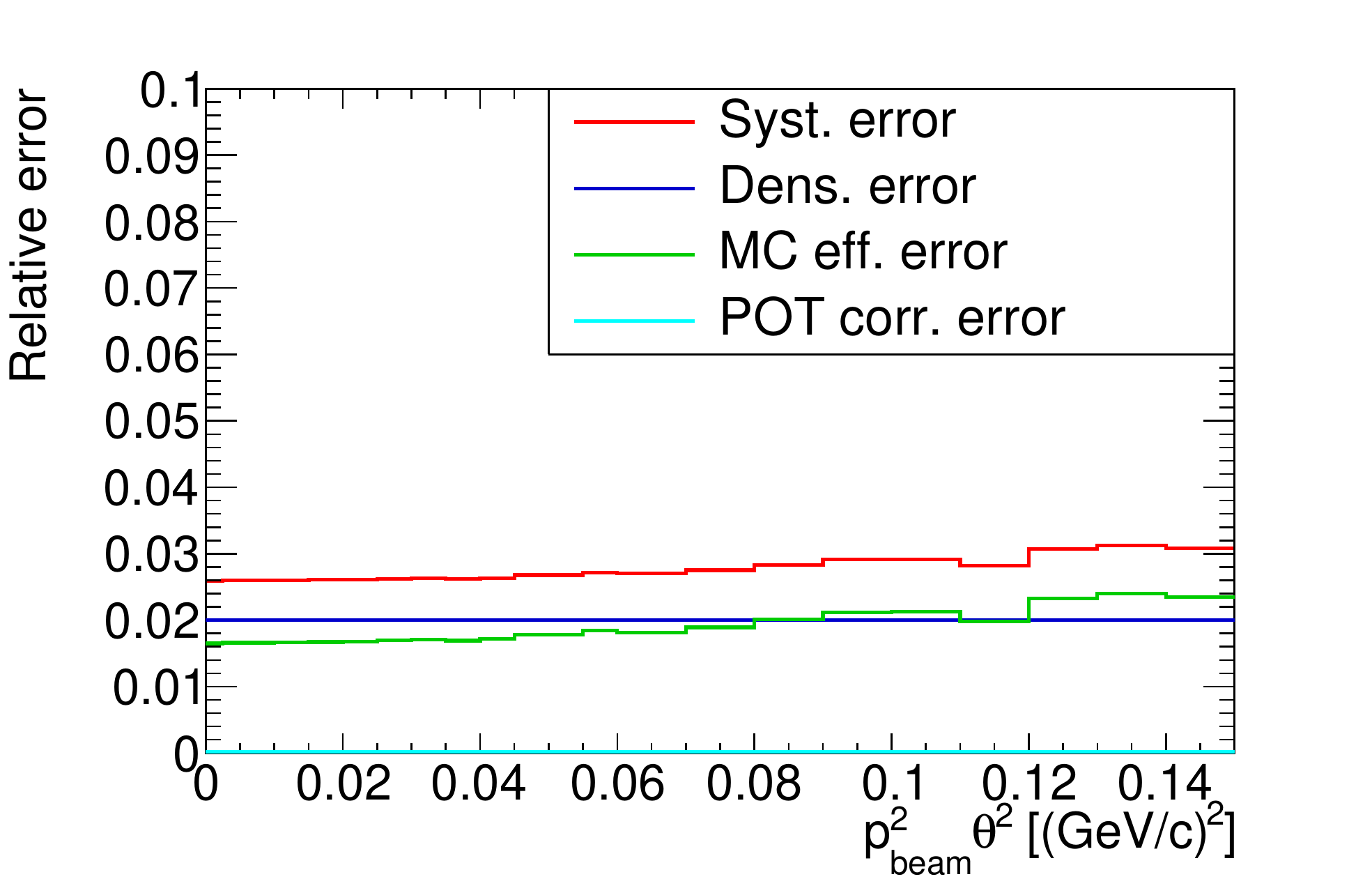}
    \caption{}
  \end{subfigure}

  \begin{subfigure}[t]{0.48\textwidth}
        \includegraphics[trim= 0 0 0 0, clip, width=1\textwidth]{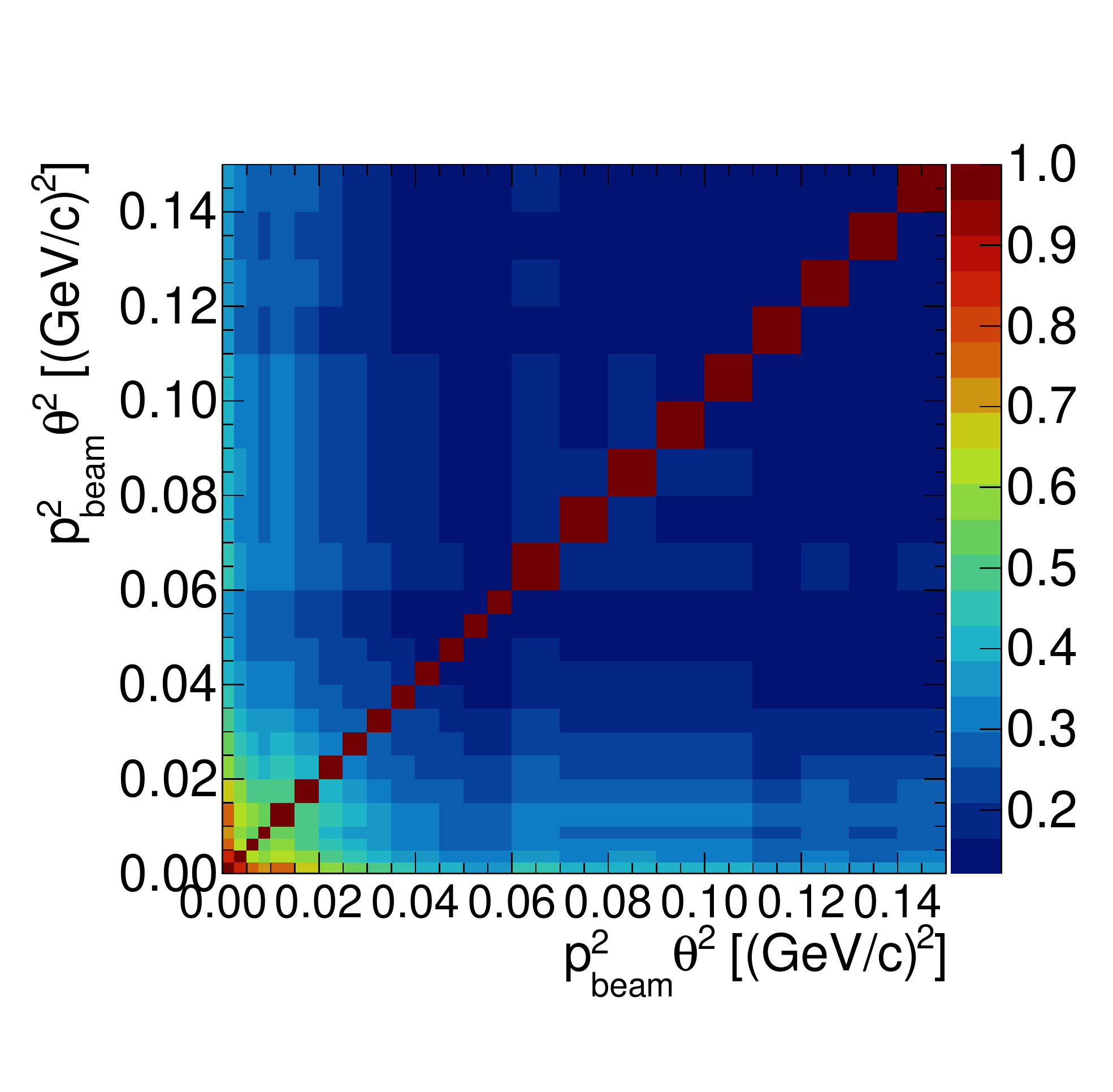}
    \caption{}
  \end{subfigure}
      
	\caption{The total, statistical and systematic uncertainty for p+C differential cross-section at $120\:$GeV/c (a) and different contributions to the systematic uncertainty (b), and correlation matrix (c).}\label{fig:pC120GeVErr}
\end{figure*}

%\subsection{Differential cross-section}
 Differential cross-section results are presented in Fig.~\ref{fig:pCRes}. Comparisons with FTFP\_BERT and QGSP\_BERT physics lists are also included. Both physics list give a similar predictions with significant differences from the data. The differences vary from $0$ to $40\%$. Total, statistical, and systematic uncertainties are also presented in Figs.~\ref{fig:pC20GeVErr}-\ref{fig:pC120GeVErr}.

%\begin{figure*}[ht]
%	\caption{The p+C differential cross-section at $30\:$GeV/c (a) and the comparison with FTFP\_BERT and QGSP\_BERT models from Geant 4.10.05.p01 (b).}\label{fig:pC30GeVRes}
%\end{figure*}

%\begin{figure*}[ht]
%	\caption{The p+C differential cross-section at $120\:$GeV/c (a) and the comparison with FTFP\_BERT and QGSP\_BERT models from Geant 4.10.05.p01 (b).}\label{fig:pC120GeVRes}
%\end{figure*}

\subsection{Model fits}
%Differential cross-section results provide the least model-dependent measurements. However, 
To compare our results with previous data it is necessary to extract total, coherent-elastic or inelastic cross-sections. Typically, the total cross-section can be extracted by doing transmission measurement in which the number of surviving beam particles is related to the total cross-section:
\begin{equation}\label{eq:surv}
    N_{S} = N_{0}e^{-nd\sigma_{tot}},
\end{equation}
where $N_{S}$ is the number of surviving beam particles, $N_{0}$ is the initial number of beam particles, $nd$ is the target number density multiplied by the target thickness, and $\sigma_{tot}$ is the total cross-section. An alternative approach is to measure the differential cross-section and use the optical theorem to extract the total cross-section. The optical theorem states that the total-cross section is proportional to the imaginary part of the scattering amplitude at $t=0\:$GeV$^{2}$.
An advantage of EMPHATIC is in the ability to do a combined measurement with both techniques, since all interacting and non-interacting events have been recorded. 

Coherent elastic and quasi-elastic proton-carbon interactions are governed by non-perturbative QCD, and there are no simple QCD predictions for our measurements. However, we have fitted a simple phenomenological model to the data to extract desired parameters. Similar models have been used in many older measurements such as~\cite{Bellettini:1966zz} and more recently in proton-proton interactions in ATLAS experiment~\cite{Adamczyk:2015gfy, Aaboud:2016ijx}. The full model used in the fit is:
\begin{widetext}
\begin{subequations}\label{eq:fullmodel}
\begin{eqnarray}
%\begin{equation}\label{eq:elastic}
M(p^2\theta^2; A,B,B_{pN},B_{I},\sigma_{tot},\sigma_{tot, pN},C, \Lambda) = \frac{1}{16\pi}\left(\frac{\sigma_{tot}}{\hbar c}\right)^2(1+\rho^2)e^{-Bp^2\theta^2}\label{eq:elastic}
%\end{equation}
\\
%\begin{equation}\label{eq:Coulomb}
+ \frac{1}{16\pi}\mathcal{H}(p^2\theta^2-\Lambda)\left(\frac{8\pi\alpha Z_{1}Z_{2}\hbar c}{p^2\theta^2}\right)^2e^{-Bp^2\theta^2}\label{eq:Coulomb}
%\end{equation}
\\
%\begin{equation}\label{eq:interference}
-2\frac{1}{16\pi}\mathcal{H}(p^2\theta^2-\Lambda)(\rho\cos{\Delta\Phi} + \sin{\Delta\Phi})\frac{\sigma_{tot}}{\hbar c}\frac{8\pi\alpha Z_{1}Z_{2}\hbar c}{p^2\theta^2}e^{-Bp^2\theta^2}\label{eq:interference}
%\end{equation}
\\
%\begin{equation}\label{eq:quasi-elastic}
+ N(A)\cdot\frac{1}{16\pi}\left(\frac{\sigma_{tot, pN}}{\hbar c}\right)^{2}e^{-B_{pN}p^2\theta^2}\label{eq:quasi-elastic}
%\end{equation}
\\
%\begin{equation}\label{eq:inelastic}
+ \frac{1}{16\pi}\left(\frac{C}{\hbar c}\right)^{2}e^{-B_{I}p^2\theta^2}\label{eq:inelastic}
%\end{equation}
\\
%\begin{equation}\label{eq:non-int}
+\mathcal{H}(10^{-4}[GeV^{2}]-p^2\theta^2)\frac{1}{nd\cdot 10^{-4}[GeV^{2}]}e^{-(\sigma_{tot}+\sigma_{C-el}+\sigma_{C})\cdot n\cdot d}\label{eq:non-int}
%\end{equation}
\end{eqnarray}
\end{subequations}
\end{widetext}
The model assumes that the nucleon distribution inside carbon nucleus follows the normal distribution and the coherent-elastic differential cross-section Eq.~\ref{eq:elastic} is an exponential function of the four-momentum transfer ($t\approx p^{2}\theta^{2}$). The normalization is determined by the optical theorem and relates to the total cross-section $\sigma_{tot}$, while the exponential parameter $B$ is proportional to the sum of squares of proton and nuclear radii. Often, it is assumed that the coherent nuclear amplitude is purely imaginary. However, this is not well measured and we assume a small real amplitude as well. The ratio of the real to imaginary part of the amplitude is denoted as $\rho$.

The Coulomb differential cross-section (Eq.~\ref{eq:Coulomb}) is a simple Rutherford formula with the exponential form factor. It is assumed that the carbon nuclear charge density follows the nucleon distribution. This assumption follows from the work by Kopeliovich and Tarasov~\cite{Kopeliovich:2000ez}. The Coulomb scattering is divergent at $t=0\:$GeV$^2$ and we include a cutoff parameter $\Lambda$ to remove the divergence. 

Possible interference between coherent-elastic and Coulomb scattering is included in the third line of Eq.~\ref{eq:fullmodel}. Interference is zero if the nuclear amplitude is purely imaginary. However, we have already assumed that this is not the case. In addition, the nuclear amplitude gains a modifying phase $\Delta\Phi$ due to the presence of Coulomb field. 

Quasi-elastic interactions are defined as elastic interactions on a single nucleon. Therefore, quasi-elastic differential cross-section (Eq.~\ref{eq:quasi-elastic}) will have the same $t$ dependence as coherent-elastic differential cross-section, but only with different parameters. The effective number of nucleons visible to a beam proton is denoted as $N(A)$

Inelastic contamination (Eq.~\ref{eq:inelastic}) in our measurements is mostly coming from events with a single forward charged particle (mostly pions). We assume that such background mostly comes from $\Delta$ resonance production. Also, Geant4 simulation suggests that the background is almost flat in $p^2\theta^2$ or has a exponential shape with a small slope. Therefore, we assume the same functional dependence as coherent-elastic and quasi-elastic differential cross-sections, only with different parameters.

Finally, non-interacting contributions (Eq.~\ref{eq:non-int}) is calculated by using Eq.~\ref{eq:surv}. Instead of using only the total nuclear proton-carbon cross-section, we have also integrated Eq.~\ref{eq:Coulomb} and Eq.~\ref{eq:interference} to get the total effective cross-section. The number of surviving particles is normalized to get the same dimension as the differential cross-section. The Heaviside step function ensures that surviving beam particles are only placed in in the first true $p^{2}\theta^{2}$ bin ($0-10^{-4}\:$GeV$^2$).

\begin{figure*}[ht]
    \begin{center}
        \includegraphics[trim= 0 0 0 0, clip, width=0.75\textwidth]{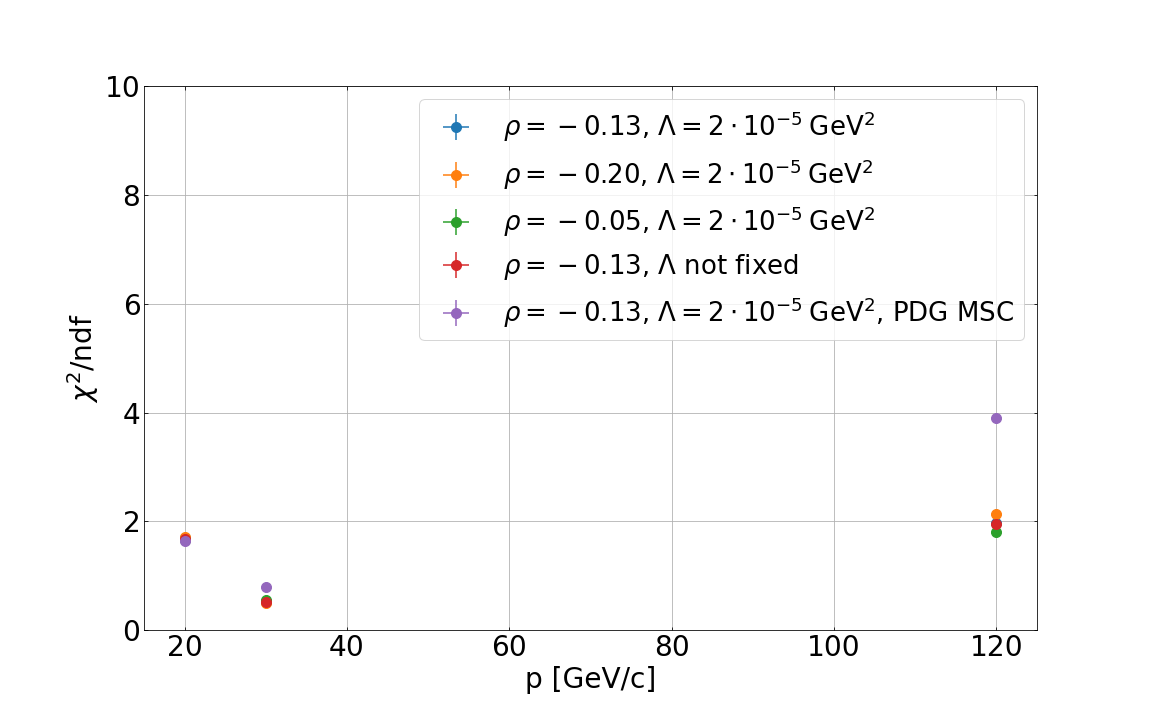}
    \end{center}
	\caption{The $\chi^2$/ndf values for all fit configurations.}\label{fig:chi2fit}
\end{figure*}

\begin{figure*}[ht]
  \begin{subfigure}[t]{0.48\textwidth}
        \includegraphics[trim= 0 0 0 0, clip, width=1\textwidth]{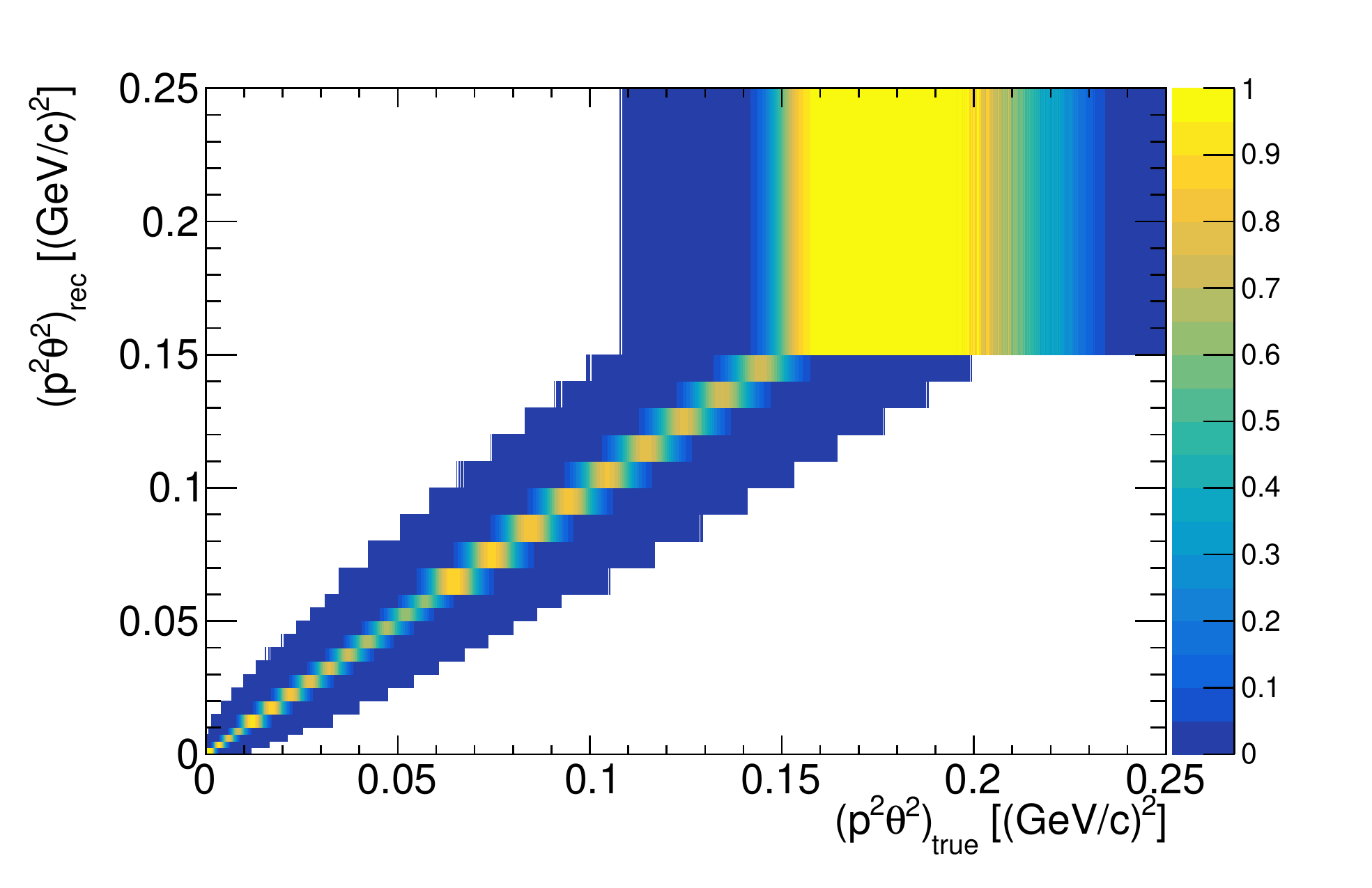}
    \caption{}
  \end{subfigure}
  \begin{subfigure}[t]{0.48\textwidth}
        \includegraphics[trim= 0 0 0 0, clip, width=1\textwidth]{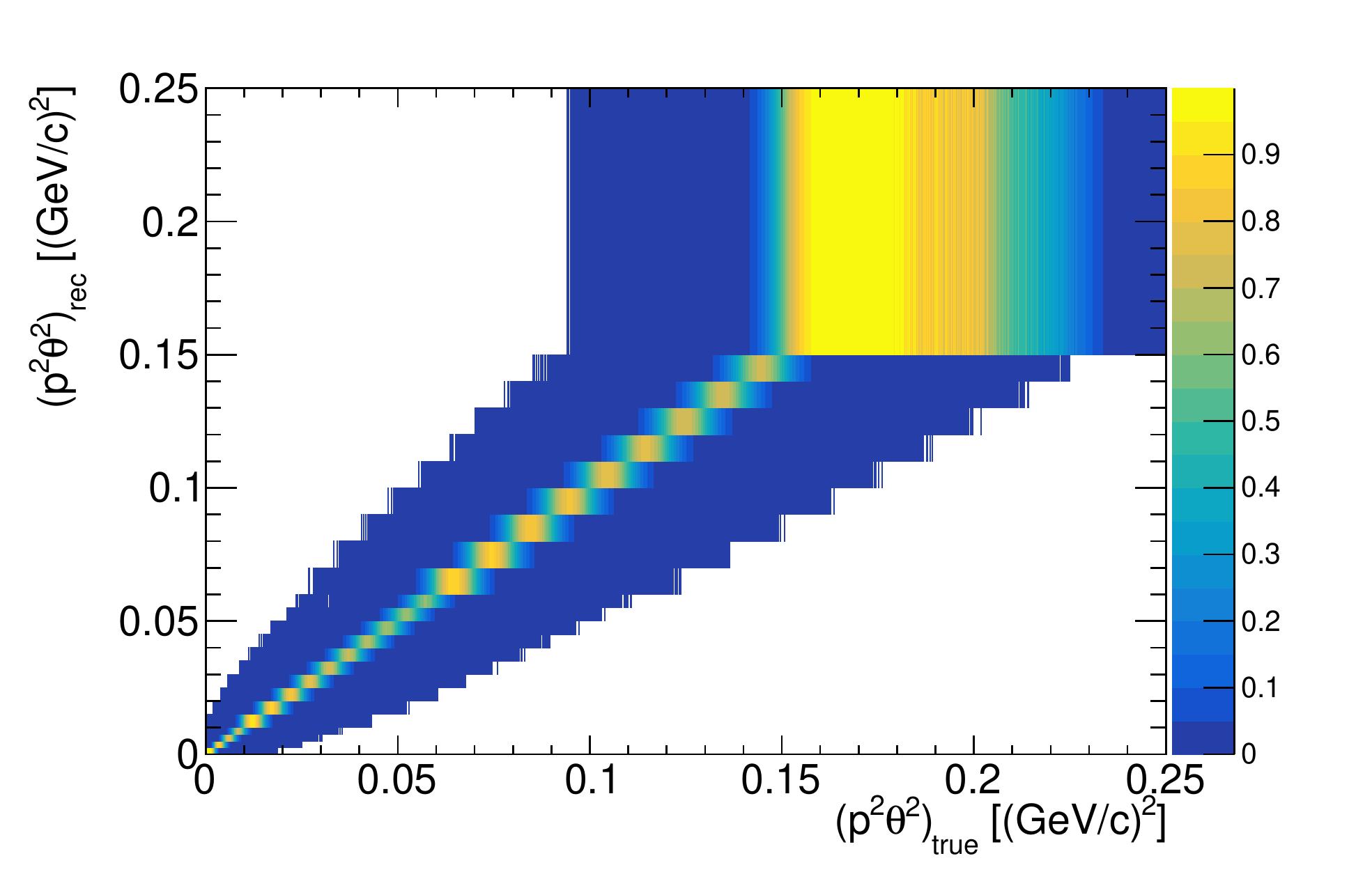}
    \caption{}
  \end{subfigure}
  \begin{subfigure}[t]{0.48\textwidth}
        \includegraphics[trim= 0 0 0 0, clip, width=1\textwidth]{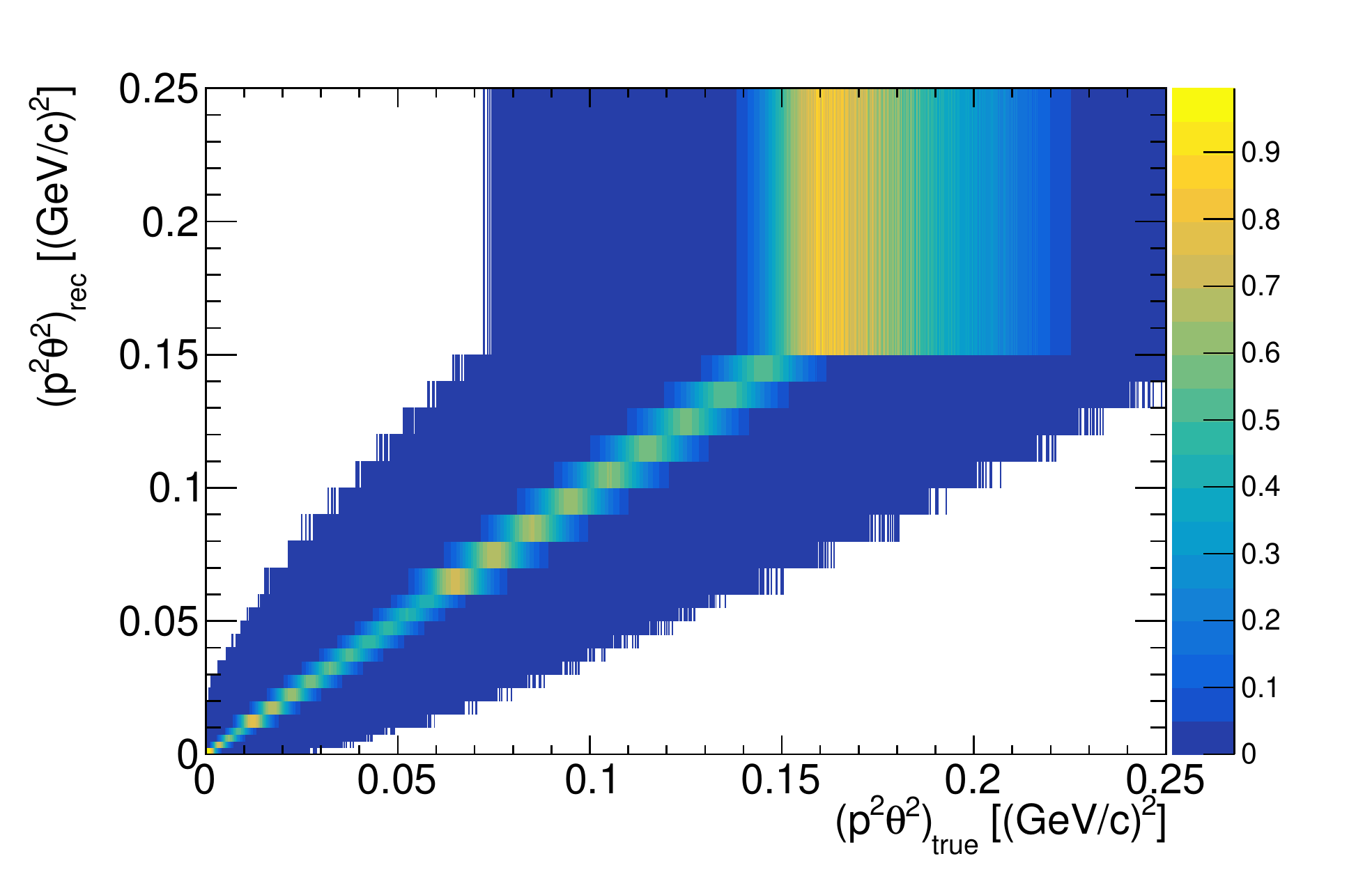}
    \caption{}
  \end{subfigure}
    
	\caption{Migration matrices for $20\:$GeV/c data (a), $30\:$GeV/c data (b), $120\:$GeV/c data (c).}\label{fig:migMat}
\end{figure*}

\begin{figure*}[ht]
    \begin{center}
        \includegraphics[trim= 0 0 0 0, clip, width=0.75\textwidth]{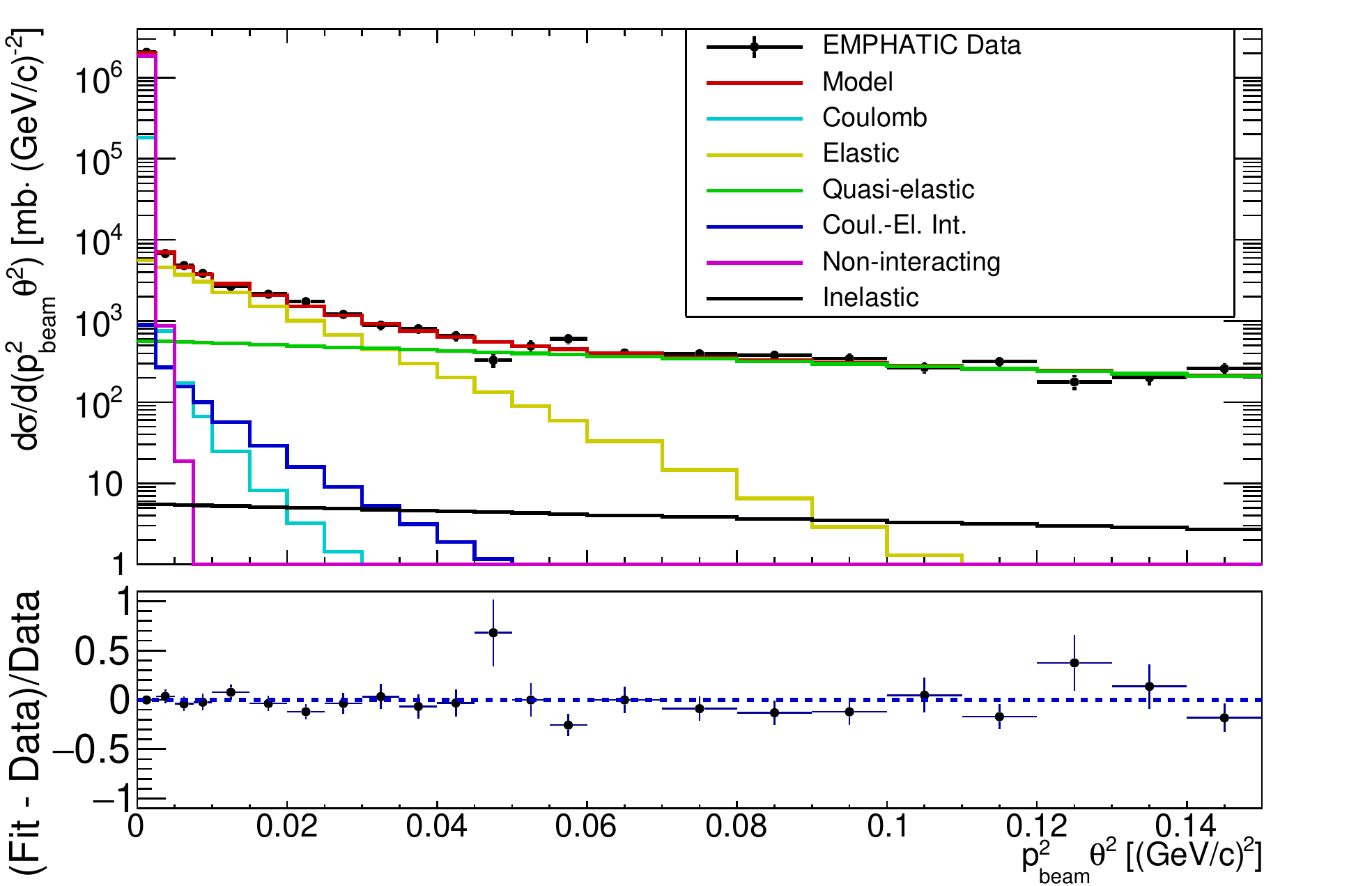}
    \end{center}
	\caption{The model fit for $20\:$GeV/c data.}\label{fig:tfit20GeV/c}
\end{figure*}

\begin{figure*}[ht]
    \begin{center}
        \includegraphics[trim= 0 0 0 0, clip, width=0.75\textwidth]{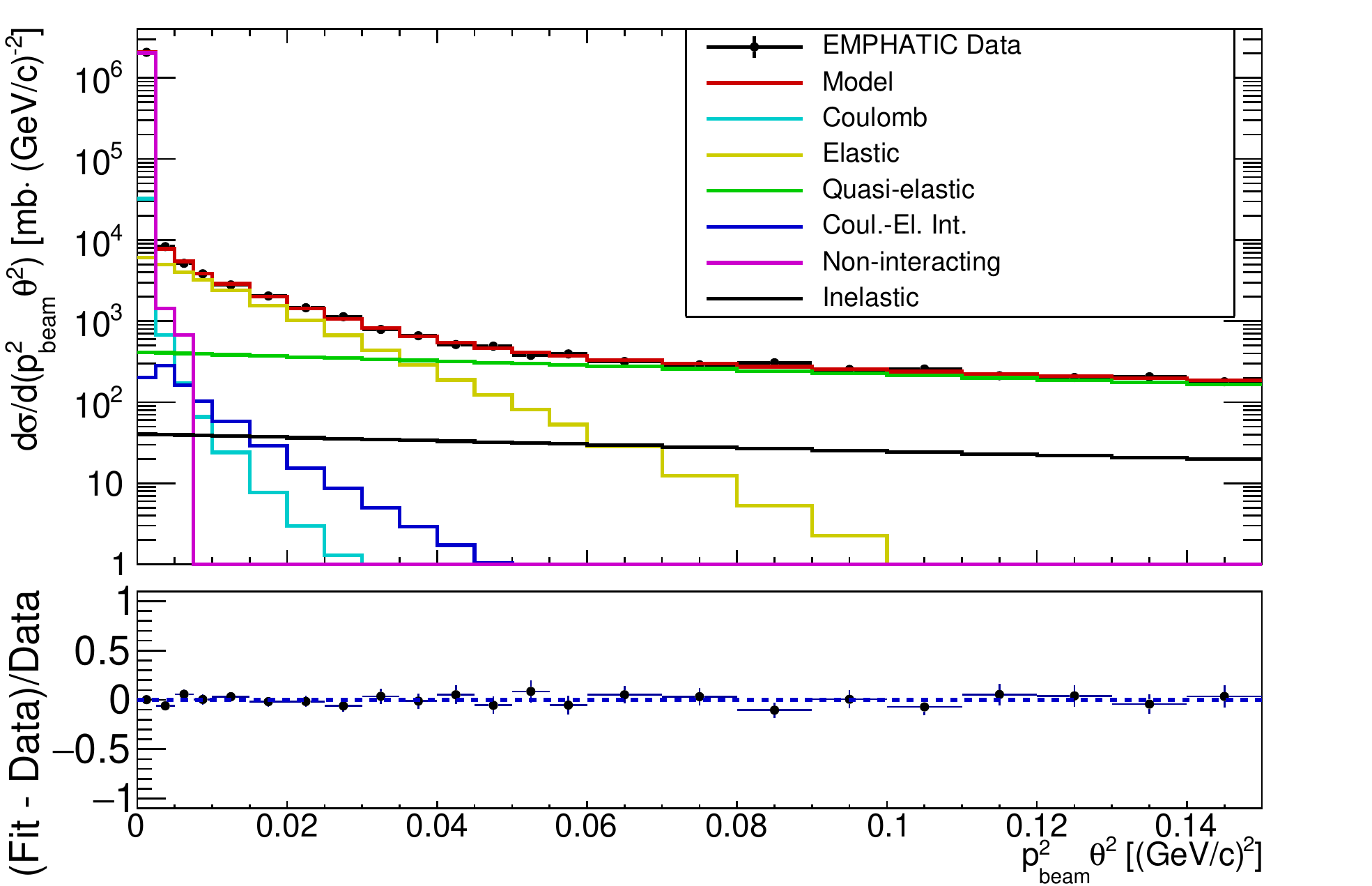}
    \end{center}
	\caption{The model fit for $30\:$GeV/c data.}\label{fig:tfit30GeV/c}
\end{figure*}

\begin{figure*}[ht]
    \begin{center}
        \includegraphics[trim= 0 0 0 0, clip, width=0.75\textwidth]{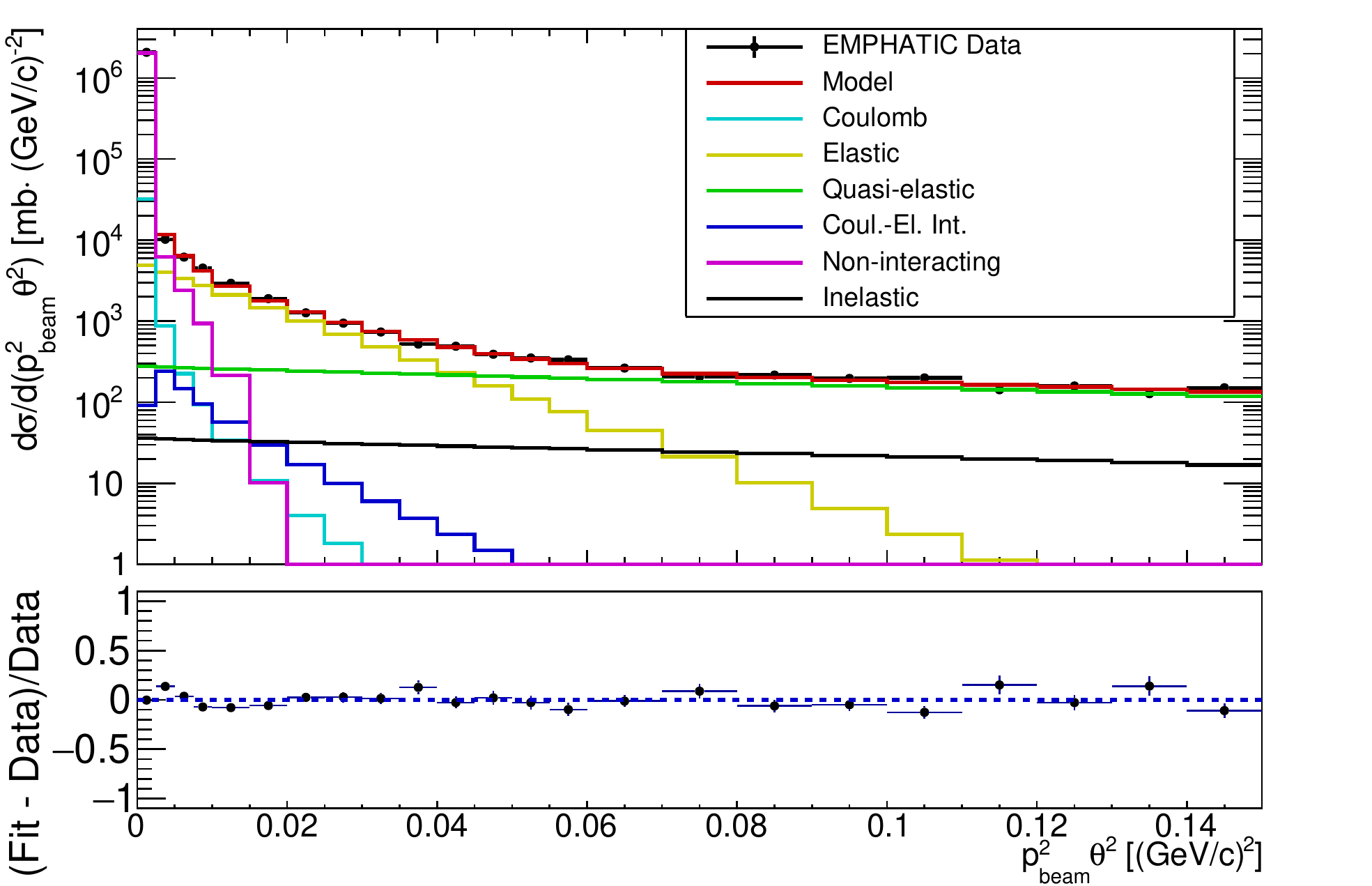}
    \end{center}
	\caption{The model fit for $120\:$GeV/c data.}\label{fig:tfit120GeV/c}
\end{figure*}

The full model is first smeared to account for the bin migration. Migration matrices are generated from the empty target data and the Geant4 simulation of the multiple scattering in the target (see Fig.~\ref{fig:migMat}). The $\chi^2$ is minimized by varying the model parameters. The relative phase $\Delta\Phi$ is not an independent parameter and it is taken from the calculation by Kopeliovich and Tarasov~\cite{Kopeliovich:2000ez}. The $\rho$ parameter is set to be a constant with a value of $-0.13$, taken from~\cite{Kopeliovich:2000kz}.
Since $N(A)$ and $\sigma_{tot, pN}$ appear only as a product, they are merged into a single parameter. 
The fits are presented in Figs.~\ref{fig:tfit20GeV/c} - ~\ref{fig:tfit120GeV/c}. All of the model contributions are also shown. The $\chi^2$/ndf values for all fit configuration are presented in Fig.~\ref{fig:chi2fit}. 
%The $\chi^2$/ndf values for $20$ and $30\:$GeV/c data are around $1$ for $6$ degrees of freedom, while the value for $120\:$GeV/c data is $2$. The increase seems to be caused by the second bin which has significantly lower value in the data. 
Since the inelastic background has the same functional $p^2\theta^2$ dependence as the quasi-elastic differential cross-section and they are similar in size, corresponding fit parameters have large uncertainties. Therefore, we are not confident in reporting these values independently. 

However, we extracted the elastic slope $B$ and the total cross-section $\sigma_{tot}$. The total coherent elastic cross-section $\sigma_{el}$ is calculated by integrating Eq.~\ref{eq:elastic}. Additionally, the total inelastic cross-section is estimated as $\sigma_{tot}-\sigma_{el}$.
To estimate systematic uncertainties, fits are repeated in different configurations. The $\rho$ parameter is varied between $-0.20$ and $-0.05$ according to~\cite{Kopeliovich:2000kz}. The migration matrix is recalculated by using the PDG approximation of the multiple scattering in the target:
\begin{equation}
\theta_{0} = \frac{13.6 \text{MeV}}{\beta cp}z\sqrt{\frac{x}{X_{0}}}
\left[1+0.038\log \left(\frac{x}{X_{0}}\right)\right]\label{eq:msc}
\end{equation}
The initial value of $\Lambda$ parameter is varied within its uncertainties and fixed during the fit. The value of $\Lambda$ is around $2\cdot10^{-5}\:$GeV$^2$ for all datasets. The values of the extracted parameters and their uncertainties for different datasets are summarized in Tab.~\ref{tab:FitParams}.

\begin{table*}[ht]
 \caption{Extracted parameters and their uncertainties}
  \centering
  \begin{small}
  \begin{tabular}{cccccc}
    \toprule
    $p$ [GeV/c] & Parameter & $B$ [(GeV)$^{-2}$] & $\sigma_{tot}$ [mb]& $\sigma_{el}$ [mb]& $\sigma_{inel}$ [mb]\\
    \midrule
    $20$ & Value & $80.9$ & $344.3$ & $74.7$ & $269.3$ \\
   &  Stat. error & $7.1$ & $12.3$ & $8.4$ & $14.9$  \\
   & Syst. error (low) & $-1.6$ & $-10.0$ & $-2.86$ & $-7.1$  \\
   & Syst. error (high) & $+2.5$ & $+11.7$ & $+3.0$ & $+10.2$ \\
    \midrule
    $30$ & Value & $85.1$  & $362.8$ & $79.1$ & $283.8$  \\
   &  Stat. error & $4.1$ & $8.2$ & $5.2$ & $9.7$  \\
   & Syst. error (low) & $-1.5$ & $-10.0$ & $-3.0$ & $-7.0$  \\
   & Syst. error (high) & $+3.1$ & $+12.2$ & $+2.8$ & $+9.6$ \\
    \midrule
    $120$ & Value & $79.3$  & $335.9$ & $72.7$ & $263.3$  \\
   &  Stat. error & $2.9$ & $6.7$ & $3.9$ & $7.8$  \\
   & Syst. error (low) & $-3.9$ & $-15.6$ & $-3.2$ & $-9.3$  \\
   & Syst. error (high) & $+2.4$ & $+9.1$ & $+2.0$ & $+7.0$ \\
    \bottomrule
  \end{tabular}
    \end{small}
  \label{tab:FitParams}
\end{table*} 

\begin{figure*}[ht]
    \begin{center}
  \begin{subfigure}[t]{0.49\textwidth}
        \includegraphics[trim= 0 0 0 0, clip, width=1\textwidth]{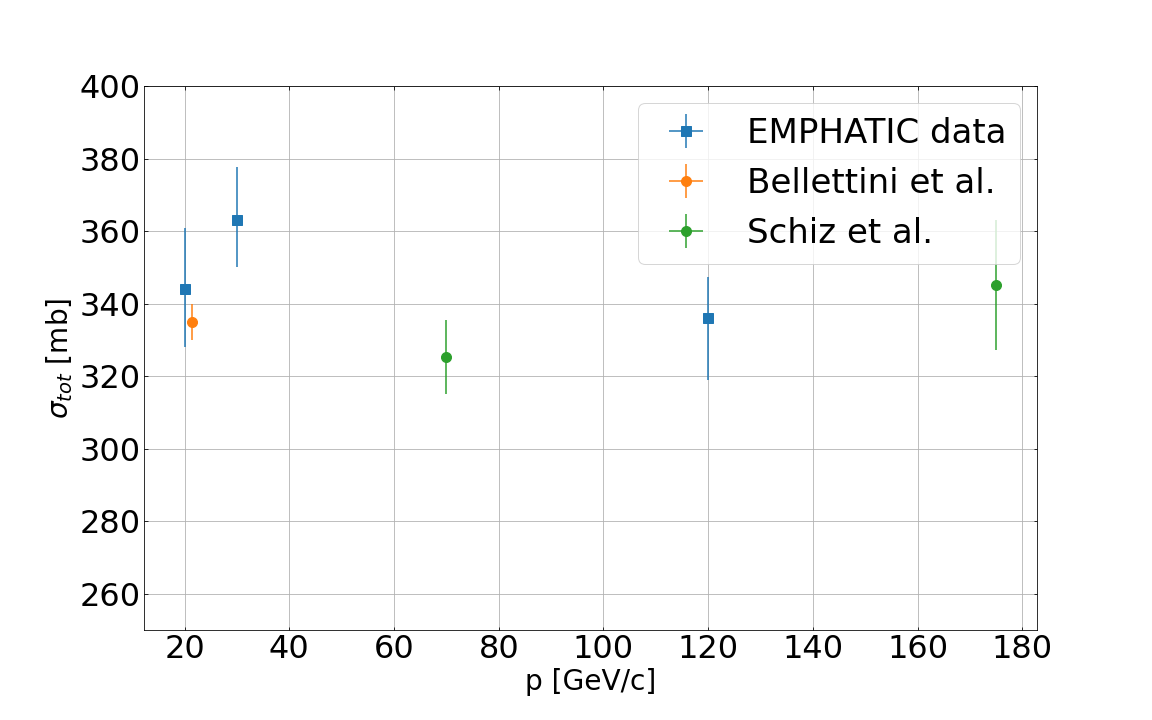}
    \caption{}
  \end{subfigure}
  \begin{subfigure}[t]{0.49\textwidth}
        \includegraphics[trim= 0 0 0 0, clip, width=1\textwidth]{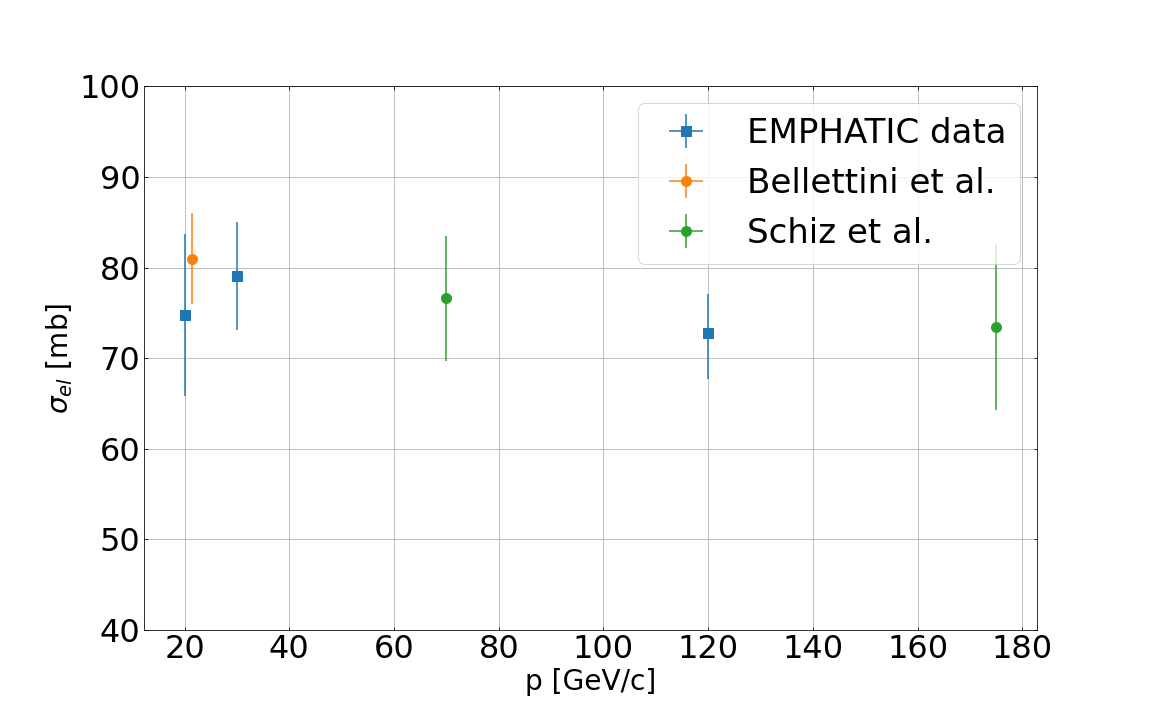}
    \caption{}
  \end{subfigure}
  \begin{subfigure}[t]{0.49\textwidth}
        \includegraphics[trim= 0 0 0 0, clip, width=1\textwidth]{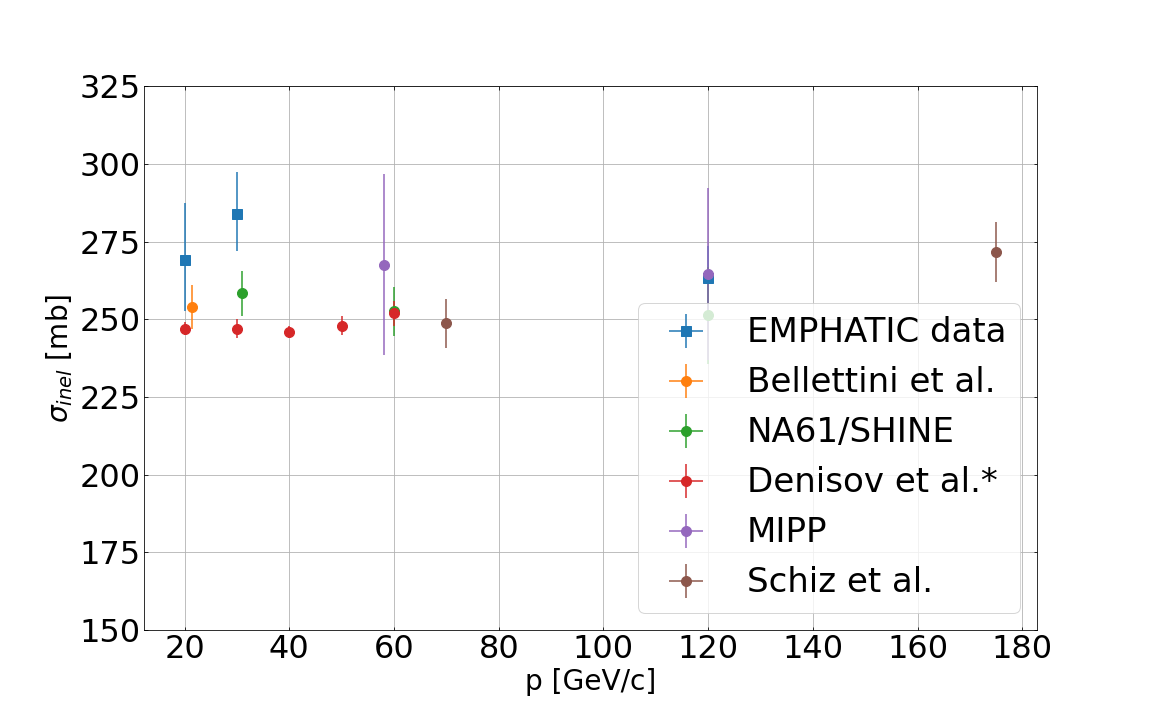}
    \caption{}
  \end{subfigure}
    \end{center} 
	\caption{Comparisons of the total (a), elastic (b), and inelastic cross-section (c) obtained from the fits with older data. The total cross-section is compared to the data from Bellettini et al.~\cite{Bellettini:1966zz}. The elastic cross-section is compared to values obtained from Bellettini et al.~\cite{Bellettini:1966zz} and Schiz et al.~\cite{Schiz:1979af}. The inelastic cross-section is compared to the results from  Bellettini et al.~\cite{Bellettini:1966zz}, NA61/SHINE collaboration ~\cite{Abgrall:2015hmv, Aduszkiewicz:2019xna}, Denisov et al.~\cite{Denisov:1973zv}, and MIPP collaboration~\cite{Mahajan:2013awa}.}\label{fig:fitParComp}
\end{figure*} 

A comparison of the cross-section results with the older measurements is presented in Fig.~\ref{fig:fitParComp}. The statistical and systematic uncertainties of the extracted cross-sections are similar in size. The main reason for the large systematic uncertainty is coming from the variation of the $\rho$ parameter. Unfortunately, our current data and methods do not allow us to extract the $\rho$ parameter with any precision due to large migration between the first four upstream bins. To improve this in the future, we will reduce the migration effect by reducing the material budget. Additionally, removing inelastic backgrounds by using a momentum measurement will allow us to measure the quasi-elastic cross-section.

\section{Summary and future measurements}
We have measured forward differential cross-section in proton-carbon at $20$, $30$, and $120\:$GeV/c with a simple setup consisting of silicon strip detectors. These results feature a novel technique that uses both transmission measurement and the optical theorem to extract the cross-section. Future EMPHATIC runs will have momentum measurements and the particle identification for secondary particles. Additionally, the material budget will be significantly reduced (by $40\%$), since the dead material (pixel telescope) will be removed. With increased statistics and these improvements, we will be able to greatly reduce our uncertainties on the extracted cross-sections. A fast DAQ rate will allow us to take many different datasets and create a cross-section table for the important interactions contributing to the neutrino flux in different experiments.

\section{Acknowledgements}
This document was prepared by members of the EMPHATIC Collaboration using the re- sources of the Fermi National Accelerator Laboratory (Fermilab), a U.S. Department of Energy, Office of Science, HEP User Facility. Fermilab is managed by Fermi Research Alliance, LLC (FRA), acting under Contract No. DE-AC02-07CH11359.  Support for participating scientists was provided by the DOE (USA), the 2017 U.S.-Japan Science and Technology Cooperation Program in High Energy Physics, and NSERC and NRC, Canada. We thank the Fermilab Test Beam Facility for its excellent technical support, use of its detectors and beam.

%\printbibliography
%\blindtext \cite{article-minimal}

\bibliographystyle{apsrev4-1} % Tell bibtex which bibliography style to use
\bibliography{references.bib} % Tell bibtex which .bib file to use (this one is some example file in TexLive's file tree)

\appendix
\section{Results}

\begin{table*}[ht]
 \caption{Measured differential cross-section for $20\:$GeV/c data.}
  \centering
  \begin{small}
  \begin{tabular}{cccccccc}
    \toprule
    $p^{2}\theta^2$ range  & $d\sigma/d(p^2\theta^2)$  & Stat. error & Stat. error & Syst. error & Syst. error & Tot. error & Tot. error \\
    
    [GeV/c] &  [mb$\cdot$(GeV/c)$^{-2}$] & [mb$\cdot$(GeV/c)$^{-2}$] & [\%] & [mb$\cdot$(GeV/c)$^{-2}$] & [\%] & [mb$\cdot$(GeV/c)$^{-2}$] & [\%] \\
    \midrule
$0.0000 - 0.0025$ & $2052289.364$ & $7126.864$ & $0.347$ & $53265.051$ & $2.595$ & $53739.724$ & $2.619$\\
$0.0025 - 0.0050$ & $6793.436$ & $412.672$ & $6.075$ & $180.526$ & $2.657$ & $450.430$ & $6.630$\\
$0.0050 - 0.0075$ & $4828.061$ & $349.346$ & $7.236$ & $132.106$ & $2.736$ & $373.489$ & $7.736$\\
$0.0075 - 0.0100$ & $3857.142$ & $312.856$ & $8.111$ & $105.574$ & $2.737$ & $330.188$ & $8.560$\\
$0.0100 - 0.0150$ & $2674.627$ & $184.129$ & $6.884$ & $74.013$ & $2.767$ & $198.447$ & $7.420$\\
$0.0150 - 0.0200$ & $2142.355$ & $165.286$ & $7.715$ & $59.941$ & $2.798$ & $175.820$ & $8.207$\\
$0.0200 - 0.0250$ & $1732.284$ & $149.091$ & $8.607$ & $49.076$ & $2.833$ & $156.961$ & $9.061$\\
$0.0250 - 0.0300$ & $1206.000$ & $124.389$ & $10.314$ & $33.697$ & $2.794$ & $128.873$ & $10.686$\\
$0.0300 - 0.0350$ & $888.948$ & $107.017$ & $12.039$ & $25.047$ & $2.818$ & $109.909$ & $12.364$\\
$0.0350 - 0.0400$ & $803.549$ & $102.051$ & $12.700$ & $23.286$ & $2.898$ & $104.674$ & $13.026$\\
$0.0400 - 0.0450$ & $655.121$ & $92.648$ & $14.142$ & $18.295$ & $2.793$ & $94.437$ & $14.415$\\
$0.0450 - 0.0500$ & $328.323$ & $65.665$ & $20.000$ & $9.419$ & $2.869$ & $66.337$ & $20.205$\\
$0.0500 - 0.0550$ & $493.315$ & $81.101$ & $16.440$ & $15.138$ & $3.069$ & $82.501$ & $16.724$\\
$0.0550 - 0.0600$ & $603.047$ & $89.897$ & $14.907$ & $18.037$ & $2.991$ & $91.689$ & $15.204$\\
$0.0600 - 0.0700$ & $403.360$ & $52.513$ & $13.019$ & $11.561$ & $2.866$ & $53.771$ & $13.331$\\
$0.0700 - 0.0800$ & $394.693$ & $52.278$ & $13.245$ & $12.9776$ & $3.288$ & $53.865$ & $13.647$\\
$0.0800 - 0.0900$ & $377.595$ & $51.867$ & $13.736$ & $12.053$ & $3.192$ & $53.249$ & $14.102$\\
$0.0900 - 0.1000$ & $344.685$ & $49.751$ & $14.434$ & $10.870$ & $3.154$ & $50.925$ & $14.774$\\
$0.1000 - 0.1100$ & $268.391$ & $44.123$ & $16.440$ & $8.946$ & $3.333$ & $45.021$ & $16.774$\\
$0.1100 - 0.1200$ & $315.072$ & $48.048$ & $15.250$ & $10.701$ & $3.396$ & $49.225$ & $15.624$\\
$0.1200 - 0.1300$ & $177.247$ & $36.180$ & $20.412$ & $5.573$ & $3.144$ & $36.607$ & $20.653$\\
$0.1300 - 0.1400$ & $199.990$ & $39.221$ & $19.612$ & $6.306$ & $3.153$ & $39.725$ & $19.864$\\
$0.1400 - 0.1500$ & $258.929$ & $45.074$ & $17.408$ & $8.222$ & $3.175$ & $45.817$ & $17.695$\\

    \bottomrule
  \end{tabular}
    \end{small}
  \label{tab:pC20GeVRes}
\end{table*}

\begin{table*}[ht]
 \caption{Measured differential cross-section for $30\:$GeV/c data.}
  \centering
  \begin{small}
  \begin{tabular}{cccccccc}
    \toprule
    $p^{2}\theta^2$ range  & $d\sigma/d(p^2\theta^2)$  & Stat. error & Stat. error & Syst. error & Syst. error & Tot. error & Tot. error \\
    
    [GeV/c] &  [mb$\cdot$(GeV/c)$^{-2}$] & [mb$\cdot$(GeV/c)$^{-2}$] & [\%] & [mb$\cdot$(GeV/c)$^{-2}$] & [\%] & [mb$\cdot$(GeV/c)$^{-2}$] & [\%] \\
    \midrule
$0.0000 - 0.0025$ & $2069050.634$ & $3828.485$ & $0.185$ & $53689.117$ & $2.595$ & $53825.446$ & $2.602$\\
$0.0025 - 0.0050$ & $8299.600$ & $244.317$ & $2.944$ & $219.998$ & $2.651$ & $328.771$ & $3.961$\\
$0.0050 - 0.0075$ & $5170.900$ & $192.708$ & $3.727$ & $137.906$ & $2.667$ & $236.969$ & $4.583$\\
$0.0075 - 0.0100$ & $3837.658$ & $166.384$ & $4.336$ & $104.313$ & $2.718$ & $196.379$ & $5.117$\\
$0.0100 - 0.0150$ & $2799.597$ & $100.695$ & $3.597$ & $75.476$ & $2.696$ & $125.841$ & $4.495$\\
$0.0150 - 0.0200$ & $2044.561$ & $86.092$ & $4.211$ & $55.046$ & $2.692$ & $102.185$ & $4.998$\\
$0.0200 - 0.0250$ & $1463.394$ & $73.078$ & $4.994$ & $39.521$ & $2.701$ & $83.080$ & $5.677$\\
$0.0250 - 0.0300$ & $1134.254$ & $64.421$ & $5.680$ & $31.196$ & $2.750$ & $71.577$ & $6.311$\\
$0.0300 - 0.0350$ & $789.365$ & $53.710$ & $6.804$ & $22.106$ & $2.800$ & $58.081$ & $7.358$\\
$0.0350 - 0.0400$ & $660.948$ & $49.402$ & $7.474$ & $18.443$ & $2.790$ & $52.732$ & $7.978$\\
$0.0400 - 0.0450$ & $514.906$ & $43.832$ & $8.513$ & $15.346$ & $2.980$ & $46.440$ & $9.019$\\
$0.0450 - 0.0500$ & $491.541$ & $42.946$ & $8.737$ & $13.618$ & $2.770$ & $45.053$ & $9.166$\\
$0.0500 - 0.0550$ & $378.509$ & $37.851$ & $10.000$ & $10.880$ & $2.875$ & $39.384$ & $10.405$\\
$0.0550 - 0.0600$ & $393.893$ & $38.624$ & $9.806$ & $11.484$ & $2.916$ & $40.296$ & $10.230$\\
$0.0600 - 0.0700$ & $317.420$ & $24.637$ & $7.762$ & $8.860$ & $2.791$ & $26.181$ & $8.248$\\
$0.0700 - 0.0800$ & $289.509$ & $23.638$ & $8.165$ & $8.024$ & $2.772$ & $24.963$ & $8.623$\\
$0.0800 - 0.0900$ & $306.383$ & $24.530$ & $8.006$ & $8.860$ & $2.892$ & $26.081$ & $8.513$\\
$0.0900 - 0.1000$ & $253.683$ & $22.250$ & $8.771$ & $7.710$ & $3.039$ & $23.547$ & $9.282$\\
$0.1000 - 0.1100$ & $256.723$ & $22.603$ & $8.805$ & $8.440$ & $3.2886$ & $24.128$ & $9.398$\\
$0.1100 - 0.1200$ & $212.069$ & $20.696$ & $9.759$ & $6.613$ & $3.118$ & $21.727$ & $10.245$\\
$0.1200 - 0.1300$ & $201.800$ & $20.180$ & $10.000$ & $6.466$ & $3.204$ & $21.191$ & $10.501$\\
$0.1300 - 0.1400$ & $205.943$ & $20.292$ & $9.853$ & $6.225$ & $3.023$ & $21.226$ & $10.307$\\
$0.1400 - 0.1500$ & $178.935$ & $19.075$ & $10.660$ & $5.725$ & $3.200$ & $19.915$ & $11.130$\\

    \bottomrule
  \end{tabular}
    \end{small}
  \label{tab:pC30GeVRes}
\end{table*} 

\begin{table*}[ht]
 \caption{Measured differential cross-section for $120\:$GeV/c data.}
  \centering
  \begin{small}
  \begin{tabular}{cccccccc}
    \toprule
    $p^{2}\theta^2$ range  & $d\sigma/d(p^2\theta^2)$  & Stat. error & Stat. error & Syst. error & Syst. error & Tot. error & Tot. error \\
    
    [GeV/c] &  [mb$\cdot$(GeV/c)$^{-2}$] & [mb$\cdot$(GeV/c)$^{-2}$] & [\%] & [mb$\cdot$(GeV/c)$^{-2}$] & [\%] & [mb$\cdot$(GeV/c)$^{-2}$] & [\%] \\
    \midrule
$0.0000 - 0.0025$ & $2073312.212$ & $2521.099$ & $0.122$ & $53750.653$ & $2.593$ & $53809.744$ & $2.595$\\
$0.0025 - 0.0050$ & $10237.337$ & $177.832$ & $1.737$ & $266.174$ & $2.600$ & $320.114$ & $3.127$\\
$0.0050 - 0.0075$ & $6173.195$ & $137.968$ & $2.235$ & $160.449$ & $2.599$ & $211.611$ & $3.428$\\
$0.0075 - 0.0100$ & $4520.413$ & $118.183$ & $2.614$ & $117.526$ & $2.600$ & $166.672$ & $3.687$\\
$0.0100 - 0.0150$ & $2933.287$ & $67.383$ & $2.297$ & $76.317$ & $2.602$ & $101.807$ & $3.471$\\
$0.0150 - 0.0200$ & $1897.459$ & $54.169$ & $2.855$ & $49.458$ & $2.607$ & $73.351$ & $3.866$\\
$0.0200 - 0.0250$ & $1268.315$ & $44.400$ & $3.501$ & $33.121$ & $2.611$ & $55.393$ & $4.367$\\
$0.0250 - 0.0300$ & $945.892$ & $38.424$ & $4.062$ & $24.812$ & $2.623$ & $45.739$ & $4.836$\\
$0.0300 - 0.0350$ & $735.341$ & $33.883$ & $4.608$ & $19.339$ & $2.630$ & $39.014$ & $5.306$\\
$0.0350 - 0.0400$ & $521.670$ & $28.630$ & $5.488$ & $13.663$ & $2.619$ & $31.723$ & $6.081$\\
$0.0400 - 0.0450$ & $490.855$ & $27.789$ & $5.661$ & $12.930$ & $2.634$ & $30.650$ & $6.244$\\
$0.0450 - 0.0500$ & $390.669$ & $24.908$ & $6.376$ & $10.461$ & $2.678$ & $27.016$ & $6.915$\\
$0.0500 - 0.0550$ & $351.783$ & $23.664$ & $6.727$ & $9.422$ & $2.678$ & $25.470$ & $7.240$\\
$0.0550 - 0.0600$ & $334.268$ & $23.233$ & $6.951$ & $9.088$ & $2.719$ & $24.947$ & $7.463$\\
$0.0600 - 0.0700$ & $263.960$ & $14.687$ & $5.564$ & $7.130$ & $2.701$ & $16.327$ & $6.185$\\
$0.0700 - 0.0800$ & $206.983$ & $13.091$ & $6.325$ & $5.698$ & $2.753$ & $14.277$ & $6.898$\\
$0.0800 - 0.090$ & $216.332$ & $13.391$ & $6.190$ & $6.132$ & $2.834$ & $14.728$ & $6.808$\\
$0.0900 - 0.1000$ & $196.612$ & $12.853$ & $6.537$ & $5.723$ & $2.911$ & $14.070$ & $7.156$\\
$0.1000 - 0.1100$ & $199.731$ & $13.199$ & $6.608$ & $5.827$ & $2.918$ & $14.428$ & $7.224$\\
$0.1100 - 0.1200$ & $142.426$ & $11.054$ & $7.762$ & $4.012$ & $2.817$ & $11.760$ & $8.257$\\
$0.1200 - 0.1300$ & $158.757$ & $11.800$ & $7.433$ & $4.873$ & $3.070$ & $12.767$ & $8.042$\\
$0.1300 - 0.1400$ & $127.473$ & $10.478$ & $8.220$ & $3.985$ & $3.126$ & $11.210$ & $8.794$\\
$0.1400 - 0.1500$ & $150.903$ & $11.407$ & $7.559$ & $4.651$ & $3.082$ & $12.319$ & $8.164$\\

    \bottomrule
  \end{tabular}
    \end{small}
  \label{tab:pC120GeVRes}
\end{table*}

\end{document}

%% file: authors.tex
\newcommand{\FNAL}{Fermi National Accelerator Laboratory, Batavia, 
Illinois 60510, USA}
\newcommand{\TRIUMF}{TRIUMF, Vancouver, BC  V6T 2A3, Canada}
\newcommand{\KEK}{Institute of
Particle and Nuclear Study (IPNS), High Energy Accelerator Research
Organization (KEK), Tsukuba, 305-0801, Japan}
\newcommand{\IPMU}{Kavli Institute for the Physics and Mathematics of the Universe (WPI), The University of Tokyo Institutes for Advanced Study, University of Tokyo, Kashiwa, Chiba, Japan}
\newcommand{\Nagoya}{Nagoya University, Nagoya, Aichi 464-8601, Japan}
\newcommand{\Winnipeg}{University of Winnipeg, Winnipeg, MB R3B 2E9, Canada}
\newcommand{\Chiba}{Department of Physics, Chiba University, Chiba, Chiba 263-8522, Japan}
\newcommand{\Kobe}{Kobe University, Kobe, Hyogo 657-8501, Japan}
\newcommand{\Regina}{Department of Physics, University of Regina, SK, S4S 0A2, Canada}
\newcommand{\York}{Department of Physics and Astronomy, York University, Toronto, Ontario, Canada}
\newcommand{\NotreDame}{Department of Physics, University of Notre Dame, Notre Dame, Indiana 46556, USA}
%\newcommand{\Wichita}{Department of Mathematics, Statistics, and Physics, Wichita State University, Wichita, Kansas 67206, USA}
%\newcommand{\Boston}{Department of Physics, Boston University, Boston, Massachusetts, USA}
%\newcommand{\WilliamMary}{Department of Physics, College of William \& Mary, Williamsburg, Virginia 23187, USA}
%\newcommand{\Cincinnati}{Department of Physics, University of Cincinnati, Cincinnati, Ohio 45221, USA}
%\newcommand{\SDSMT}{South Dakota School of Mines and Technology, Rapid City, South Dakota 57701, USA}
%\newcommand{\UTA}{Department of Physics, University of Texas at Austin, Austin, TX 78712, USA}

%\affiliation{\Boston}
\affiliation{\Chiba}
%\affiliation{\Cincinnati}
\affiliation{\FNAL}
\affiliation{\KEK}
%\affiliation{\Sinica}
\affiliation{\IPMU}
\affiliation{\Kobe}
%\affiliation{\Kyoto}
\affiliation{\Nagoya}
%\affiliation{\OsakaCity}
\affiliation{\NotreDame}
%\affiliation{\Osaka}
%\affiliation{\RCNP}
\affiliation{\Regina}
%\affiliation{\Tohoku}
%\affiliation{\ELPH}
%\affiliation{\RIKEN}
%\affiliation{\SDSMT}
%\affiliation{\UTA}
\affiliation{\TRIUMF}
%\affiliation{\Wichita}
%\affiliation{\WilliamMary}
\affiliation{\Winnipeg}
\affiliation{\York}

\author{M.~Pavin}
\affiliation{\TRIUMF}

%\author{T.~Akaishi}
%\affiliation{\Osaka}

\author{L.~Aliaga-Soplin}
\affiliation{\FNAL}

%\author{H.~Asano}
%\affiliation{\RIKEN}

%\author{A.~Aurisano}
%\affiliation{\Cincinnati}

\author{M.~Barbi}
\affiliation{\Regina}

\author{L.~Bellantoni}
\affiliation{\FNAL}

\author{S.~Bhadra}
\affiliation{\York}

%\author{W-C.~Chang} 
%\affiliation{\York}

\author{B.~Ferrazzi}
\affiliation{\Regina}

\author{L.~Fields}
\affiliation{\FNAL}
\affiliation{\NotreDame}

\author{A.~Fiorentini} 
\thanks{now at South Dakota School of Mines and Technology}
\affiliation{\York}

%\author{M.~Friend}
%\affiliation{\KEK}

\author{T.~Fukuda}
\affiliation{\Nagoya}

%\author{D.~Harris}
%\affiliation{\FNAL}
\author{K.~Gameil}
\thanks{now at Carleton University, Ottawa, Ontario, Canada}
\affiliation{\TRIUMF}

\author{Y.~Al Hakim}
\thanks{now at Imperial College London, London, England}
\affiliation{\TRIUMF}

\author{M.~Hartz}
\affiliation{\TRIUMF}
\affiliation{\IPMU}

%\author{R.~Honda} 
%\affiliation{\Tohoku}

%\author{T.~Ishikawa} 
%\affiliation{\ELPH}

\author{B.~Jamieson}
\affiliation{\Winnipeg}

%\author{E.~Kearns}
%\affiliation{\Boston}
\author{M.~Kiburg}
\affiliation{\FNAL}

\author{N.~Kolev}
\affiliation{\Regina}

%\author{M.~Komatsu} 
%\affiliation{\Nagoya}

%\author{Y.~Komatsu} 
%\affiliation{\KEK}
\author{H.~Kawai}
\affiliation{\Chiba}

\author{A.~Konaka}
\affiliation{\TRIUMF}

%\author{M.~Kordosky}
%\affiliation{\WilliamMary}

%\author{K.~Lang}
%\affiliation{\UTA}

\author{P.~Lebrun}
\affiliation{\FNAL}

\author{T.~Lindner}
%\affiliation{\Winnipeg}
\affiliation{\TRIUMF}

%\author{Y.~Ma}
%\affiliation{\RIKEN}

%\author{D.~A.~Martinez Caicedo}
%\affiliation{\SDSMT}

%\author{M.~Muether}
%\affiliation{\Wichita}

\author{T.~Mizuno} 
\affiliation{\Chiba}

\author{N.~Naganawa} 
\affiliation{\Nagoya}

%\author{M.~Naruki} 
%\affiliation{\Kyoto}

%\author{E.~Niner} 
%\affiliation{\FNAL}

%\author{H.~Noumi}
%\affiliation{\RCNP}

%\author{K.~Ozawa} 
%\affiliation{\KEK}

\author{J.~Paley}
\affiliation{\FNAL}

%\author{P.~de Perio}
%\affiliation{\TRIUMF}

%\author{M.~Proga}
%\affiliation{\UTA}

%\author{F.~Sakuma}
%\affiliation{\RIKEN}
\author{R.~Rivera}
\affiliation{\FNAL}

\author{G.~Santucci} 
\affiliation{\York}

%\author{T.~Sawada} 
%\affiliation{\OsakaCity}

\author{O.~Sato} 
\affiliation{\Nagoya}

\author{T.~Sekiguchi}
\affiliation{\KEK}

\author{A.~Sikora}
\thanks{now at McGill University, Montreal, Quebec, Canada}
\affiliation{\Winnipeg}

\author{G.~Slater}
\affiliation{\TRIUMF}

%\author{K.~Shirotori}
%\affiliation{\RCNP}

\author{A.~Suzuki} 
\affiliation{\Kobe}

\author{T.~Sugimoto}
\affiliation{\Kobe}

\author{M.~Tabata}
\affiliation{\Chiba}

%\author{T.~Takahashi}
%\affiliation{\RCNP}

%\author{N.~Tomida}
%\affiliation{\RCNP}

%\author{R.~Wendell}
%\affiliation{\Kyoto}

\author{L.~Uplegger}
\affiliation{\FNAL}

\author{T.~Vladisavljevic}
\thanks{now at STFC, Rutherford Appleton Laboratory, Harwell Oxford, England}
\affiliation{\IPMU}

%\author{T.~Yamaga}
%\affiliation{\RIKEN}

%\collaboration{The EMPHATIC Collaboration}
%\noaffiliation

%\author{Author List to be included later}
%\affiliation{Institutional List}

%% file: main.bbl
%merlin.mbs apsrev4-1.bst 2010-07-25 4.21a (PWD, AO, DPC) hacked
%Control: key (0)
%Control: author (72) initials jnrlst
%Control: editor formatted (1) identically to author
%Control: production of article title (-1) disabled
%Control: page (0) single
%Control: year (1) truncated
%Control: production of eprint (0) enabled
\begin{thebibliography}{24}%
\makeatletter
\providecommand \@ifxundefined [1]{%
 \@ifx{#1\undefined}
}%
\providecommand \@ifnum [1]{%
 \ifnum #1\expandafter \@firstoftwo
 \else \expandafter \@secondoftwo
 \fi
}%
\providecommand \@ifx [1]{%
 \ifx #1\expandafter \@firstoftwo
 \else \expandafter \@secondoftwo
 \fi
}%
\providecommand \natexlab [1]{#1}%
\providecommand \enquote  [1]{``#1''}%
\providecommand \bibnamefont  [1]{#1}%
\providecommand \bibfnamefont [1]{#1}%
\providecommand \citenamefont [1]{#1}%
\providecommand \href@noop [0]{\@secondoftwo}%
\providecommand \href [0]{\begingroup \@sanitize@url \@href}%
\providecommand \@href[1]{\@@startlink{#1}\@@href}%
\providecommand \@@href[1]{\endgroup#1\@@endlink}%
\providecommand \@sanitize@url [0]{\catcode `\\12\catcode `\$12\catcode
  `\&12\catcode `\#12\catcode `\^12\catcode `\_12\catcode `\%12\relax}%
\providecommand \@@startlink[1]{}%
\providecommand \@@endlink[0]{}%
\providecommand \url  [0]{\begingroup\@sanitize@url \@url }%
\providecommand \@url [1]{\endgroup\@href {#1}{\urlprefix }}%
\providecommand \urlprefix  [0]{URL }%
\providecommand \Eprint [0]{\href }%
\providecommand \doibase [0]{http://dx.doi.org/}%
\providecommand \selectlanguage [0]{\@gobble}%
\providecommand \bibinfo  [0]{\@secondoftwo}%
\providecommand \bibfield  [0]{\@secondoftwo}%
\providecommand \translation [1]{[#1]}%
\providecommand \BibitemOpen [0]{}%
\providecommand \bibitemStop [0]{}%
\providecommand \bibitemNoStop [0]{.\EOS\space}%
\providecommand \EOS [0]{\spacefactor3000\relax}%
\providecommand \BibitemShut  [1]{\csname bibitem#1\endcsname}%
\let\auto@bib@innerbib\@empty
%</preamble>
\bibitem [{\citenamefont {Abe}\ \emph {et~al.}(2018)\citenamefont {Abe} \emph
  {et~al.}}]{H2KDesignReport}%
  \BibitemOpen
  \bibfield  {author} {\bibinfo {author} {\bibfnamefont {K.}~\bibnamefont
  {Abe}} \emph {et~al.} (\bibinfo {collaboration} {Hyper-Kamokande
  Proto-Collaboration}),\ }\href@noop {} {\  (\bibinfo {year} {2018})},\
  \Eprint {http://arxiv.org/abs/1805.04163} {arXiv:1805.04163
  [physics.ins-det]} \BibitemShut {NoStop}%
\bibitem [{\citenamefont {Abi}\ \emph {et~al.}(2020{\natexlab{a}})\citenamefont
  {Abi} \emph {et~al.}}]{DUNETDR}%
  \BibitemOpen
  \bibfield  {author} {\bibinfo {author} {\bibfnamefont {B.}~\bibnamefont
  {Abi}} \emph {et~al.},\ }\href@noop {} {\bibfield  {journal} {\bibinfo
  {journal} {JINST}\ }\textbf {\bibinfo {volume} {15}},\ \bibinfo {pages}
  {T08008} (\bibinfo {year} {2020}{\natexlab{a}})}\BibitemShut {NoStop}%
\bibitem [{\citenamefont {Abe}\ \emph {et~al.}(2013)\citenamefont {Abe} \emph
  {et~al.}}]{Abe:2012av}%
  \BibitemOpen
  \bibfield  {author} {\bibinfo {author} {\bibfnamefont {K.}~\bibnamefont
  {Abe}} \emph {et~al.} (\bibinfo {collaboration} {T2K}),\ }\href {\doibase
  10.1103/PhysRevD.87.012001} {\bibfield  {journal} {\bibinfo  {journal} {Phys.
  Rev. D}\ }\textbf {\bibinfo {volume} {87}},\ \bibinfo {pages} {012001}
  (\bibinfo {year} {2013})},\ \bibinfo {note} {[Addendum: Phys.Rev.D 87, 019902
  (2013)]},\ \Eprint {http://arxiv.org/abs/1211.0469} {arXiv:1211.0469
  [hep-ex]} \BibitemShut {NoStop}%
\bibitem [{\citenamefont {Abgrall}\ \emph
  {et~al.}(2016{\natexlab{a}})\citenamefont {Abgrall} \emph
  {et~al.}}]{Abgrall:2015hmv}%
  \BibitemOpen
  \bibfield  {author} {\bibinfo {author} {\bibfnamefont {N.}~\bibnamefont
  {Abgrall}} \emph {et~al.} (\bibinfo {collaboration} {NA61/SHINE}),\ }\href
  {\doibase 10.1140/epjc/s10052-016-3898-y} {\bibfield  {journal} {\bibinfo
  {journal} {Eur. Phys. J.}\ }\textbf {\bibinfo {volume} {C76}},\ \bibinfo
  {pages} {84} (\bibinfo {year} {2016}{\natexlab{a}})},\ \Eprint
  {http://arxiv.org/abs/1510.02703} {arXiv:1510.02703 [hep-ex]} \BibitemShut
  {NoStop}%
%%CITATION = ARXIV:1510.02703;%%
\bibitem [{\citenamefont {Abgrall}\ \emph
  {et~al.}(2016{\natexlab{b}})\citenamefont {Abgrall} \emph
  {et~al.}}]{Abgrall:2016jif}%
  \BibitemOpen
  \bibfield  {author} {\bibinfo {author} {\bibfnamefont {N.}~\bibnamefont
  {Abgrall}} \emph {et~al.} (\bibinfo {collaboration} {NA61/SHINE}),\ }\href
  {\doibase 10.1140/epjc/s10052-016-4440-y} {\bibfield  {journal} {\bibinfo
  {journal} {Eur. Phys. J.}\ }\textbf {\bibinfo {volume} {C76}},\ \bibinfo
  {pages} {617} (\bibinfo {year} {2016}{\natexlab{b}})},\ \Eprint
  {http://arxiv.org/abs/1603.06774} {arXiv:1603.06774 [hep-ex]} \BibitemShut
  {NoStop}%
%%CITATION = ARXIV:1603.06774;%%
\bibitem [{\citenamefont {Aduszkiewicz}\ \emph {et~al.}(2019)\citenamefont
  {Aduszkiewicz} \emph {et~al.}}]{Aduszkiewicz:2019xna}%
  \BibitemOpen
  \bibfield  {author} {\bibinfo {author} {\bibfnamefont {A.}~\bibnamefont
  {Aduszkiewicz}} \emph {et~al.} (\bibinfo {collaboration} {NA61/SHINE}),\
  }\href {\doibase 10.1103/PhysRevD.100.112001} {\bibfield  {journal} {\bibinfo
   {journal} {Phys. Rev. D}\ }\textbf {\bibinfo {volume} {100}},\ \bibinfo
  {pages} {112001} (\bibinfo {year} {2019})},\ \Eprint
  {http://arxiv.org/abs/1909.03351} {arXiv:1909.03351 [hep-ex]} \BibitemShut
  {NoStop}%
\bibitem [{\citenamefont {Abgrall}\ \emph {et~al.}(2019)\citenamefont {Abgrall}
  \emph {et~al.}}]{Berns:2018tap}%
  \BibitemOpen
  \bibfield  {author} {\bibinfo {author} {\bibfnamefont {N.}~\bibnamefont
  {Abgrall}} \emph {et~al.} (\bibinfo {collaboration} {NA61/SHINE}),\ }\href
  {\doibase 10.1140/epjc/s10052-019-6583-0} {\bibfield  {journal} {\bibinfo
  {journal} {Eur. Phys. J.}\ }\textbf {\bibinfo {volume} {C79}},\ \bibinfo
  {pages} {100} (\bibinfo {year} {2019})},\ \Eprint
  {http://arxiv.org/abs/1808.04927} {arXiv:1808.04927 [hep-ex]} \BibitemShut
  {NoStop}%
%%CITATION = ARXIV:1808.04927;%%
\bibitem [{\citenamefont {Apollonio}\ \emph
  {et~al.}(2009{\natexlab{a}})\citenamefont {Apollonio} \emph
  {et~al.}}]{Apollonio:2009bu}%
  \BibitemOpen
  \bibfield  {author} {\bibinfo {author} {\bibfnamefont {M.}~\bibnamefont
  {Apollonio}} \emph {et~al.} (\bibinfo {collaboration} {HARP}),\ }\href
  {\doibase 10.1016/j.nuclphysa.2009.01.080} {\bibfield  {journal} {\bibinfo
  {journal} {Nucl. Phys.}\ }\textbf {\bibinfo {volume} {A821}},\ \bibinfo
  {pages} {118} (\bibinfo {year} {2009}{\natexlab{a}})},\ \Eprint
  {http://arxiv.org/abs/0902.2105} {arXiv:0902.2105 [hep-ex]} \BibitemShut
  {NoStop}%
%%CITATION = ARXIV:0902.2105;%%
\bibitem [{\citenamefont {Apollonio}\ \emph
  {et~al.}(2009{\natexlab{b}})\citenamefont {Apollonio} \emph
  {et~al.}}]{Apollonio:2009en}%
  \BibitemOpen
  \bibfield  {author} {\bibinfo {author} {\bibfnamefont {M.}~\bibnamefont
  {Apollonio}} \emph {et~al.} (\bibinfo {collaboration} {HARP}),\ }\href
  {\doibase 10.1103/PhysRevC.80.035208} {\bibfield  {journal} {\bibinfo
  {journal} {Phys. Rev.}\ }\textbf {\bibinfo {volume} {C80}},\ \bibinfo {pages}
  {035208} (\bibinfo {year} {2009}{\natexlab{b}})},\ \Eprint
  {http://arxiv.org/abs/0907.3857} {arXiv:0907.3857 [hep-ex]} \BibitemShut
  {NoStop}%
%%CITATION = ARXIV:0907.3857;%%
\bibitem [{\citenamefont {Paley}\ \emph {et~al.}(2014)\citenamefont {Paley}
  \emph {et~al.}}]{MIPPNuMI}%
  \BibitemOpen
  \bibfield  {author} {\bibinfo {author} {\bibfnamefont {J.~M.}\ \bibnamefont
  {Paley}} \emph {et~al.} (\bibinfo {collaboration} {MIPP}),\ }\href {\doibase
  10.1103/PhysRevD.90.032001} {\bibfield  {journal} {\bibinfo  {journal} {Phys.
  Rev. D}\ }\textbf {\bibinfo {volume} {90}},\ \bibinfo {pages} {032001}
  (\bibinfo {year} {2014})},\ \Eprint {http://arxiv.org/abs/1404.5882}
  {arXiv:1404.5882 [hep-ex]} \BibitemShut {NoStop}%
\bibitem [{\citenamefont {Lebedev}(2007)}]{Lebedev:2007zz}%
  \BibitemOpen
  \bibfield  {author} {\bibinfo {author} {\bibfnamefont {A.~V.}\ \bibnamefont
  {Lebedev}},\ }\emph {\bibinfo {title} {{Ratio of pion kaon production in
  proton carbon interactions}}},\ \href {\doibase 10.2172/948174} {Ph.D.
  thesis},\ \bibinfo  {school} {Harvard U.} (\bibinfo {year}
  {2007})\BibitemShut {NoStop}%
\bibitem [{\citenamefont {Seun}(2007)}]{Seun:2007zz}%
  \BibitemOpen
  \bibfield  {author} {\bibinfo {author} {\bibfnamefont {S.~M.}\ \bibnamefont
  {Seun}},\ }\emph {\bibinfo {title} {{Measurement of $\pi-K$ ratios from the
  NuMI target}}},\ \href {\doibase 10.2172/935004} {Ph.D. thesis},\ \bibinfo
  {school} {Harvard U.} (\bibinfo {year} {2007})\BibitemShut {NoStop}%
\bibitem [{\citenamefont {Abi}\ \emph {et~al.}(2020{\natexlab{b}})\citenamefont
  {Abi} \emph {et~al.}}]{DUNESensStudies}%
  \BibitemOpen
  \bibfield  {author} {\bibinfo {author} {\bibfnamefont {B.}~\bibnamefont
  {Abi}} \emph {et~al.},\ }\href@noop {} {\bibfield  {journal} {\bibinfo
  {journal} {Eur. Phys. J. C}\ }\textbf {\bibinfo {volume} {80}},\ \bibinfo
  {pages} {978} (\bibinfo {year} {2020}{\natexlab{b}})}\BibitemShut {NoStop}%
\bibitem [{\citenamefont {Akaishi}\ \emph {et~al.}(2019)\citenamefont {Akaishi}
  \emph {et~al.}}]{Akaishi:2019dej}%
  \BibitemOpen
  \bibfield  {author} {\bibinfo {author} {\bibfnamefont {T.}~\bibnamefont
  {Akaishi}} \emph {et~al.} (\bibinfo {collaboration} {EMPHATIC}),\ }\href@noop
  {} {\  (\bibinfo {year} {2019})},\ \Eprint {http://arxiv.org/abs/1912.08841}
  {arXiv:1912.08841 [hep-ex]} \BibitemShut {NoStop}%
\bibitem [{\citenamefont {Burmistrov}\ \emph {et~al.}(2020)\citenamefont
  {Burmistrov} \emph {et~al.}}]{Burmistrov:2020dvn}%
  \BibitemOpen
  \bibfield  {author} {\bibinfo {author} {\bibfnamefont {L.}~\bibnamefont
  {Burmistrov}} \emph {et~al.},\ }\href {\doibase 10.1016/j.nima.2019.05.073}
  {\bibfield  {journal} {\bibinfo  {journal} {Nucl. Instrum. Meth. A}\ }\textbf
  {\bibinfo {volume} {958}},\ \bibinfo {pages} {162232} (\bibinfo {year}
  {2020})}\BibitemShut {NoStop}%
\bibitem [{\citenamefont {Agostinelli}\ \emph {et~al.}(2003)\citenamefont
  {Agostinelli} \emph {et~al.}}]{Geant4}%
  \BibitemOpen
  \bibfield  {author} {\bibinfo {author} {\bibfnamefont {S.}~\bibnamefont
  {Agostinelli}} \emph {et~al.},\ }\href {\doibase
  https://doi.org/10.1016/S0168-9002(03)01368-8} {\bibfield  {journal}
  {\bibinfo  {journal} {Nucl. Instrum. Meth. A}\ }\textbf {\bibinfo {volume}
  {506}},\ \bibinfo {pages} {250} (\bibinfo {year} {2003})}\BibitemShut
  {NoStop}%
\bibitem [{\citenamefont {Bellettini}\ \emph {et~al.}(1966)\citenamefont
  {Bellettini}, \citenamefont {Cocconi}, \citenamefont {Diddens}, \citenamefont
  {Lillethun}, \citenamefont {Matthiae}, \citenamefont {Scanlon},\ and\
  \citenamefont {Wetherell}}]{Bellettini:1966zz}%
  \BibitemOpen
  \bibfield  {author} {\bibinfo {author} {\bibfnamefont {G.}~\bibnamefont
  {Bellettini}}, \bibinfo {author} {\bibfnamefont {G.}~\bibnamefont {Cocconi}},
  \bibinfo {author} {\bibfnamefont {A.}~\bibnamefont {Diddens}}, \bibinfo
  {author} {\bibfnamefont {E.}~\bibnamefont {Lillethun}}, \bibinfo {author}
  {\bibfnamefont {G.}~\bibnamefont {Matthiae}}, \bibinfo {author}
  {\bibfnamefont {J.}~\bibnamefont {Scanlon}}, \ and\ \bibinfo {author}
  {\bibfnamefont {A.}~\bibnamefont {Wetherell}},\ }\href {\doibase
  10.1016/0029-5582(66)90267-7} {\bibfield  {journal} {\bibinfo  {journal}
  {Nucl. Phys.}\ }\textbf {\bibinfo {volume} {79}},\ \bibinfo {pages} {609}
  (\bibinfo {year} {1966})}\BibitemShut {NoStop}%
\bibitem [{\citenamefont {Adamczyk}(2015)}]{Adamczyk:2015gfy}%
  \BibitemOpen
  \bibfield  {author} {\bibinfo {author} {\bibfnamefont {L.}~\bibnamefont
  {Adamczyk}} (\bibinfo {collaboration} {ATLAS}),\ }\href {\doibase
  10.22323/1.247.0061} {\bibfield  {journal} {\bibinfo  {journal} {PoS}\
  }\textbf {\bibinfo {volume} {DIS2015}},\ \bibinfo {pages} {061} (\bibinfo
  {year} {2015})}\BibitemShut {NoStop}%
\bibitem [{\citenamefont {Aaboud}\ \emph {et~al.}(2016)\citenamefont {Aaboud}
  \emph {et~al.}}]{Aaboud:2016ijx}%
  \BibitemOpen
  \bibfield  {author} {\bibinfo {author} {\bibfnamefont {M.}~\bibnamefont
  {Aaboud}} \emph {et~al.} (\bibinfo {collaboration} {ATLAS}),\ }\href
  {\doibase 10.1016/j.physletb.2016.08.020} {\bibfield  {journal} {\bibinfo
  {journal} {Phys. Lett. B}\ }\textbf {\bibinfo {volume} {761}},\ \bibinfo
  {pages} {158} (\bibinfo {year} {2016})},\ \Eprint
  {http://arxiv.org/abs/1607.06605} {arXiv:1607.06605 [hep-ex]} \BibitemShut
  {NoStop}%
\bibitem [{\citenamefont {Kopeliovich}\ and\ \citenamefont
  {Tarasov}(2001)}]{Kopeliovich:2000ez}%
  \BibitemOpen
  \bibfield  {author} {\bibinfo {author} {\bibfnamefont {B.}~\bibnamefont
  {Kopeliovich}}\ and\ \bibinfo {author} {\bibfnamefont {A.}~\bibnamefont
  {Tarasov}},\ }\href {\doibase 10.1016/S0370-2693(00)01316-2} {\bibfield
  {journal} {\bibinfo  {journal} {Phys. Lett. B}\ }\textbf {\bibinfo {volume}
  {497}},\ \bibinfo {pages} {44} (\bibinfo {year} {2001})},\ \Eprint
  {http://arxiv.org/abs/hep-ph/0010062} {arXiv:hep-ph/0010062} \BibitemShut
  {NoStop}%
\bibitem [{\citenamefont {Kopeliovich}\ and\ \citenamefont
  {Trueman}(2001)}]{Kopeliovich:2000kz}%
  \BibitemOpen
  \bibfield  {author} {\bibinfo {author} {\bibfnamefont {B.}~\bibnamefont
  {Kopeliovich}}\ and\ \bibinfo {author} {\bibfnamefont {T.}~\bibnamefont
  {Trueman}},\ }\href {\doibase 10.1103/PhysRevD.64.034004} {\bibfield
  {journal} {\bibinfo  {journal} {Phys. Rev. D}\ }\textbf {\bibinfo {volume}
  {64}},\ \bibinfo {pages} {034004} (\bibinfo {year} {2001})},\ \Eprint
  {http://arxiv.org/abs/hep-ph/0012091} {arXiv:hep-ph/0012091} \BibitemShut
  {NoStop}%
\bibitem [{\citenamefont {Schiz}(1979)}]{Schiz:1979af}%
  \BibitemOpen
  \bibfield  {author} {\bibinfo {author} {\bibfnamefont {A.~M.}\ \bibnamefont
  {Schiz}},\ }\emph {\bibinfo {title} {{Hadron - Nucleus Scattering at
  70-GeV/c, 125-GeV/c and 175-GeV/c and a High Statistics Study of Hadron -
  Proton Elastic Scattering at 200-GeV/c}}},\ \href {\doibase 10.2172/1434691}
  {Ph.D. thesis},\ \bibinfo  {school} {Yale U.} (\bibinfo {year}
  {1979})\BibitemShut {NoStop}%
\bibitem [{\citenamefont {Denisov}\ \emph {et~al.}(1973)\citenamefont {Denisov}
  \emph {et~al.}}]{Denisov:1973zv}%
  \BibitemOpen
  \bibfield  {author} {\bibinfo {author} {\bibfnamefont {S.}~\bibnamefont
  {Denisov}} \emph {et~al.},\ }\href {\doibase 10.1016/0550-3213(73)90351-9}
  {\bibfield  {journal} {\bibinfo  {journal} {Nucl.\ Phys.}\ }\textbf {\bibinfo
  {volume} {B61}},\ \bibinfo {pages} {62} (\bibinfo {year} {1973})}\BibitemShut
  {NoStop}%
%%CITATION = NUPHA,B61,62;%%
\bibitem [{\citenamefont {Mahajan}\ and\ \citenamefont
  {Raja}(2013)}]{Mahajan:2013awa}%
  \BibitemOpen
  \bibfield  {author} {\bibinfo {author} {\bibfnamefont {S.}~\bibnamefont
  {Mahajan}}\ and\ \bibinfo {author} {\bibfnamefont {R.}~\bibnamefont {Raja}}
  (\bibinfo {collaboration} {MIPP}),\ }in\ \href@noop {} {\emph {\bibinfo
  {booktitle} {{Meeting of the APS Division of Particles and Fields}}}}\
  (\bibinfo {year} {2013})\ \Eprint {http://arxiv.org/abs/1311.2258}
  {arXiv:1311.2258 [hep-ex]} \BibitemShut {NoStop}%
\end{thebibliography}%
